\documentclass[letterpaper,english,aps,prb,preprint]{revtex4-1}
\usepackage[T1]{fontenc}
\usepackage[latin9]{inputenc}
\usepackage{amsmath}
\usepackage{amssymb}
\usepackage{graphicx}

\usepackage{babel}
\begin{document}

\title{Non-local spin valve in Van der Pauw cross geometry with four ferromagnetic
electrodes. }

\author{K.-V. Pham}

\email{pham@lps.u-psud.fr}

\selectlanguage{english}%

\affiliation{Laboratoire de Physique des Solides, Univ. Paris-Sud, CNRS, UMR 8502,
F-91405 Orsay Cedex, France}
\begin{abstract}
We consider a non-local spin valve in a Van der Pauw cross geometry
with four ferromagnetic electrodes. Two antiparallel ferromagnets
are used as (charge) source and drain while the detector circuit involves
measuring the voltage between two collinear ferromagnets with parallel
or antiparallel magnetizations. We find a potentially large increase
of the non-local spin voltage. The setup displays several additional
interesting properties: (i) infinite GMR for the non-local resistance
(if a symmetry requirement for the device is met); (ii) ON-OFF switch
effect, when the injector electrodes are parallel instead of antiparallel;
(iii) insensitivity to offset voltages. The device can additionally
be used as a Direct Spin Hall Effect probe and as a reprogrammable
magneto-logic gate implementing basic operations (NOR, NAND, inverter,
AND, OR, etc).
\end{abstract}


\maketitle

\section{Introduction.\label{sec:Introduction.}}

Pure spin manipulation is an important topic in spintronics due to
possible applications for programmable logic and memory. Non-local
spin valves \cite{johnson_interfacial_1985,johnson_spin_1993,johnson_bipolar_1993,jedema_electrical_2001,jedema_electrical_2002,takahashi_spin_2006}
provide an example of pure spin current generation. In the latter
a pure spin current generated at a ferromagnetic - paramagnetic interface
reaches a ferromagnetic probe in the absence of charge current; the
spin accumulation created in the probe vicinity can then be detected
as a charge voltage by virtue of Johnson-Silsbee charge-spin coupling\cite{johnson_interfacial_1985,johnson_thermodynamic_1987,johnson_coupling_1988}.
Such pure spin currents have already proved useful to switch magnetizations\cite{kimura_switching_2006,yang_giant_2008}.
This is opening promises for further applications in logic, sensing
and memory devices but to that end it is desirable to increase the
signals: the spin voltages are typically in the $\mu V$ range while
$mV$ would be more suitable to ensure sufficient $SNR$ (signal to
noise ratio). 

While spin valves are in everyday use in hard-drives, in order to
reach or even go beyond the $1\; Tbit/inch^{2}$ density in hard-drives\cite{takagishi_magnetoresistance_2010,nagasaka_cpp-gmr_2009,wood_2009},
novel spin valves are required with a low $RA$ resistance times area
product (typically $RA\leq0.1\;\Omega\mu m^{2}$)\cite{nagasaka_cpp-gmr_2009,katine_device_2008,wood_2009}
while sustaining a $mA$ current and $mV$ voltage; this is beyond
TMR (tunneling magnetoresistance) sensors capability since they are
too resistive. Metallic spin valves are therefore more appropriate.
However in order for them to have a suitable $SNR$ it is also necessary
that the read-heads function with a large enough contrast\cite{nagasaka_cpp-gmr_2009}:
$\Delta R\, A\geq5\; m\Omega\mu m^{2}$. Metallic spin valves with
larger GMR ratio $\Delta R/R$ are therefore required.

At first sight non-local metallic spin valves are not obvious candidates
for larger MR (magnetoresistance) ratios: they indeed underperform
when compared to their local counterparts; using the same materials
and dimensions a local $CPP$ spin valve is expected to have a larger
$\Delta R\, A$ since spin confinement is better\cite{fert_spin_2002,george_direct_2003,fert_semiconductors_2007}. 

The goal of this paper is to discuss a non-local spintronics device
with potentially:
\begin{itemize}
\item enhanced spin voltage in the $mV$ range for currents $\sim mA$ (so
that the non-local resistance variation $\Delta R_{nl}$ is in the
Ohm range) with realistic density currents $j<10^{8}\; A/cm^{2}$
addressing the needs of industry.
\item enhanced non-local $GMR$ ratio: $\Delta R_{nl}/R_{nl}$ (up to $100\%$
for the pessimistic ratio; or up to infinity for the optimistic ratio),
helping quite generally for better SNR and perhaps making them suitable
candidates as sensors or read-heads for hard-drive areal densities
larger than $1\; Tbit/inch^{2}$.
\end{itemize}
Regarding the enhancement of the non-local signal, spin valves with
tunnel junctions have been reported with non-local resistance in the
$\Omega$ range but due to a polarization decreasing rapidly when
the current is larger than $\sim\mu A$ , the spin voltage remains
small in the usual $\mu V$ range\cite{valenzuela_spin_2005,valenzuela_spin-polarized_2004}.
However much progress has been reported recently in pure metallic
lateral spin valves ($\sim10\;\mu V$ for nanopillars\cite{yang_giant_2008})
or lateral valves with very low resistance tunnel junctions (using
a thin $nm$ $MgO$ layer)\cite{wakamura_large_2011,fukuma_giant_2011,fukuma_enhanced_2010},
reaching in the latter case the $100\;\mu V$ range with $RA\sim0.2\;\Omega\mu m^{2}$
so that already $\Delta R\, A$ is of the order of a few $1\; m\Omega\mu m^{2}$. 

We propose to go even further in the improvement by relying on two
ideas: (i) use two injectors instead of a single one, which should
at face value double the signal; (ii) enhance the spin confinement
by making good use of tunnel barriers, thin enough to stay close to
the metallic regime but resistive enough to hinder spin leakage. The
idea of minimizing the spin relaxation volume has been expressed in
particular in\cite{jaffres_spin_2010,fert_semiconductors_2007} and
explains the large signals seen in spin valves using carbon nanotubes\cite{hueso_transformation_2007}. 

Our basic setup applies these ideas by using four collinear ferromagnetic
terminals. In a standard lateral spin valve\cite{johnson_spin_1993,johnson_bipolar_1993,jedema_spin_2003}
the charge current flows from a ferromagnetic electrode to a paramagnetic
drain; the injector electrode is connected by a lateral wire to another
ferromagnetic electrode used as a detector. In our setup we propose
to replace the paramagnetic drain by a ferromagnet antiparallel to
the terminal acting as current source; the two antiparallel electrodes
are connected by a paramagnetic metal with thin tunnel barriers in
order to better confine spin. The two antiparallel ferromagnets act
as spin sources although in terms of charge one is a source and the
other a drain: this effectively doubles the spin accumulation in the
lateral wire while the tunnel junctions make sure spin is confined. 

We further change the standard detection setup by using a ferromagnetic
counter-electrode instead of a paramagnetic one. The advantage of
using two ferromagnets is evidenced when the two detector electrodes
are placed symmetrically with respect to the injectors (source and
drain): provided they are otherwise identical terminals this implies
that when their magnetizations are parallel, their voltage difference
should be identically zero by symmetry. This is how we reach an infinite
non-local $GMR$ ratio. 

We will also address the issue of voltage offsets plaguing non-local
setups\cite{johnson_interfacial_1985,johnson_spin_1993,jedema_electrical_2001,fert_spin_2002,jedema_electrical_2002,george_direct_2003}:
while voltages generated by spin accumulation are clearly observed,
some additional voltages of various origins are also usually seen.
These offset (or baseline) voltages have been credited to charge current
inhomogeneities\cite{ichimura_geometrical_2004,hamrle_current_2005,hamrle_three-dimensional_2005,johnson_calculation_2007,bass_spin-diffusion_2007}
(which impact the calculations done for non-local setups since they
usually assume one dimensional drift-diffusion equations\cite{johnson_thermodynamic_1987,valet_theory_1993,rashba_diffusion_2002}),
or to heating (notably Joule and Peltier heating\cite{garzon_temperature-dependent_2005,casanova_control_2009,bakker_interplay_2010}).
They may or may not be a nuisance but at any rate they prevent observation
of pure non-local voltages. The device we discuss in this paper can
be made insensitive to these offset voltages when the two detector
electrodes are identical and symmetric since the offsets will cancel
out when the voltage difference is measured. This is an additional
advantage of our device.

The geometry of our device is that of a Van der Pauw cross as in the
Jedema and coll. seminal experiments\cite{jedema_electrical_2001}.
A close device within a pure lateral geometry will be discussed elsewhere\cite{kv}. 

In Section \ref{sec:Van-der-Pauw} we introduce the Van der Pauw geometry
with four ferromagnetic terminals and give general expressions for
the non-local voltage. The basic functionalities of the device are
discussed, and notably we will show that the device can perform logic
operations (notably as a NOR or NAND gate), be reprogrammed to perform
other functions (AND, XOR and inverter gates), displays a potentially
interesting ON-OFF switch effect, and when used as a standard 1-bit
read-head shows an infinite $GMR$ for the non-local resistance. Use
as a Direct Spin Hall Effect probe will also be discussed. 

The next section \ref{sec:2f} studies in detail the impact of the
transparency of interfaces and of the number of ferromagnetic terminals
(two or three out of four) on spin confinement, resulting in small
or large non-local signals. The signals expected are systematically
compared to those in the standard lateral geometry. 

The last section \ref{sec:4f} discusses the main setup with four
ferromagnetic symmetric terminals since it displays the previously
mentioned properties of (i) immunity to offset voltages and (ii) infinite
$GMR$ ratio for the non-local resistance. Issues pertaining to the
use as a sensor are briefly touched upon.

The bulk of calculations are relegated to the Appendices. Appendix
\ref{sec:Revisiting-the-bipolar} revisits the bipolar spin switch
calculations by including spin leakage in the measuring electrodes.

\section{Van der Pauw setup.\label{sec:Van-der-Pauw}}

\subsection{Geometry and notations.\label{sub:Geometry-and-notations.}}

\subsubsection{Geometry\label{sub:Geometry.}.}

We consider in this section a four-terminal device in a Van der Pauw
geometry (see Fig. \ref{fig:Van-der-Pauw4Funequal}). The terminals
are ferromagnets $F1-F4$ positioned as in the Figure \ref{fig:Van-der-Pauw4Funequal};
we allow for arms of unequal lengths. In sections \ref{sec:2f}-\ref{sub:3f}
we will allow some of these terminals to be paramagnetic through a
suitable choice of parameters.

\textbf{One-dimensional assumption. }We will assume that width and
thickness of all arms are much smaller than their length. Experimentally,
current inhomogeneities due to departures from strict one-dimensional
flow can arise; however the basic functionalities of our device are
for the most part independent of that assumption although quantitative
predictions may accordingly lose accuracy. 

\textbf{Injector electrodes.} $F1$ injects a charge current which
is collected in terminal $F2$. We will designate them collectively
as injector electrodes; when the need to differentiate them shows
up, we will say that $F1$ is the source or injector electrode while
$F2$ is the drain or collector electrode.

\textbf{Detector electrodes.} The detection sub-setup consists in
terminals $F3$ and $F4$ hooked to a voltmeter (or a potentiometer
or an ammeter). The latter will measure the non-local voltage as a
function of the magnetization orientations of each terminal.

\textbf{Orientation.} We define points $O(x=0;\; z=0)$ the origin
and center of the Van der Pauw cross, $A(x=0;\: z=L_{1})$, $B(x=0;\: z'=L_{2})$
, $C(x=L_{3};\: z=0)$ and $D(x=0;\: z'=L_{4})$ where each arm has
been for later convenience \emph{oriented away} from $O$ (axis $Ox$,
$Ox'$, $Oz$ and $Oz'$ ) following Jedema and coll.\cite{jedema_electrical_2001}.

The four paramagnetic arms are: $I-IV$ (resp. $OA$, $OB$, $OC$,
$OD$).

\begin{center}
\begin{figure}
\begin{centering}
\includegraphics[width=1\columnwidth]{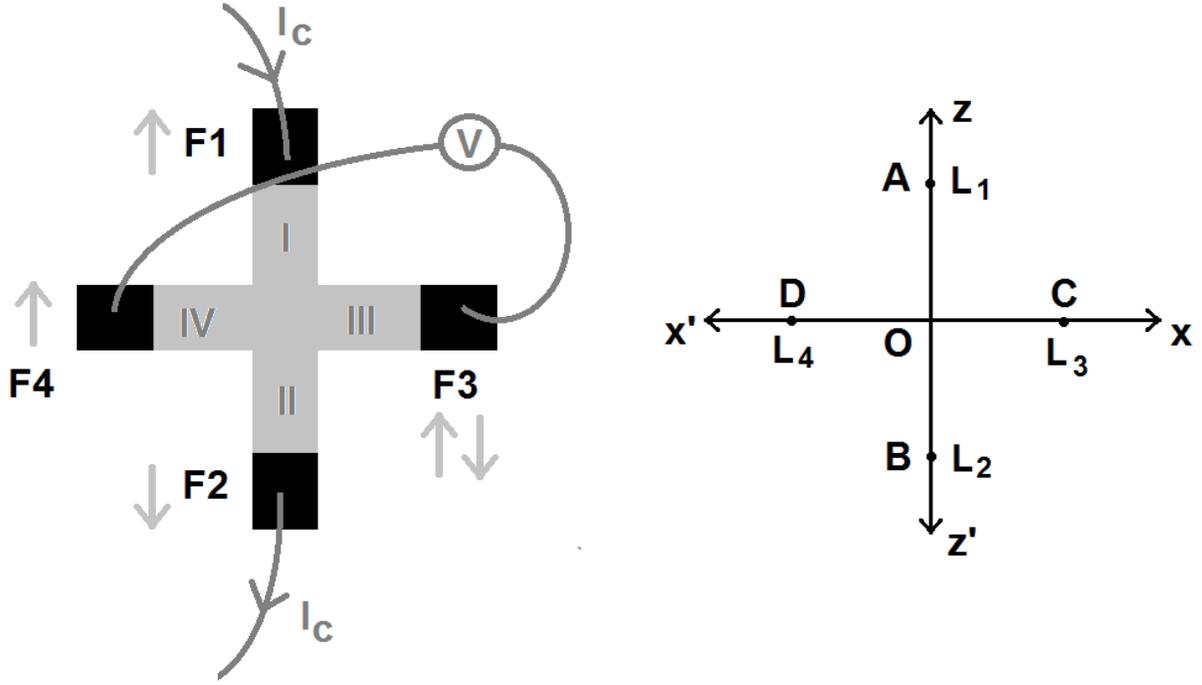}
\par\end{centering}

\caption{Van der Pauw cross with four ferromagnetic terminals. The arm lengths
are respectively $L_{1}$, $L_{2}$, $L_{3}$ and $L_{4}$. The arrows
represent the magnetization direction. When the device is used as
a basic spin-valve, only one terminal can switch its magnetization
(here $F3$). The arms are oriented \emph{away} from origin $O$.
\label{fig:Van-der-Pauw4Funequal}}
\end{figure}

\par\end{center}

The charge current flowing through $F1-I-II-F2$ is $I_{c}$ and flows
from top to bottom (is therefore negative relative to arm $I$, but
positive, relative to arm $II$).

The spin accumulation is defined as: 
\begin{equation}
\Delta\mu=\frac{\left(\mu_{\uparrow}-\mu_{\downarrow}\right)}{2e}
\end{equation}
 (where for later convenience we have divided by the electron charge
$e$).

The spin currents are oriented away from origin $O$ (therefore are
counted positive on a given arm when flowing \emph{away} from $O$).

\subsubsection{Spin parameters\label{sub:Spin-parameters.}.}

\textbf{Arms parameters.} The central cross is a normal metal. Its
parameters are its \emph{spin resistance} $R_{N}=\rho_{N}^{*}l_{N}/A_{N}$
where $l_{N}$ is the spin diffusion length, $A_{N}$ is the cross-section
and $\rho_{N}^{*}$ is the resistivity. We define the lengths of each
arm relative to the spin diffusion length as (for $i=1-4$):
\begin{equation}
l_{i}=\frac{L_{i}}{l_{N}}.
\end{equation}

\textbf{Ferromagnet parameters.} For each ferromagnetic terminal $F1-F4$
($i=1-4$), one defines the conductivity polarization $P_{F,i}$ ($=\beta_{i}$
in Valet-Fert notation\cite{valet_theory_1993}), \emph{spin resistance}
$R_{F,i}=\rho_{F,i}^{*}l_{F,i}/A_{F,i}$ where $l_{F,i}$ is the spin
diffusion length, $A_{F,i}$ is the cross-section and $\rho_{F,i}^{*}=\left(\rho_{i\uparrow+}\rho_{i\downarrow}\right)/4$
.

(NB: note on terminology; we will call throughout the paper 'spin
resistance' the characteristic resistance found as the product of
the resistivity times the spin diffusion length divided by the cross-section
).

\textbf{Interface parameters.} At the interface between the ferromagnets
and the paramagnetic arms we assume there is a spin dependent interface
resistance so that one can define for each interface $F_{i}-N$ ($i=1-4$)
a conductance polarization $P_{ci}$ ($=\gamma_{i}$ in Valet-Fert
notation), a \emph{spin resistance} $R_{ci}=\left(R_{i\uparrow+}R_{i\downarrow}\right)/4$.
For simplicity we will neglect all spin flips at interfaces so that
spin relaxation occurs solely in the bulk of the device.

\subsubsection{Spin resistance mismatch\label{subsec:Spin-resistance-mismatch.}.}

We define spin resistance mismatch parameters at F - N interfaces
as:
\begin{equation}
X=\frac{R_{F}+R_{c}}{R_{N}}.
\end{equation}

Three limits can be singled out:

$X<1$: this corresponds to the limit of a transparent junction, which
as we will see later in detail (Appendix \ref{sub:spin_currents-detectors})
is very leaky in terms of spin: this favors large spin currents at
the cost of reduced spin accumulations in the central paramagnet.

$X>1$: spin confining or tunneling regime, for which spin accumulation
increases but spin current decreases (in magnitude) (see Appendix
\ref{sub:spin_currents-detectors}). For $X\sim10$, for which the
contact resistance is moderate (about $10\;\Omega$) we will say that
we are in the weak tunneling limit. This is the most interesting limit
in terms of applications to all-metallic read-heads or sensors. For
$R_{c}=10^{2}-10^{4}\;\Omega$ which are usual values in tunnel junctions,
$X\sim10^{2}-10^{4}$ which we will qualify as strong tunneling limit.
Although the strong tunneling regime can be described by our equations,
we will focus primarily in the discussions on the transparent and
weak tunneling regime where resistances are in the metallic range
which interests us for sensor applications.

$X=1$: spin impedance matching. The naming for this border situation
will be justified below in the discussion on effective spin resistance
(section \ref{sub:Effective-spin-resistances.}).

\subsubsection{Effective polarizations\label{sub:Effective-polarizations.}. }

The following definition will also prove useful. We define for each
terminal ($i=1-4$) an effective polarization as:
\begin{equation}
P_{eff,i}=\frac{\widetilde{PR_{i}}}{\delta_{i}^{+}}
\end{equation}
where: 
\begin{eqnarray}
\widetilde{PR_{i}} & = & \left(P_{Fi}R_{Fi}+P_{ci}R_{ci}\right)/R_{N}
\end{eqnarray}

and: 
\begin{equation}
\delta_{i}^{\pm}=\frac{\left(X_{i}+1\right)}{2}\;\exp l_{i}\pm\frac{\left(X_{i}-1\right)}{2}\;\exp-l_{i}.\label{eq:delta}
\end{equation}

The effective polarization can be rewritten as:
\begin{equation}
P_{eff,i}=\frac{P_{Fi}R_{Fi}+P_{ci}R_{ci}}{R_{N}\sinh l_{i}+\left(R_{Fi}+R_{ci}\right)\cosh l_{i}};
\end{equation}
 clearly, $\left|P_{eff,i}\right|\leq1$.

Upon magnetization reversal of the electrode, the effective polarization
is an odd function: 
\[
P_{eff,i}\longrightarrow-P_{eff,i}.
\]

In the limit of short arm length $l_{i}\ll1$: 
\begin{equation}
P_{eff,i}\longrightarrow\frac{P_{Fi}R_{Fi}+P_{ci}R_{ci}}{R_{Fi}+R_{ci}}
\end{equation}
which is a weighted average of the electrode bulk and interface polarizations. 

When $l_{i}\longrightarrow\infty$ the effective polarization vanishes
exponentially which translates the complete spin relaxation in the
arm: 
\begin{equation}
P_{eff,i}=2\frac{P_{Fi}R_{Fi}+P_{ci}R_{ci}}{R_{N}+\left(R_{Fi}+R_{ci}\right)}\;\exp-l_{i}.
\end{equation}
The effective polarization therefore varies between $0$ and its maximum
value $\frac{P_{Fi}R_{Fi}+P_{ci}R_{ci}}{R_{Fi}+R_{ci}}$ which is
bounded from above by $\sup\left(P_{Fi};\; P_{ci}\right)$.

\subsubsection{Effective spin resistances\label{sub:Effective-spin-resistances.}. }

We also define an effective spin resistance for each arm of length
$l_{i}$ ($i=1-4$) as:

\begin{eqnarray}
R_{eff,i}(l_{i},\; X_{i}) & = & R_{N}\;\frac{\delta_{i}^{+}}{\delta_{i}^{-}}\\
 & = & R_{N}\;\frac{X_{i}\,\cosh l_{i}+\sinh l_{i}}{X_{i}\,\sinh l_{i}+\cosh l_{i}};
\end{eqnarray}

$R_{eff,i}(X_{i})$ is an increasing function of spin resistance mismatch
$X_{i}$.

It is shown in Appendix \ref{sub:Spin-currents} that the spin current
$I_{s,i}(O)$ at the cross center $O$ on arm $i$ and the spin accumulation
there are related by $\Delta\mu(O)=-R_{eff,i}\; I_{s,i}(O)$ (for
arms $III-IV$; for arms $I-II$ a more general relation taking into
account the spin injection at $F1$ and $F2$ holds). This is an analog
of Ohm's law for spin which explains our identification of $R_{eff,i}$
as a spin resistance. (Note that the analogy is not complete: the
relation for spin is a local one (expressed here at point $O$), while
Ohm's law holds for a voltage difference and is therefore non-local.
This results of course from the non-conservation of spin current.)

We also define a total effective spin resistance for the device:

\begin{equation}
R_{eff}=\frac{1}{\sum_{i=1-4}\;\frac{1}{R_{eff,i}}}.
\end{equation}

The previous expression admits obvious generalization to an arbitrary
number $n\geq4$ of arms.

As can be seen from its definition, the total effective resistance
$R_{eff}$ is related to the arms spin resistances $R_{eff,i}$ by
the analog of a parallel resistance addition law. We will show later
(section \ref{subsec:Spin-voltage-andnonlocalresistance}) that the
spin voltage is proportional to the total effective spin resistance
$R_{eff}$ so that large effective resistances are desirable; this
will also effectively demonstrate for our geometry the parallel addition
law for spin resistances. (For a discussion of spin resistance addition
law at nodes we refer the reader to \cite{kimura_estimation_2005}.)
Since the total effective spin resistance is the sum of four resistances
in parallel, whenever one is much smaller than the others, it will
short the other arms: spin leakage will be stronger so that spin accumulation
will be reduced. 

\begin{figure}
\begin{centering}
\includegraphics[width=1\columnwidth]{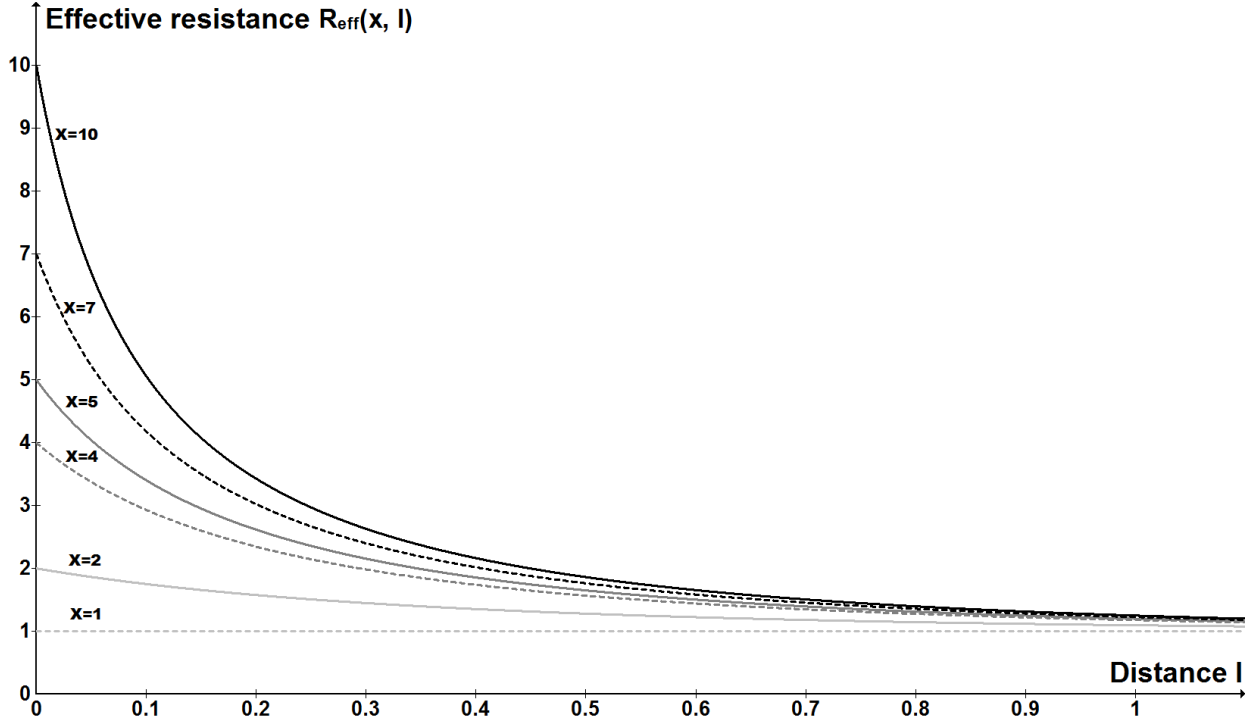}
\par\end{centering}

\caption{Effective resistance for one terminal (normalized to $R_{N}$) as
a function of arms length $l$ (relative to paramagnet spin relaxation
length) for various impedance mismatches $X=1-10$ in the (weak) tunneling
regime\label{fig:Effective-resistance1}.}

\end{figure}

\begin{figure}
\begin{centering}
\includegraphics[width=1\columnwidth]{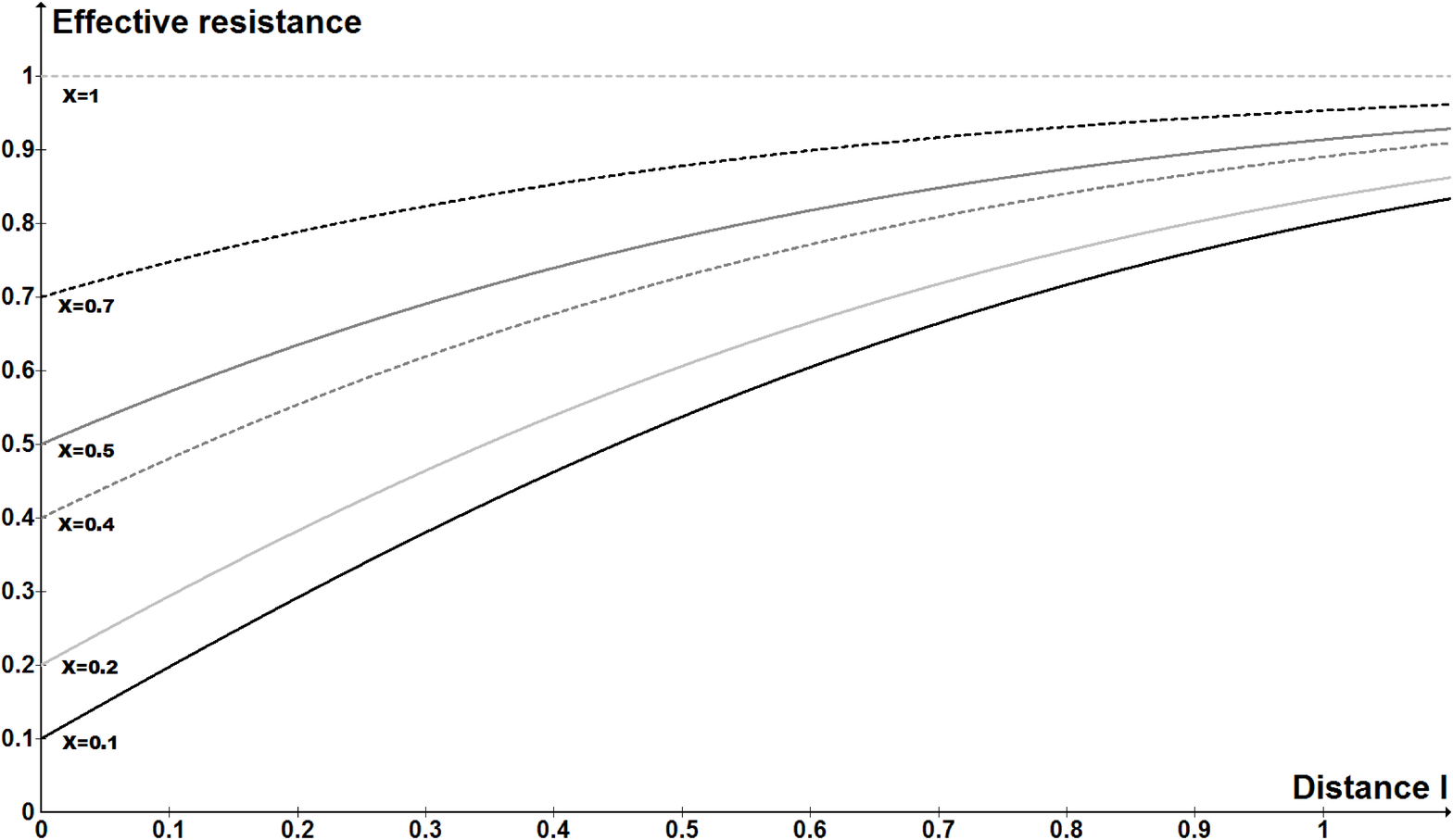}
\par\end{centering}

\caption{Effective resistance for one terminal (normalized to $R_{N}$) as
a function of arms length $l$ (relative to paramagnet spin relaxation
length) for various impedance mismatches $X=0.1-1$ in the transparent
regime ($X<1$)\label{fig:Effective-resistance2}.}
\end{figure}

The length dependence of the effective resistance for various values
of the spin resistance mismatch $X$ is shown in Fig. \ref{fig:Effective-resistance1}-\ref{fig:Effective-resistance2}
. At large distance the effective resistance converges exponentially
fast to $R_{n}$ the spin resistance of the paramagnetic arm:
\begin{equation}
R_{eff,i}\longrightarrow R_{N}
\end{equation}
which reflects the fact that the spin relaxation is dominated by the
paramagnet bulk. In the limit $l_{i}\longrightarrow0$ 
\begin{equation}
R_{eff,i}\longrightarrow R_{Fi}+R_{ci}
\end{equation}
(which is sensible since the paramagnet is then too short for spin
relaxation to occur). 

When $l_{i}\ll1$, the effective resistance remains close to $R_{Fi}+R_{ci}$
if the distance $l$ obeys:
\[
l\;\left|X^{2}-1\right|<X.
\]
When $X\gg1$, the condition becomes $l\ll1/X$ which reflects a steeper
exponential decrease.

The effective spin resistance for a given arm is therefore comprised
between $R_{N}$ and $R_{F,i}+R_{c,i}$: 
\begin{equation}
R_{N}\leq R_{eff,i}\leq R_{Fi}+R_{ci}
\end{equation}
(or the reverse inequality if $R_{N}\geq R_{Fi}+R_{ci}$).

As a rule the effective spin resistance will be larger for large interface
or ferromagnet spin resistance ($R_{c}$ and $R_{F}$); since the
ferromagnet spin resistance $R_{F}$ is in general much smaller than
the paramagnet spin resistance $R_{N}$ due to short spin diffusion
lengths, large interface resistances $R_{c}$ are required to achieve
large effective resistances $R_{eff,i}$.

For $X=1$, one observes that the effective spin resistance does not
depend any more on the arm length $l_{i}$ and is equal to $R_{N}$.
One then has (on arms $III$ or $IV$) $\Delta\mu(O)=-R_{N}\; I_{s,i}(O)$
which is the relation one would get from an infinite arm. Everything
happens as if the interface had been washed away: this is the reason
why we qualified the case $X=1$ as corresponding to spin impedance
matching in \ref{subsec:Spin-resistance-mismatch.}.

In the transparent regime ($X<1$), the effective spin resistance
is larger at large distance, which is an interesting feature for the
design of large non-local circuits. This advantage is circumvented
by the exponential decrease of the effective polarization (the non-local
resistance will be shown to be proportional to both in \ref{subsec:Spin-voltage-andnonlocalresistance})
so that in terms of large signals the transparent regime is not interesting,
neither in the short-distance nor the large-distance limit.

\subsection{General expression of the spin voltage in Van der Pauw geometry.\label{sub:General-expression}}

\subsubsection{Spin accumulation at cross center.\label{subsec:Spin-accumulation.}}

Solving the one-dimensional drift-diffusion equations (see Appendix
\ref{sec:Computing-the-Spin} for details of the calculations) leads
to the following results.

The spin accumulation in the center of the cross is found as:

\begin{equation}
\Delta\mu(O)=\left(P_{eff,1}-P_{eff,2}\right)\; R_{eff}\; I_{c}\label{eq:delta_mu}
\end{equation}

We can define a total effective polarization at injector electrodes
$F1$ and $F2$: 
\begin{equation}
P_{injector}=P_{eff,1}-P_{eff,2}.
\end{equation}
The minus sign in front of $P_{eff,2}$ comes from the fact that $F2$
is a charge drain: in terms of spin accumulation it therefore acts
contrariwise to electrode $F1$. This also shows that to ensure maximum
signal it is better to have opposite orientations for $F1$ and $F2$
(antiparallel injector electrodes). This is easy to understand: when
injector electrodes are parallel, $F2$ acts as a spin sink for the
spins injected by $F1$; therefore the spin accumulation should decrease.
But when $F2$ is antiparallel to $F1$, spin leakage is frustrated;
although $F2$ is a charge drain, it acts as an additional spin source.

It is noteworthy that the spin accumulation does not depend on the
magnetization orientations of detector electrodes $F3$ and $F4$
(although it does depend on its parameters through $R_{eff}$). This
is sensible: spin injection is ensured by $F1$ and $F2$ not $F3$
and $F4$. 

The general structure of the spin accumulation in terms of $F1$ and
$F2$ relative orientation can be understood simply. Suppose we flip
all spins of the setup. Then: $\Delta\mu\longrightarrow-\Delta\mu$
since we have exchanged spin up and spin down electrons. This implies
that the spin accumulation must be an odd function of electrode polarizations.
This is easily checked on Eq. (\ref{eq:delta_mu}): when both $F1$
and $F2$ are flipped, their effective polarizations get reversed
$P_{eff}\longrightarrow-P_{eff}$ and therefore the spin accumulation
is reversed. Another way to reach the same conclusion is to notice
that the physics of the setup should be invariant when both current
and magnetizations are reversed. Flipping all spins or reversing the
charge current should lead to the same spin accumulation, namely a
reversed one.

An important consequence of the structure of Eq. (\ref{eq:delta_mu})
is that the spin accumulation at the cross center vanishes when $F1$
and $F2$ are identical and parallel so that $P_{eff,1}=P_{eff,2}$.
This follows clearly from symmetry: when $F1$ and $F2$ are symmetric
with respect to the cross center and have identical parameters, the
spin accumulation on the line $F1-F2$ should be antisymmetric and
the spin current should be symmetric when $F1$ and $F2$ are parallel
(the opposite when they are antiparallel). This can be understood
by reversing the current: if one reverses the current, on the one
hand, the spin accumulation at the cross center should not change
since $F1$ and $F2$ are identical and parallel (by watching Fig.
\ref{fig:Van-der-Pauw4Funequal} after a $\pi$ rotation); on the
other hand, reversing the current must reverse the spin accumulation
since reversing the current is equivalent to reversing all spins.
The only way out is for the spin accumulation at the cross center
to vanish. This result does not depend on the assumption of one-dimensional
flow and follows directly from symmetry.

We will see later that this enables an ON-OFF switch function onto
the device.

\subsubsection{Spin voltage and non-local resistance.\label{subsec:Spin-voltage-andnonlocalresistance}}

\textbf{Spin voltage.} The spin voltage (or non-local voltage) is
the voltage drop between terminals $F3$ and $F4$:
\begin{equation}
V_{nl}=-\left[\mu_{F3}(+\infty)-\mu_{F4}(+\infty)\right]/e.
\end{equation}

Straightforward calculations (see Appendix \ref{sub:Spin-voltage-appendix})
lead to: 
\begin{equation}
V_{nl}=-\Delta\mu(O)\;\left(P_{eff,3}-P_{eff,4}\right).\label{eq:vnl}
\end{equation}
The behaviour under magnetization reversal is easy to understand:
when the magnetizations of both detector electrodes are switched,
their coupling to the spin accumulation gets reversed so that the
non-local spin voltage should change sign. The spin voltage is therefore
odd under magnetization switching of both detector electrodes.

Inserting the expression of the spin accumulation in Eq. (\ref{eq:delta_mu}):
\begin{eqnarray}
V_{nl} & = & \left(P_{eff,1}-P_{eff,2}\right)\;\left(P_{eff,3}-P_{eff,4}\right)\; R_{eff}\: I_{c}
\end{eqnarray}

This expression factors out neatly in three contributions:

(i) the geometry dependent effective spin resistance $R_{eff}=\left(\sum_{i=1-4}\; R_{eff,i}^{-1}\right)^{-1}$; 

(ii) a total effective polarization for injector and collector electrodes
($F1$ and $F2$) 
\begin{equation}
P_{injector}=P_{eff,1}-P_{eff,2};
\end{equation}

(iii) and a total effective polarization for the two detector electrodes
($F3$ and $F4$): 
\begin{equation}
P_{detector}=P_{eff,3}-P_{eff,4}.
\end{equation}

\textbf{Non-local resistance.} The non-local resistance (sometimes
called a transresistance) is defined as the ratio of the non-local
voltage to the charge current flowing through the injector electrode
$F1$ to the collector electrode $F2$ :

\begin{equation}
R_{nl}=\frac{V_{nl}}{I_{c}}=R_{eff}\; P_{eff,injector}\; P_{eff,detector}.\label{eq:non-local res}
\end{equation}
To achieve a large signal it is therefore necessary to have a large
total effective spin resistance for the cross $R_{eff}$ and to have
on the one hand antiparallel source and drain terminals, on the other
hand antiparallel detector electrodes.

$R_{eff}$ is largest when all arms effective spin resistances are
also large, which is the case if spin resistance mismatches are in
the tunneling regime according to the discussion in \ref{sub:Effective-spin-resistances.}
and if the arms length is short enough. Two situations may arise:

i) one or several spin resistance mismatches are in the transparent
regime ($X\leq1$). Then $R_{eff}\sim R_{N}$ or smaller. No enhancement
of the spin voltage is to be expected when compared with the usual
lateral setup. We will say that we have an open geometry\cite{jaffres_spin_2010}
which leaks spin (larger spin currents but smaller spin accumulations).

ii) all spin resistance mismatches are in the spin confining regime
($X>1$). Then $R_{eff}\sim R_{c}$ in the limit of short length for
the arms (and assuming for simplicity mismatches roughly equal $X_{i}\sim X\sim R_{c}/R_{N}$.
In such a geometry which will be qualified as closed\cite{jaffres_spin_2010}
the signal is therefore potentially much larger than in an open geometry.

Let us compare the non-local resistance to local resistances in the
same device.

i) Firstly a local resistance can be measured between source and drain
($F1$ and $F2$); as shown in Appendix \ref{sub:rlocal}: 
\begin{equation}
R_{local,12}=R_{eff}\:\left[P_{eff,1}-P_{eff,2}\right]^{2}+R_{0}\left\{ P_{1-2}\right\} 
\end{equation}
where 
\begin{eqnarray}
R_{0}\left\{ P_{1-2}\right\}  & = & \sum_{i=1-2}\rho_{Fi}^{*}\,\left(1-P_{Fi}^{2}\right)\: z_{i}+\rho_{N}^{*}\: l_{i}\nonumber \\
+ & R_{ci} & +\frac{\left[P_{Fi}^{2}R_{ci}R_{Fi}-P_{ci}^{2}R_{ci}^{2}-2P_{Fi}P_{ci}R_{Fi}R_{ci}\right]}{R_{Fi}+R_{ci}}.
\end{eqnarray}
($z_{i}$ are the locations of probes in the electrodes, see Appendix
\ref{sub:rlocal}). In general $R_{local,12}\approx R_{0}$ since
the GMR effect is a few percents; the non-local resistance which is
commensurate with $R_{local,12}-R_{0}$ is therefore much smaller
than $R_{local,12}$. 

It is more meaningful to compare the variations upon magnetization
switching $\Delta R=R_{AP}-R_{P}$; for the non-local signal, we have
set in the following $F1$ and $F2$ antiparallel while switching
$F4$ magnetization:
\begin{equation}
\frac{\Delta R_{nl}}{\Delta R_{local,12}}=\frac{\left(P_{eff,1}+P_{eff,2}\right)\: P_{eff,4}}{2P_{eff,1}\, P_{eff,2}}
\end{equation}
(where for $R_{local}$, either terminal $F1$ or $F2$ have been
switched). In the case of identical source and drain electrodes ($F1$
and $F2$) this reduces to:
\begin{equation}
\frac{\Delta R_{nl}}{\Delta R_{local,12}}=\frac{P_{eff,4}}{P_{eff,1}};
\end{equation}
if the ratio is larger than unity ($P_{eff,4}>P_{eff,1}$) there is
an amplification of non-local MR variation versus local MR (or the
converse if $P_{eff,4}>P_{eff,1}$). 

ii) If we then compare to the local resistance found when current
flows from $F3$ to $F4$, one gets instead:
\begin{equation}
\frac{\Delta R_{nl}}{\Delta R_{local,34}}=\frac{\left(P_{eff,1}+P_{eff,2}\right)}{2P_{eff,3}};
\end{equation}
One can again get an MR amplification (or reduction if $P_{eff,3}$
is sufficiently large).

But if all electrodes are identical, all these MR ratios are then
equal to unity 
\begin{equation}
\frac{\Delta R_{nl}}{\Delta R_{local}}=1
\end{equation}
which means the non-local measurement performs equally well as local
measurements in terms of raw resistance variation, with one proviso:
non-local resistances have smaller baselines. Indeed for $R_{nl}$,
the baseline or smallest signal is: 
\[
\left|R_{nl,min}\right|=R_{eff}\;\left|\left(\left|P_{eff,1}\right|-\left|P_{eff,2}\right|\right)\;\left(\left|P_{eff,3}\right|-\left|P_{eff,4}\right|\right)\right|
\]
while for $R_{local,12}$: 
\[
\left|R_{local,min}\right|=R_{0}\left\{ P_{1-2}\right\} +R_{eff}\;\left(\left|P_{eff,1}\right|-\left|P_{eff,2}\right|\right)^{2}
\]
which is larger on account of the $R_{0}$ term.

iii) It is worthwhile to compare to CPP GMR for a spin valve with
similar dimensions:
\begin{eqnarray*}
\Delta R_{CPP} & =\\
 & \frac{4R_{N}\;\left(P_{c}R_{c}+P_{F}R_{F}\right)^{2}}{2R_{N}\left(R_{F}+R_{c}\right)\cosh l+\left[\left(R_{F}+R_{c}\right)^{2}+R_{N}^{2}\right]\sinh l}
\end{eqnarray*}
for identical and infinite ferromagnetic layers separated by the same
distance $L=l\; l_{N}$. In terms of $X=\left(R_{F}+R_{c}\right)/R_{N}$
this can be recast as:
\begin{eqnarray*}
\Delta R_{CPP} & = & \frac{8R_{N}^{-1}\;\left(P_{c}R_{c}+P_{F}R_{F}\right)^{2}}{\left(X+1\right)^{2}\;\exp l-\left(X-1\right)^{2}\;\exp-l}
\end{eqnarray*}
so that for a non-local device with identical terminals: 
\begin{equation}
\frac{\Delta R_{nl}}{\Delta R_{CPP}}=\frac{1}{2}.
\end{equation}
 The non-local signal is smaller by a factor 2 which is easy to understand:
in the cross geometry the spin relaxation volume is doubled when compared
with a spin valve with a paramagnetic layer of identical length because
of the side arms. 

A systematic comparison of the non-local resistance in the cross geometry
with the standard lateral geometry is left to sections \ref{sub:Van-der-Pauw2f}-\ref{sec:4f}.

\subsubsection{Non-local charge current.}

The voltage between the detector electrodes $F3$ and $F4$ actually
acts as an electromotive force (emf) of magnetic origin; when $F3$
and $F4$ are shorted, a charge current therefore appears. This non-local
charge current is induced by the spin accumulation generated by the
remote source and drain $F1$ and $F2$. Note that since the spin
voltage is an electromotive force, it can be measured either through
a (nano-)voltmeter or with a potentiometer: in the latter case, the
advantage is that there is no current at all during the measurement. 

If however it proves advantageous that the signal be a current (for
chaining the non-local device to a bipolar transistor for instance
rather than a MOSFET), the non-local current is found (see Appendix
\ref{sub:nonlocalcurrent}) as:
\begin{equation}
I_{nl}=-\frac{R_{nl}}{R_{local,34}}\; I_{c}
\end{equation}
where $R_{local,34}$ is the local resistance measured when one drives
a current between $F3$ and $F4$. This can be rewritten as:
\begin{equation}
I_{nl}=-\frac{R_{eff}\;\left(P_{eff,1}-P_{eff,2}\right)\;\left(P_{eff,3}-P_{eff,4}\right)}{R_{eff}\:\left[P_{eff,3}-P_{eff,4}\right]^{2}+R_{0}\left\{ P_{3-4}\right\} }\; I_{c}
\end{equation}
where:
\begin{eqnarray*}
R_{0}\left\{ P_{3-4}\right\}  & = & \sum_{i=3-4}\rho_{Fi}^{*}\,\left(1-P_{Fi}^{2}\right)\: z_{i}+\rho_{N}^{*}\: l_{i}+R_{ci}
\end{eqnarray*}
\begin{equation}
+\frac{\left[P_{Fi}^{2}R_{ci}R_{Fi}-P_{ci}^{2}R_{ci}^{2}-2P_{Fi}P_{ci}R_{Fi}R_{ci}\right]}{R_{Fi}+R_{ci}}.
\end{equation}
(it can be checked that $R_{0}\geq0$); $z_{i}$ ($i=3-4$) are the
locations on $F3$ and $F4$ of the wires which short them together
(in the following we have chosen to focus on $R_{nl}$ rather than
$I_{nl}$).

\subsection{Main properties of the device.\label{sub:Properties}}

The device functionalities depend on the terminals for which the magnetization
has been fixed (prevented from switching); additional symmetry requirements
can also add properties. We first discuss functionalities pertaining
to sensing or data storage.

As explained in the introduction the non-local resistance and spin
voltage are contaminated by offsets, the origin of which is still
under debate (current inhomogeneities; thermal origin). This may adversely
affect the measured signals and in the following we will take care
to indicate the potential impact of these offsets. 

\textbf{1-bit reading.} This is the basic functionality of the spin
valve as a sensor. Suppose the orientations of all terminals but one
(say $F3$) are pinned (for definiteness we assume: $F1:\uparrow$,
$F2:\downarrow$ and $F4:\downarrow$ ) so that $F3$ acts as a sensing
electrode. We make no special assumptions on the terminals (later
on some conditions will be imposed for further functionalities). One
recovers a spin-valve behaviour (see Fig. \ref{fig:Non-local-resistance-variation-0})
with two distinct values of the non-local resistance which uniquely
determine the orientation of terminal $F3$: 
\begin{eqnarray*}
R_{P} & = & R_{eff}\;\left(P_{eff,1}+P_{eff,2}\right)\;\left(P_{eff,3}-P_{eff,4}\right)\\
R_{AP} & = & R_{eff}\;\left(P_{eff,1}+P_{eff,2}\right)\;\left(P_{eff,3}+P_{eff,4}\right)
\end{eqnarray*}
(the $P/AP$ index refer to $F3$ magnetization orientation relative
to $F4$). The spin valve can therefore be used to read a 1-bit information.

In the standard non-local spin valve and if we neglect voltage offsets,
the spin voltage changes sign when one terminal is flipped so that
for antiparallel and parallel alignment $R_{AP}=-R_{P}$ (this is
recovered here in the limit $P_{eff,4}=0$, when $F4$ is a paramagnet);
this is not the case here although depending on the relative values
of $P_{eff,3}$ and $P_{eff,4}$ there can still be a change of sign
(Fig. \ref{fig:Non-local-resistance-variation-0}-b, when $P_{eff,3}>P_{eff,4}$).

For the Van der Pauw cross under the most general conditions the difference
with the standard non-local setup is therefore minor; it remains to
see if larger signals can be achieved. This will be the topic of sections
\ref{sec:2f}-\ref{sec:4f} where we will show that tunnel contacts
at the four terminals can greatly enhance the non-local resistance. 

Furthermore this functionality is clearly affected by offset voltages
which will shift the signals and change the GMR ratios (adversely
if the offset is positive). We now discuss a simple way to circumvent
these offsets.

\begin{center}
\begin{figure}
\begin{centering}
\includegraphics[width=1\columnwidth]{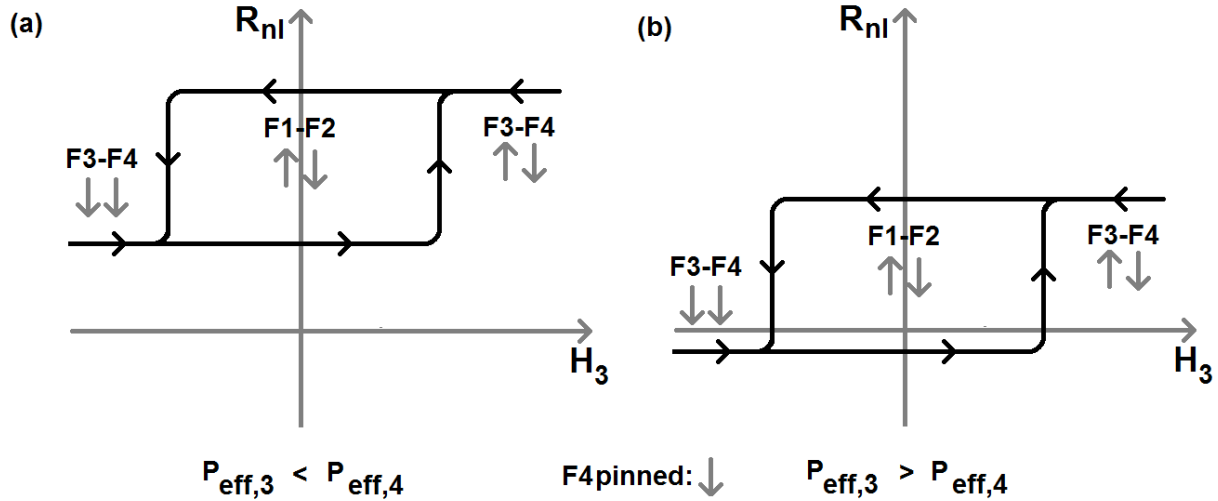}
\par\end{centering}

\caption{Non-local resistance variation when the orientation of a single terminal
(here $F3$) is switched by an external magnetic field (here $H_{3}$).
All other terminals are pinned. Injector $F1$ and collector $F2$
are antiparallel with orientation as shown on the graph. (a) case
$P_{eff,3}<P_{eff,4}$: standard spin valve effect (1-bit reading);
(b) case $P_{eff,3}>P_{eff,4}$: spin valve effect with spin voltage
change of sign.). \label{fig:Non-local-resistance-variation-0}}
\end{figure}

\par\end{center}

\textbf{Offset free 1-bit reading with infinite non-local GMR. }

The offset issue is easily fixed if we assume $F3$ and $F4$ are
identical electrodes placed symmetrically with respect to the rest
of the setup. Due to symmetry offset voltages are neutralized (assuming
offset voltages are spin independent) since they will shift both voltages
$V_{3}$ and $V_{4}$ in the same manner so that the spin voltage
$V_{nl}=V_{3}-V_{4}$ is free from offsets.

Additionally: $R_{P}=R_{eff}\;\left(P_{eff,1}+P_{eff,2}\right)\;\left(P_{eff,3}-P_{eff,4}\right)=0$
in parallel alignment since $P_{eff,3}=P_{eff,4}$ while $R_{nl}\neq0$
in antiparallel alignment (we have assumed that $F1$ and $F2$ are
antiparallel and pinned).

One has achieved an infinite GMR for the non-local resistance since
the ratio $GMR=\left(R_{AP}-R_{P}\right)/R_{P}\longrightarrow\infty$
(using the optimistic ratio; the pessimistic ratio would be $100\:\%$).
There is an intrinsic contrast which is protected by symmetry from
the voltage offsets.

Such a maximized MR ratio is clearly helpful for SNR in terms of Johnson
noise (or shot noise in the strong tunneling limit) since the latter
scales as $\Delta R/\sqrt{R}$; indeed if we compare with a CPP spin
valve with a $\Delta R/R=1-10\;\%$ GMR ratio, this would imply at
identical $\Delta R$ ($\Delta R=\Delta R_{local}=\Delta R_{nl}$)
an increase of SNR by $10-20\; dB$. (Indeed: SNR for non-local device
would be $\propto1/\sqrt{\Delta R}$ so that $SNR_{non-local}/SNR_{CPP}\propto\sqrt{R_{local}/\Delta R}$).
The difference is quite significant given that under operation one
expects in general at least $30\; dB$ SNR.

In terms of geometry requirements, note that the symmetry between
detector electrodes is required only on the scale of a few spin relaxation
lengths $l_{F}$ (on the ferromagnet side) since the spin accumulation
is washed at larger distance. 

The property is also clearly independent of the precise geometric
arrangement, does not depend on the assumption of one-dimensional
flow and will be valid for other geometries than the cross studied
in this paper, provided the two detectors are arranged symmetrically
with respect to the injectors. Experimentally this is very useful
since this gives a lot of flexibility in terms of design.

\textbf{3-bit reading or storage.} For that function one terminal
is pinned, while the other three are free and play the role of input
signals (the non-local resistance measured between $F3$ and $F4$
is as previously the output signal). Eq. (\ref{eq:non-local res})
shows the non-local resistance can assume 8 different values when
the electrodes orientation are changed and the maximum value for $\left|R_{nl}\right|$
is reached when on the one hand $F1$ and $F2$ are antiparallel,
and on the other hand $F3$ and $F4$ are also antiparallel. The orientations
of the three non-pinned terminals are uniquely determined by the 8
distinct values of the spin voltage. This implies that the device
encodes 3 bits in principle (3 bit spin valve). 

Fig. \ref{fig:Non-local-resistance-variation} shows the eight outputs
signals as a function of the input variables (for illustrative purposes
$F3$ is varied by an external field $H3$ on the graphs, showing
the output variation with the change of one input).

The property survives offset voltages which come as an additive contribution
to the voltage. However resolution may be adversely affected by offsets.
Note that although there are four ferromagnets and a priori 4 bits
could be stored, the spin voltage can assume only 8 values, not $2^{4}=16$.
This is because when all spins are reversed, the spin voltage is unchanged
in Eq. (\ref{eq:non-local res}).

\begin{center}
\begin{figure}
\begin{centering}
\includegraphics[width=1\columnwidth]{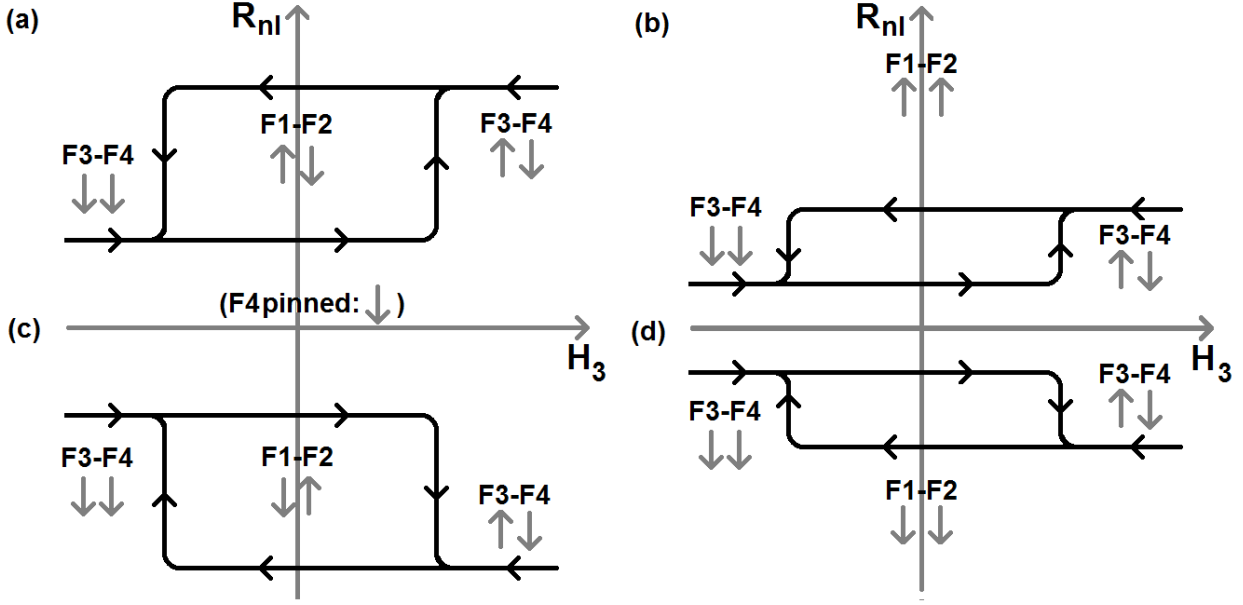}
\par\end{centering}

\caption{Non-local resistance variation as a function of terminals magnetization
orientation showing 3-bit sensor or storage. For illustrative purposes
$F3$ magnetization is varied in each figure (a-d) (through an applied
field $H3$) to show the output signal variation with the change of
one input. (a) Injector $F1$ and collector $F2$ are antiparallel.
(b) They are switched to parallel: the signal collapses (to some extent).
(c) Signal is reversed when injector and collector are both switched
from antiparallel alignment in (a). (d) Injector and Collector are
parallel (but with directions opposite to (b) ).\label{fig:Non-local-resistance-variation}}

\end{figure}

\par\end{center}

\textbf{Offset-free 2-bit reading}. By the same token applied previously
to detectors for the 1-bit reading it is possible to convert the 3-bit
reading function into a protected 2-bit reading function: suppose
that detector terminals $F3$ and $F4$ are symmetric (namely, have
identical parameters and are placed symmetrically with respect to
the device) but that $F1$ and $F2$ differ ($P_{eff,1}\neq P_{eff,2}$).
Let us pin terminals $3$ and $4$ in antiparallel orientation ($F3:\uparrow$
and $F4:\downarrow$), while only $F1$ and $F2$ are allowed to switch
magnetizations. The non-local spin voltage then takes 4 distinct values
depending on the orientations of terminals $1$ and $2$: $R_{nl}=R_{eff}\;\left(P_{eff,3}+P_{eff,4}\right)\;\left(\pm P_{eff,1}\pm P_{eff,2}\right)$.
These values uniquely determine the 2-bit state of $F1$ and $F2$.
The main advantage when compared with the 3-bit function discussed
previously is that since terminals $F3$ and $F4$ are symmetric,
there can be no offset voltages since they automatically cancel out
when measuring the voltage drop between $F3$ and $F4$ (if offset
voltages are spin-independent; if it is not the case, they will add
up to the spin voltage). Note that if $F1$ and $F2$ are pinned while
$F3$ and $F4$ are not pinned but identical (to avoid offsets), the
spin voltage only assumes three values since $R_{P}=0$; to store
2 bits without offsets it is therefore necessary to pin $F3$ and
$F4$, not the other way around. 

\textbf{ON-OFF switch for the 1-bit read-out or storage function.}
When source $F1$ and drain $F2$ are identical ferromagnets (same
distance from origin, same conductivities, polarizations, interface
resistances, spin diffusion length), $P_{eff,1}=P_{eff,2}$. This
then implies that when $F1$ and $F2$ are parallel, the spin accumulation
at the cross center vanishes 
\begin{equation}
\Delta\mu(O)=0
\end{equation}
 so that whatever the orientation of $F3$ and $F4$, $R_{P}=0=R_{AP}.$
The spin voltage has been killed and we have disabled the read-head
or 1-bit storage. The property is clearly interesting in terms of
logic if the device is chained to another device for instance a MOSFET
whose gate is controlled by the spin voltage (after suitable amplification). 

That property should survive offset voltages since symmetry protects
it. 

But one might wonder if the spin voltage measured at $F3$ and $F4$
will still vanish if we take into account departures from strict one-dimensional
charge and spin flow. It is clear indeed that even if we take into
account the 3-dimensional nature of the device but remain in a quasi-one-dimensional
approximation, the spin accumulation on the side arms (zones $III-IV$)
will remain small (it may be non zero at edges) and will never diffuse
as far as $F3$ and $F4$ provided the width of each arm is much smaller
than its length ($w\ll l$): indeed when $F1$ and $F2$ are parallel,
the spin accumulation is an odd function of $z$ in the direction
$F1-F2$; therefore, on the side arms ($III-IV$), there will be some
spilling of spin accumulation with opposite signs on opposite edges,
close to the origin $O$. But if $w\ll l$, spin diffusion will mix
these opposite spin accumulations which will cancel out when one reaches
the detector terminals ($F3$ and $F4$). The spin voltage measured
at $F3$ and $F4$ should therefore still vanish.

\textbf{Direct Spin Hall Effect probe.} The geometry lends itself
easily to probing the Spin Hall Effect\cite{dyakonov_1971,hirsch_spin_1999,takahashi_spin_2006}.
Imagine $F1$ and $F2$ are normal electrodes and that the magnetizations
of $F3$ and $F4$ are perpendicular to the plane of the cross. On
the arm $AB$ ($F1-I-O-II-F2$ or $zOz'$) a spin accumulation may
appear due to Hall coupling $\alpha_{H}=\sigma_{xx}/\sigma_{xy}$
on the width $w$ (for $-w/2\leq x\leq w/2$):
\begin{equation}
\Delta\mu(x,\; z)=\frac{\alpha_{H}\;\Delta V}{L_{1}+L_{2}}\; x.
\end{equation}
When we reach the cross center, there can be a spilling of charge
current lines to the side arms, but we will neglect this effect by
assuming that the arms width are much smaller than their lengths ($w\ll L_{i}$
for $i=1-4$). When we move on the side arms ($III-IV$), the spin
accumulation decreases in magnitude. This decrease can be estimated
using drift-diffusion equations as:
\begin{eqnarray}
\Delta\mu_{III}(C) & = & \frac{X_{3}}{\delta_{3}^{+}}\;\Delta\mu(x=\frac{w}{2},\; z=0)
\end{eqnarray}
{[}see Appendix \ref{sub:Spin-voltage-appendix}, Eq. (\ref{eq:mu_c}){]}
with a similar expression for arm $IV$:
\begin{equation}
\Delta\mu_{IV}(D)=\frac{X_{4}}{\delta_{4}^{+}}\;\Delta\mu(x=-\frac{w}{2},\; z=0)
\end{equation}
 $\Delta\mu_{III}(C)$ can be re-expressed as:
\begin{equation}
\Delta\mu_{III}(C)=\frac{X_{3}}{\delta_{3}^{+}}\;\frac{\alpha_{H}\;\Delta V}{L_{1}+L_{2}}\;\frac{w}{2}
\end{equation}
and:
\begin{equation}
\Delta\mu_{IV}(D)=-\frac{X_{4}}{\delta_{4}^{+}}\;\frac{\alpha_{H}\;\Delta V}{L_{1}+L_{2}}\;\frac{w}{2}.
\end{equation}

The spin voltage is therefore (see Appendix \ref{sub:Spin-voltage-appendix}):
\begin{eqnarray}
V_{nl} & = & -\left(P_{eff,3}+P_{eff,4}\right)\;\Delta\mu(x=\frac{w}{2})\\
 & = & -\left(P_{eff,3}+P_{eff,4}\right)\;\frac{\alpha_{H}\;\Delta V}{L_{1}+L_{2}}\;\frac{w}{2}.
\end{eqnarray}
Note that for identical and antiparallel electrodes $F3$ and $F4$,
the signal therefore vanishes since $P_{eff,3}=-P_{eff,4}$ and that
for maximal non-local signal, parallel magnetizations are required.

Let us go back to our initial setup with four ferromagnetic electrodes.
If magnetizations are perpendicular to the plane of the Van der Pauw
cross, the spin Hall effect will add up to the non-local spin voltage
arising from spin injection. But if we choose a symmetric setup with
identical parallel electrodes $F3$ and $F4$, offsets cancel out
in $V_{nl}=V_{3}-V_{4}$ and $V_{P}=0$ as explained previously. Therefore
the only remaining signal is that of the Direct Spin Hall Effect.
This provides an interesting all-electrical alternative to the observation
of Direct Spin Hall Effect, which initially was observed optically
in semiconductors\cite{kato_observation_2004,wunderlich_experimental_2005}
although later on some electrical detection schemes have been used
in metallic systems to investigate both Direct and Inverse Spin Hall
Effect\cite{valenzuela_direct_2006,kimura_spinhall_2007,vila_2007}.

\subsection{Implementation of magneto-logic gates.}

Many proposals exist in the literature for the use of magnetoelectronics
circuits as logic gates \cite{zutic_spintronics:_2005}; this has
prompted a lot of activity in the field of semiconductor spintronics
since integration to existing processes would be optimal \cite{fabian_semicon_2007}.
Our device is metallic and therefore not the best candidate as a spin
transistor since there is no amplification: this renders the chaining
of gates more delicate for instance (unless one uses hybrid designs
combining pure magnetoelectronics devices with conventional transistors).
However, the device still possesses obvious capability as a programmable
magneto-logic gate as we now demonstrate.

Let us associate bit $0$ with down $\downarrow$ magnetization and
bit $1$ with up $\uparrow$ magnetization. $F1$ and $F2$ are assumed
to be identical; $F3$ and $F4$ are also identical (same arm length,
same polarizations, spin diffusion length, etc); only the magnetization
orientations are allowed to differ. The non-local resistance can then
assume only three values: $0$ and $\pm R_{0}=\pm4R_{eff}\; P_{eff,1}\; P_{eff,3}$.
For logic operations we will consider two conventions:

(i) associate bit 0 to zero resistance, and bit 1 to $R_{nl}=+R_{0}$. 

(ii) or (opposite convention) associate bit 1 to zero resistance,
and bit 0 to $R_{nl}=+R_{0}$.

Note that in what follows we have discarded all configurations for
which $R_{nl}=-R_{0}$ to avoid ambiguities in the bit association
and keep only those for which there are only two possible outputs:
$0$ and $R_{0}$. 

For the first convention assigning the bit content of $R_{nl}$, the
expression for $R_{nl}$ {[}see Eq. (\ref{eq:non-local res}){]} can
therefore be rewritten in terms of bits as the Boolean equation: 
\begin{equation}
\left(F1-F2\right)\left(F3-F4\right)=R_{nl}\label{eq:logic}
\end{equation}
where to simplify notation we have conflated $R_{nl}$ and its bit
content. 

With the opposite convention, 
\begin{equation}
\left(F1-F2\right)\left(F3-F4\right)=\overline{R_{nl}}\label{eq:logic2}
\end{equation}
Note that in order to have well defined HIGH and LOW states the associated
voltages (or non-local resistances) must be sufficiently different
which is not ensured in the presence of offset voltages. That's why
we choose symmetric terminals to get rid of these offsets.

We first try to reproduce basic binary Boolean functions. We need
therefore to pin two terminals; the other two terminals will represent
the variables treated by the device in the following manner (for instance):
imagine each terminal is screened to prevent any magnetic field applied
to a given electrode to influence any other electrode; we then apply
external magnetic fields to either of the two non-pinned terminals
but not to the other twos, whose magnetizations are therefore fixed
during the whole operation.

\textbf{NOR gate.} We choose convention of Eq. (\ref{eq:logic}) for
bit coding. We pin $F2$ and $F4$ ($F2:\uparrow$ and $F4:\uparrow$)
so that $F2=1$ and $F4=1$ in bit terms. To have a non-zero resistance
among the four possible bit configurations of $F1$ and $F3$, only
$F1=0$ and $F3=0$ are allowed. The non-local resistance is then
$R_{nl}=+R_{0}$ which we associate with HIGH state or bit 1: this
means that $\overline{F1}\;\overline{F3}=R_{nl}$; this can also be
recovered through algebra by using Eq. (\ref{eq:logic}) which in
our case is:
\begin{equation}
\left(F1-1\right)\left(F3-1\right)=R_{nl}
\end{equation}
so that:
\begin{equation}
\overline{F1}\;\overline{F3}=R_{nl}.
\end{equation}
 This is precisely a NOR gate since $\overline{F1}\;\overline{F3}=\overline{F1+F3}$
by De Morgan theorem. This is a very important property since NOR
has functional completeness: any Boolean function can be implemented
by using a combination of NOR gates (only NAND possesses the same
property).

\textbf{OR gate.} In the same configuration as for the NOR gate, if
LOW state (bit 0) is now associated with $R_{nl}=+R_{0}$ and HIGH
state (bit 1) to $R_{nl}=0$ {[}convention of Eq. (\ref{eq:logic2}){]},
one obviously gets an OR gate. Of course since the association of
HIGH and LOW has been reversed with the previous case, the two settings
are incompatible since they correspond to opposite bit assignment. 

\textbf{AND gate.} We still pin $F2$ and $F4$ ($F2:\downarrow$
and $F4:\downarrow$) so that $F2=0$ and $F4=0$ in bit terms. Using
Eq. (\ref{eq:logic}) this implies $F1\; F3=R_{nl}$ which is an AND
operation (using the first convention for $R_{nl}$ bit content).
This gate is compatible with the NOR gate but not the OR gate which
is produced with a different convention for bit coding.

\textbf{NAND gate.} By changing the bit coding of $R_{nl}$, the AND
gate turns into a NAND gate with the same configuration for $F2$
and $F4$. As mentioned previously, the achievement of a NAND is quit
noteworthy since it has functional completeness in Boolean algebra.

\textbf{$\overline{A}B$ or $A\nLeftarrow B$ ($A$ not implied by
$B$) gate}. We now pin $F2$ and $F3$ and choose $F2=1$ and $F3=0$
in bit terms. Then Eq. (\ref{eq:logic}) (first convention) turns
into $\overline{F1}\; F4=R_{nl}$. 

\textbf{$A+\overline{B}$ or $A\Leftarrow B$ gate}. Using the same
configuration for $F2$ and $F3$ as previous gate but exchanging
the conventions for the bit content for $R_{nl}$ leads to an inverted
gate: indeed $\overline{F1}\; F4=\overline{R_{nl}}$ implies $F1+\overline{F4}=R_{nl}$
by De Morgan theorem.

\textbf{$A\overline{B}$ or $A\nRightarrow B$ gate.} We pin $F2$
and $F3$ and choose $F2=0$ and $F3=1$ in bit terms. Then Eq. (\ref{eq:logic})
(first convention) turns into $F1\;\overline{F4}=R_{nl}$. 

\textbf{$\overline{A}+B$ or $A\Rightarrow B$ gate}. Choosing the
second convention of Eq. (\ref{eq:logic2}) with $F2=0$ and $F3=1$
leads to $F1\;\overline{F4}=\overline{R_{nl}}$ which is equivalent
to $\overline{F1}+F4=R_{nl}$ by De Morgan theorem.

\textbf{FALSE gate.} We pin $F1$ and $F2$ in parallel configuration.
Then $R_{nl}=0$ (in terms of resistance not bit) whatever the configuration
of $F3$ and $F4$. If we adopt the first convention, the zero resistance
translates into bit 0. This is therefore a FALSE gate.

\textbf{TRUE gate.} In the same terminals configuration as previous
gates, if we adopt the opposite convention for the bit content of
$R_{nl}$, one then gets a TRUE function.

The six other possible binary operations (out of sixteen) can not
be built as easily using a single cross device; this leaves out the
NOR and XNOR gates from the list of the basic gates in use in electronic
logic (AND, OR, NOR, NAND, XOR, XNOR, buffer, inverter). A way to
achieve them would be to chain gates since all Boolean functions can
be recovered using only a NOR (or NAND) gate. 

We now turn to the two other basic gates, buffer and inverter.

\textbf{Buffer.} For that function only one terminal is not pinned.
We choose to pin $F1=1$, $F2=0$ and $F4=0$. Using Eq. (\ref{eq:logic})
this implies $F3=R_{nl}$. This can also be realized by pinning $F1=0$,
$F2=1$ and $F4=1$. Using Eq. (\ref{eq:logic2}) this implies $\overline{F3}=\overline{R_{nl}}$.

\textbf{Inverter.} When pinning $F1=1$, $F2=0$ and $F4=0$, using
Eq. (\ref{eq:logic2}) implies $F3=\overline{R_{nl}}$. This can also
be realized with Eq. (\ref{eq:logic}) by pinning $F1=0$, $F2=1$
and $F4=1$ which leads to $\overline{F3}=R_{nl}$.

The NOR, AND, $A\nLeftarrow B$, $A\nRightarrow B$ , FALSE, buffer
and inverter gates use the same bit convention for $R_{nl}$ (HIGH
state for $R_{nl}=+R_{0}$, LOW state for $R_{nl}=0$) and can be
implemented together in the same circuit; since they differ only by
the assignment of pinned terminals, this means that one can easily
turn a gate into another one (reprogrammable logic) through local
application of an external magnetic field or through spin transfer
torque. 

In summary, the non-local cross with four ferromagnetic terminals
is a versatile spintronics device which can be used either as a standard
spin valve: when some symmetry requirements are met, it displays the
important property of an infinite GMR for the non-local resistance;
it can be immunized against offset voltages observed in many non-local
setups. Finally, it can be used as a magneto-logic gate as well as
probe the Direct Spin Hall Effect.

\section{Van der Pauw cross with two or three ferromagnets.\label{sec:2f}}

Which principles should guide us in order to achieve large signals
in non-local setups? As stressed time and again in the literature\cite{fert_semiconductors_2007,jaffres_spin_2010,fert_spin_2002},
they are quite simple: larger spin accumulations in the paramagnet
can be generated if (i) the paramagnet volume in which spins can relax
is small (in comparison with $l_{N}^{3}$) and if (ii) spin back-flow
to the ferromagnets is hindered by large enough interface resistances. 

One can classify geometries as open or closed according to the (non-)fulfillment
of these prescriptions\cite{jaffres_spin_2010}: it has been argued
in the latter reference that at small enough volume and for large
enough tunnel barriers the spin accumulation can be greatly enhanced
because it then scales with the (large) interface resistances $R_{c}$.
But whenever the geometry is open, the spin accumulation scales with
the much smaller spin resistance of the paramagnetic channel $R_{N}$
which in practice is often in the Ohm range. Although prescriptions
(i) and (ii) have been stressed repeatedly, spin leakage in the current
or voltage probes is often overlooked although they may alter significantly
the signal (an example is provided in Appendix \ref{sec:Revisiting-the-bipolar}).

The goal of this section is to examine the transition from open to
closed geometries as a strategy to enhance the non-local signal used
for instance to read or sense 1 bit, and to study its interaction
with the doubling of spin injector electrodes. The signal generated
in our cross geometry will be systematically compared with the spin
voltage in the standard lateral device. 

The following general features will be a guide: a large signal requires
a large total effective spin resistance for the device $R_{eff}$.
Since the latter can be interpreted as resulting from the addition
of four spin resistances $R_{eff,i}$ ($i=1-4$) in parallel corresponding
to the four arms of the cross, one will expect a small signal whenever
one or several of these spin resistances are significantly smaller
than the others since they will short the other arms. If all effective
spin resistances $R_{eff,i}$ are commensurate, in order to have a
large $R_{eff,i}$, it is then necessary that $R_{c,i}\gg R_{N}$
simultaneously for all four terminals as explained in section \ref{subsec:Spin-voltage-andnonlocalresistance}.

(As an aside remark we wish to bring to the reader's attention that
the expressions derived in the literature\cite{fert_theory_1996,hershfield_charge_1997}
for the bipolar spin switch transistor of Johnson remarkably show
the enhancements characteristic of closed geometries over open geometries
with a typical scaling with interface resistance $R_{c}$ at small
enough distance between the ferromagnetic terminals. As discussed
in Appendix \ref{sec:Revisiting-the-bipolar} this is due to neglecting
spin leakage to the current drain; when this leakage is taken into
account, the signal is actually found to scale with $R_{N}$ as in
open geometries.)

\subsection{Van der Pauw cross with two ferromagnets.\label{sub:Van-der-Pauw2f}}

\subsubsection{Open geometry with two ferromagnets.\label{subsec:Standard-(open)-geometry}}

For the sake of comparison we first consider the case when the detector
electrodes are paramagnets as in the cross arms $I-IV$. When $F1$
and $F3$ are identical ferromagnets at the same distance of origin
($l_{1}=l_{3})$ while $F2$ and $F4$ are identical paramagnets with
spin resistance $R_{N}$, the spin resistance mismatches at each terminal
are 
\begin{equation}
X_{1}=X_{3}=X;\; X_{2}=X_{4}=1
\end{equation}
while: 
\begin{eqnarray}
\widetilde{PR_{1}} & = & \widetilde{PR_{3}}=\widetilde{PR}\\
\widetilde{PR_{2}} & = & \widetilde{PR_{4}}=0
\end{eqnarray}
and defining $\delta_{F}^{\pm}$ for arms $I$ and $III$:
\begin{equation}
\delta_{F}^{\pm}=\delta_{1/3}^{\pm}=\frac{\left(X+1\right)}{2}\;\exp l_{1}\pm\frac{\left(X-1\right)}{2}\;\exp-l_{1}.
\end{equation}

\begin{equation}
\delta_{2/4}^{\pm}=\;\exp l_{2/4}
\end{equation}

so that:
\begin{eqnarray}
R_{nl} & = & \frac{\sigma_{1}\sigma_{3}\; R_{N}\;\left(\widetilde{PR}\right)^{2}}{2\left[\delta_{F}^{-}\delta_{F}^{+}+\left(\delta_{F}^{+}\right)^{2}\right]}\nonumber \\
 & = & \frac{\sigma_{1}\sigma_{3}\; R_{N}\;\left(\widetilde{PR}\right)^{2}}{\left[\left(1+X\right)^{2}\;\exp l+\left(X^{2}-1\right)\right]}
\end{eqnarray}
where $l=2L_{1}/l_{N}$ the total length separating the ferromagnetic
terminals (in units of the spin diffusion length in the paramagnet)
and where we have defined $\sigma_{1}=\pm1,\;\sigma_{3}=\pm1$ to
index the majority spin directions of electrodes $F1$ and $F3$ (relative
to an absolute axis).

In the standard geometry, since one has a spin valve it is customary
to quote the resistance variation when one ferromagnet magnetization
is switched; this is twice the maximum value $R_{P}$:

\begin{equation}
\delta R_{nl,0}=R_{P}-R_{AP}=\frac{2\; R_{N}\;\left(\widetilde{PR}\right)^{2}}{\left[\left(1+X\right)^{2}\;\exp l+\left(X^{2}-1\right)\right]}.\label{eq:rnl}
\end{equation}
 This generalizes the expression found in the literature for the non-local
resistance variation: more precisely the result quoted by Jedema and
coll.\cite{jedema_electrical_2001} corresponds to the case of vanishing
interface resistance, so that $X=R_{F}/R_{N}$ ($=1/M$ using Jedema
and coll. notations\cite{jedema_electrical_2001}).

Although the geometry is open (in the sense that spin current can
leak easily since terminals $F2/F4$ do not hinder its flow ($X_{2}=X_{4}=1$)
), it is interesting to observe that the larger the resistance mismatch
$X$ at the ferromagnets, the larger the signal (for instance, if
we set $P=1$ in Eq. (\ref{eq:rnl}) one gets that $\delta R_{nl,0}\propto X^{2}/\left[\left(1+X\right)^{2}\;\exp l+\left(X^{2}-1\right)\right]$
which is an increasing function of $X$). This means that large resistance
mismatches are already beneficial and increase the spin accumulation
although there is some spin leakage. 

The largest value is at short distance when the denominator of Eq.
(\ref{eq:rnl}) is $2X\left(X+1\right)$. In the tunneling regime
$X\gg1$, $\delta R_{nl,0}\propto R_{N}$. In the opposite limit $X\ll1$,
the signal will be even smaller since it scales as $R_{c}+R_{F}\ll R_{N}$:
\[
\delta R_{nl,0}(l\longrightarrow0)\geq\inf\left(P_{c},\; P_{F}\right)^{2}\left(R_{c}+R_{F}\right)
\]
and 
\[
\delta R_{nl,0}(l\longrightarrow0)\leq\sup\left(P_{c},\; P_{F}\right)^{2}\left(R_{c}+R_{F}\right).
\]
 Such a scaling which is at most $\sim R_{N}$ or even below ($\ll R_{N}$)
is characteristic of open geometries.

Let us compare to the non-local resistance variation for the lateral
geometry with identical parameters (same distance $l$ between injector
and detector, same spin resistance mismatch $X$ at ferromagnets,
same cross-section for the paramagnet channel connecting the ferromagnets)
\cite{takahashi_spin_2006}:

\begin{equation}
\delta R_{nl,lateral}=\frac{4R_{n}\;\widetilde{PR}^{2}}{\left[\left(2X+1\right)^{2}\exp l-\exp-l\right]}.
\end{equation}
We plot on Fig. \ref{fig:lateral} the ratio of the cross signal versus
the one in the lateral geometry: 
\begin{equation}
m(X,\; l)=\frac{\delta R_{nl,0}}{\delta R_{nl,lateral}}=\frac{\left[\left(2X+1\right)^{2}\exp l-\exp-l\right]}{2\left[\left(1+X\right)^{2}\;\exp l+\left(X^{2}-1\right)\right]}.
\end{equation}

In the short distance limit:
\[
m(X,\; l\longrightarrow0)\longrightarrow1
\]
 which is expected since the cross and the lateral geometries are
then identical. At large distance, the signal is larger in the cross
geometry whenever $X>1/\sqrt{2}$ (which includes the tunneling regime
at the ferromagnets and also part of the transparent regime). The
impact of a large value of $X$ is moderate (about $10\;\%$ between
$X=10-100$) in stark contrast to what we will observe in a closed
geometry. The general behaviour is easy to understand: in the tunneling
regime spin current leakage is less pronounced in the cross geometry
since the paramagnetic drain is further away (while the detector ferromagnet
and paramagnetic counter-electrodes are at the same distance). In
the transparent regime, by a similar reasoning one gets the opposite.

\begin{center}
\begin{figure}
\begin{centering}
\includegraphics[width=1\columnwidth]{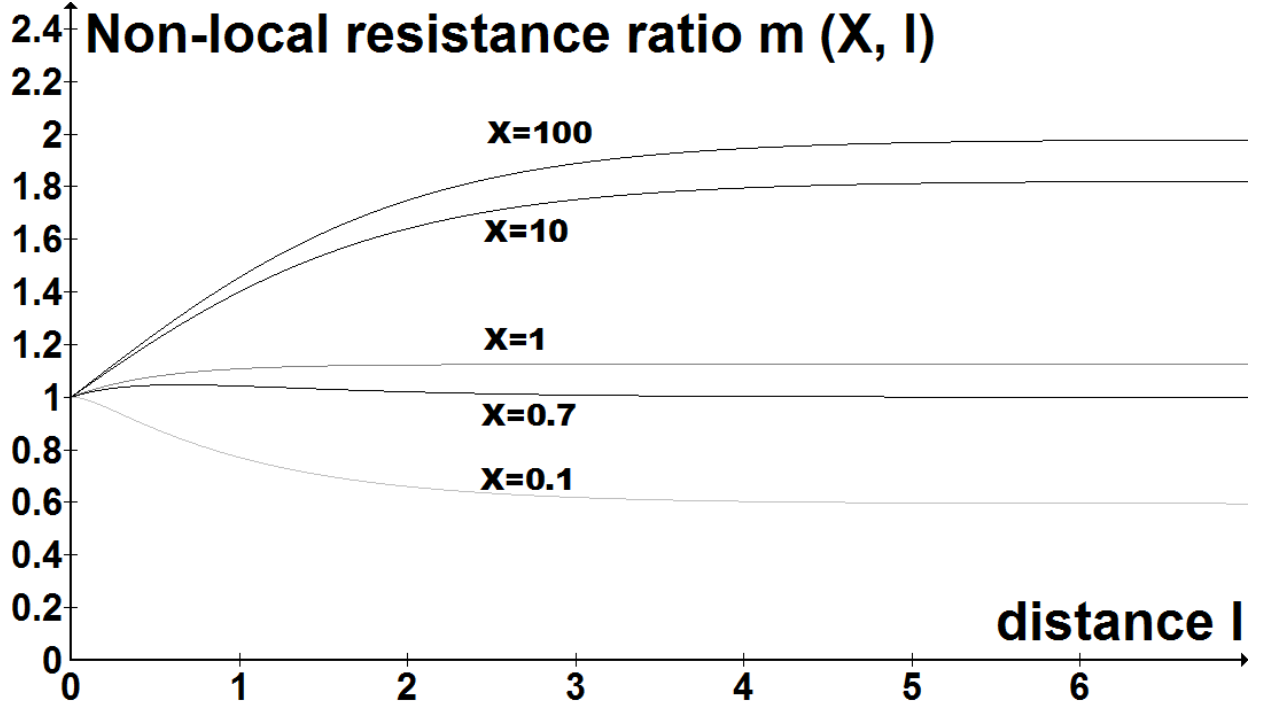}
\par\end{centering}

\caption{Non-local resistance in cross setup compared with lateral setup. $l$
is the distance in units of spin diffusion length between injector
and detector in either setup. In tunneling regime ($X>1$) and in
part of the transparent regime, the signal is larger in the cross
geometry.\label{fig:lateral}}

\end{figure}

\par\end{center}

\subsubsection{Spin confining geometry with two ferromagnets.\label{subsec:Spin-confining-(closed)}}

It has been argued by Jaffres and coll.\cite{jaffres_spin_2010} that
spin confinement tends to increase non-local signals (resistance or
voltage) since spin accumulation is stronger whenever spin leaking
is hindered by large tunnel barriers. We can study this by considering
that electrodes $F2$ and $F4$ are normal paramagnets but that there
is a resistance mismatch 
\begin{equation}
Y=\frac{R_{N,0}+R_{c,0}}{R_{N}}
\end{equation}
 where $R_{N,0}$ and $R_{c,0}$ are the spin resistance of $F2$
and $F4$ and the interface resistance between them and the central
cross. One can therefore interpolate between an \emph{open geometry}
(the standard geometry, i.e. $Y=1$) and a \emph{closed geometry}
($Y\gg1$ and $X\gg1$). We will now study the transition from one
regime to the other.

Let us assume that terminals $F1$ and $F3$ are identical ferromagnets
at identical distance $l_{1}=l_{3}=l_{0}$ of the cross center $O$,
that $F2$ and $F4$ are identical paramagnets also at the same distance
$l_{2}=l_{4}=l_{0}$ from $O$ ($ABCD$ is then a square); the total
distance from the injector $F1$ to the detector $F3$ is $l=2l_{0}$. 

We also set:

\begin{equation}
X_{1}=X_{3}=X;\; X_{2}=X_{4}=Y
\end{equation}
and: 
\begin{eqnarray}
\widetilde{PR_{1}} & = & \widetilde{PR_{3}}=\widetilde{PR},\\
\widetilde{PR_{2}} & = & \widetilde{PR_{4}}=0.
\end{eqnarray}

There are several obvious ways to create this spin resistance mismatch
$Y$: one is to deposit a tunnel barrier between the terminals $F2$
and $F4$ and the central cross ($R_{c,0}\neq0$); another is to use
the same paramagnet for both $F2$, $F4$ and the central cross but
have a different cross-section ($R_{c,0}=0$ but $R_{N,0}=\rho_{N}^{*}l_{N}/A_{N,0}\neq R_{N}=\rho_{N}^{*}l_{N}/A_{N}$).

One ends up with:
\begin{equation}
R_{nl}=\frac{\sigma_{1}\sigma_{3}\; R_{N}\;\left(\widetilde{PR}\right)^{2}}{2\left[\delta_{F}^{-}\delta_{F}^{+}+\frac{\delta_{N}^{-}}{\delta_{N}^{+}}\left(\delta_{F}^{+}\right)^{2}\right]}
\end{equation}
where 
\[
\delta_{F}^{\pm}=\delta_{1/3}^{\pm},\;\;\delta_{N}^{\pm}=\delta_{2/4}^{\pm}.
\]
{[}The definitions for $\delta_{i}^{\pm}$ are given in Eq. (\ref{eq:delta}).{]} 

\begin{widetext}

Let us define
\begin{eqnarray}
\Delta_{2F}(X,\; Y,\; l)= & 2\left[\left(Y+1\right)\;\exp l_{0}+\left(Y-1\right)\;\exp-l_{0}\right]^{-1}\nonumber \\
 & \left\{ \left(Y+1\right)\;\exp l_{0}\;\right.\left[\left(X+1\right)^{2}\;\exp2l_{0}+\left(X^{2}-1\right)\right]\nonumber \\
 & -\left(Y-1\right)\;\exp-l_{0}\left.\left[\left(X-1\right)^{2}\;\exp-2l_{0}+\left(X^{2}-1\right)\right]\right\} 
\end{eqnarray}

\end{widetext}

and 
\begin{eqnarray}
\Delta_{0}(X,\; l) & = & \left[\left(2X+1\right)^{2}\;\exp2l_{0}-\exp(-l)\right]
\end{eqnarray}
so that: 
\begin{equation}
\delta R_{nl}=\frac{4\; R_{N}\;\left(\widetilde{PR}\right)^{2}}{\Delta_{2F}}
\end{equation}

The signal will be largest at short distance for which: 
\[
\inf\left(P_{c},\; P_{F}\right)^{2}\; R_{N}\;\frac{XY}{X+Y}\leq\delta R_{nl}(l\longrightarrow0)
\]
and:
\[
\delta R_{nl}(l\longrightarrow0)\leq\sup\left(P_{c},\; P_{F}\right)^{2}\; R_{N}\;\frac{XY}{X+Y}
\]
 so that $\delta R_{nl}(l\longrightarrow0)$ scales as $R_{N}\frac{XY}{X+Y}$.

Therefore in the strong tunneling regime ($X\gg1$ and $Y\gg1$),
$\delta R_{nl}$ scales as 
\[
\inf(X,\; Y)\; R_{N}\gg R_{N}
\]
 yielding much larger signals than in the lateral geometry. But whenever
either of $X$ or $Y$ is in the transparent regime, $\delta R_{nl}$
will scale as the smaller of the two and so will be at most at the
scale of $R_{N}$ as predicted for open geometries (see \ref{subsec:Spin-voltage-andnonlocalresistance}).
\begin{subequations}
In order to check the impact of increasing the spin resistance mismatches
at the various terminals to the signal in the lateral geometry we
consider a non-local resistance ratio $m(X,\; Y,\; l)$ per:
\begin{eqnarray}
m(X,\; Y,\; l) & = & \frac{\delta R_{nl}(X,\; Y,\; l)}{\delta R_{nl,lateral}(X,\; l)}\\
 & = & \frac{\Delta_{0}(X,\; l)}{\Delta_{2F}(X,\; Y,\; l)}.
\end{eqnarray}
The relative increase of the signal is then $m(X,\; Y,\; l)-1$.
\end{subequations}
At $l=0$:
\begin{equation}
m(X,\; Y,\; l=0)\longrightarrow\frac{Y\;\left(X+1\right)}{X+Y}.
\end{equation}
At large distance:
\begin{equation}
m(X,\; Y,\; l)\longrightarrow\frac{\left(2X+1\right)^{2}}{2\left(X+1\right)^{2}}
\end{equation}
which is larger than $1$ whenever $x\geq1/\sqrt{2}$ and more generally
belongs to the interval $\left[0.5;\;2\right]$. The large distance
behaviour is of course less interesting since the non-local signal
is weaker, so we will discuss in detail only the short distance behaviour.

At short distances, large values are achieved whenever both parameters
$X$ and $Y$ are large (spin confining regime, closed geometry).
For $X\gg Y\gg1$, $m(X,\; Y,\; l=0)\sim Y$; while for $Y\gg X\gg1$,
$m(X,\; Y,\; l=0)\sim X+1$; finally for $X\sim Y\gg1$, $m(X,\; Y,\; l=0)\sim\left(X+1\right)/2$.
This third situation is probably the easiest to achieve: indeed if
we want to gain at least an order of magnitude the other cases would
imply that one of the mismatches is at least two order of magnitudes
larger while when $X\sim Y\gg1$ they both have the same magnitude.
For a realistic value $X\sim10$ in the weak tunneling regime (where
the contact resistance is usually metallic) this yields an enhancement
by a factor up to $5$ (see Section \ref{sec:4f} for a discussion
of realistic parameters $X\sim10$).

We plot in the next figures (Fig. \ref{fig:2f1}-\ref{fig:2f3}) the
length dependence for various values of $X$ and $Y$ in the weak
tunneling regime (keeping moderate values $\left(X,\, Y\right)\leq10$)
covering the three situations described in the previous paragraph
(larger $X$ , larger $Y$ or equal $X=Y$). 

\begin{center}
\begin{figure}
\begin{centering}
\includegraphics[width=1\columnwidth]{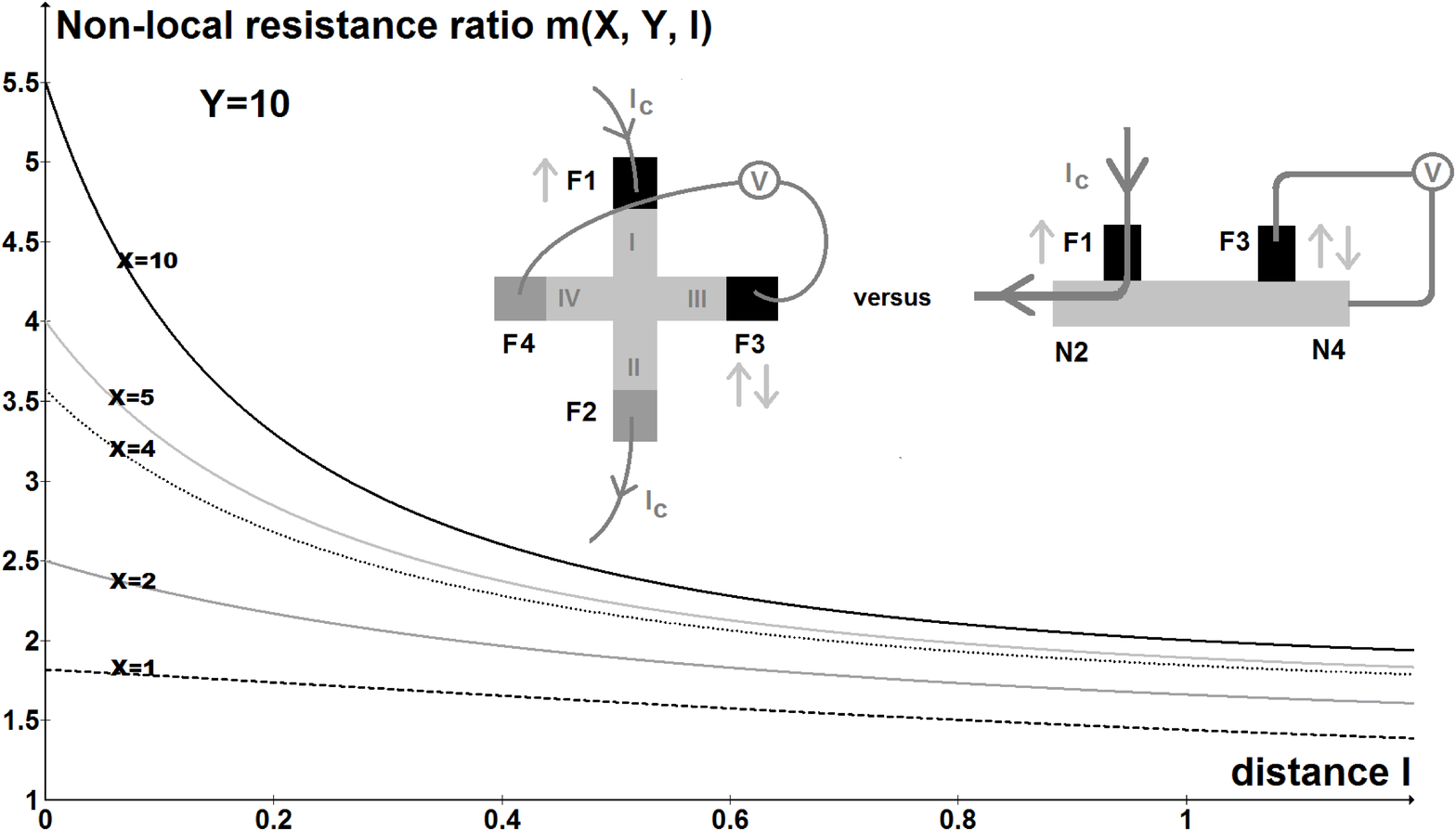}
\par\end{centering}

\caption{Ratio of non-local resistance variation in cross versus lateral setup
as a function of distance $l$ between injector $F1$ and detector
$F3$ (in units of spin diffusion length $l_{N}$) for spin resistance
mismatches $Y=10$ (at terminals $F2/F4$) and $X=1-10$ (at terminals
$F1/F3$). $F1$ and $F3$ are identical ferromagnetic metals while
$F2$ and $F4$ are paramagnets in this section with a spin resistance
mismatch with the central paramagnet.\label{fig:2f1}}

\end{figure}

\par\end{center}

\begin{center}
\begin{figure}
\begin{centering}
\includegraphics[width=1\columnwidth]{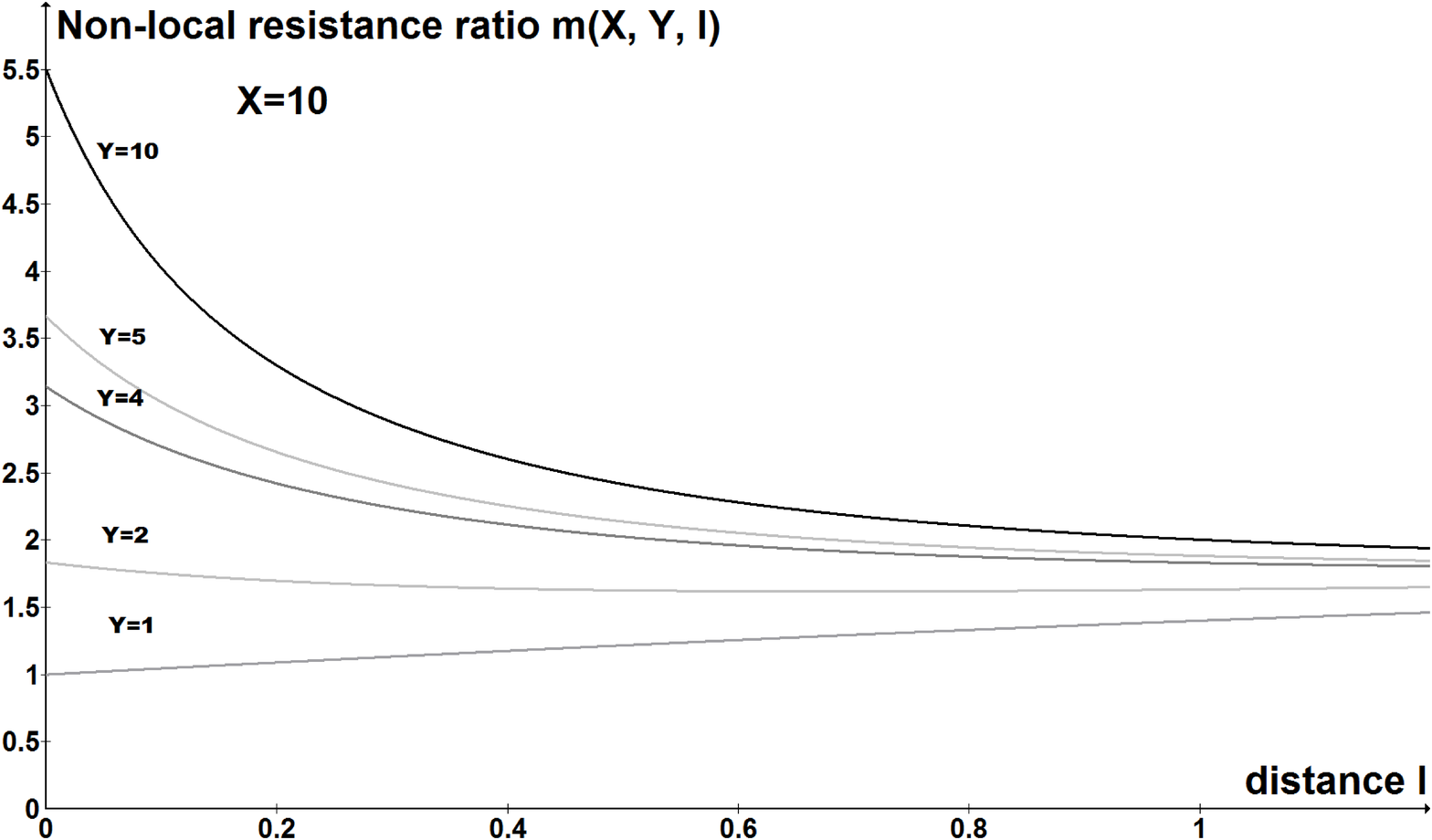}
\par\end{centering}

\caption{Ratio of non-local resistance variation in cross versus lateral setup
as a function of distance $l$ between injector $F1$ and detector
$F3$ (in units of spin diffusion length $l_{N}$) for spin resistance
mismatches $Y=1-10$ (at terminals $F2/F4$) and $X=10$ (at terminals
$F1/F3$). $F1$ and $F3$ are identical ferromagnetic metals while
$F2$ and $F4$ are paramagnets in this section with a spin resistance
mismatch with the central paramagnet.\label{fig:2f2}}
\end{figure}

\par\end{center}

\begin{center}
\begin{figure}
\begin{centering}
\includegraphics[width=1\columnwidth]{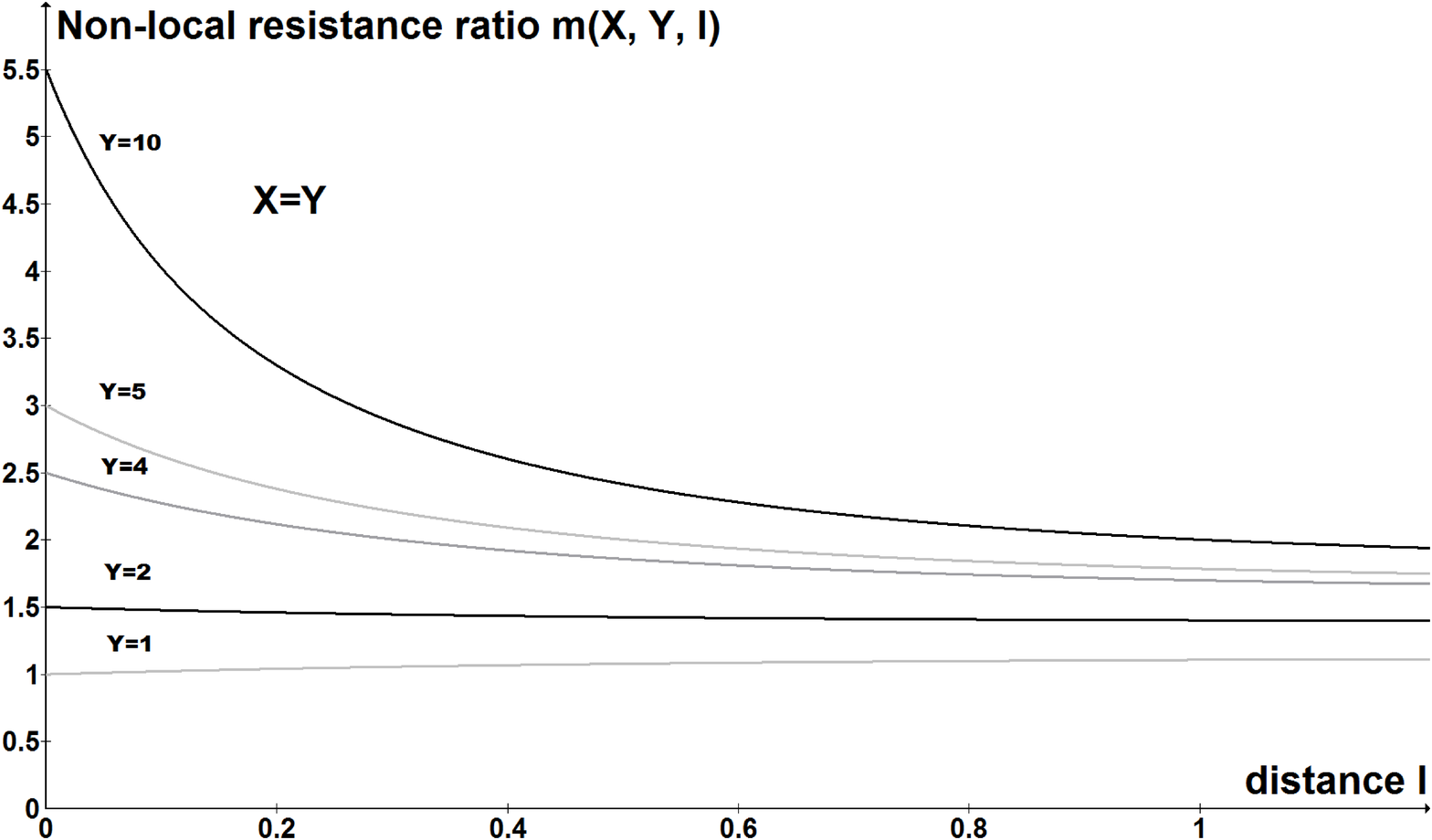}
\par\end{centering}

\caption{Ratio of non-local resistance variation in cross versus lateral setup
as a function of distance $l$ between injector $F1$ and detector
$F3$ (in units of spin diffusion length $l_{N}$) for spin resistance
mismatches $Y=1-10$ (at terminals $F2/F4$) and $X=Y$ (at terminals
$F1/F3$). $F1$ and $F3$ are identical ferromagnetic metals while
$F2$ and $F4$ are paramagnets in this section.\label{fig:2f3}.}
\end{figure}

\par\end{center}

Still in the weak tunneling regime we also plot on Fig. \ref{fig:f2fig4}
the non-local resistance ratio for fixed distance $l=0.1-0.3$ as
a function of spin resistance mismatch $X$ (assuming $Y=X$). Values
in the range $l=0.1-0.3$ are quite reasonable experimentally (for
instance for $Cu$ at ambient temperature $l_{N}=300\; nm$ while
a distance $L=100\; nm$ is within reach of lithography). At such
distances the enhancement can still be several hundred of percents
as can be seen in Fig. \ref{fig:f2fig4} unless $X<1$ (transparent
regime).

\begin{center}
\begin{figure}
\begin{centering}
\includegraphics[width=1\columnwidth]{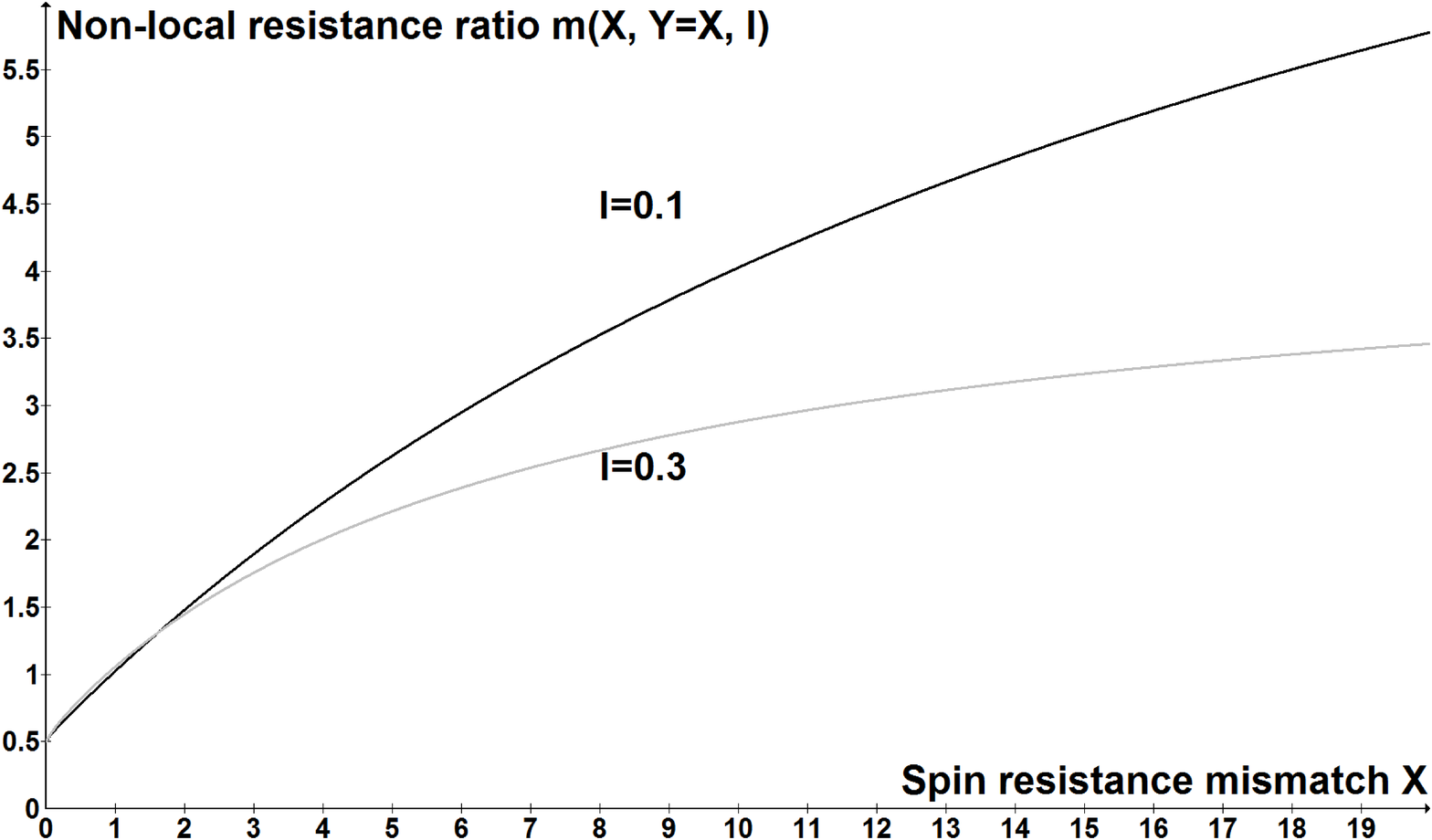}
\par\end{centering}

\caption{Non-local resistance ratio in cross versus lateral setup at fixed
distances $l=0.1-0.3$ between injector $F1$ and detector $F3$ (in
units of spin diffusion length $l_{N}$) for spin resistance mismatches
$X=Y$.\label{fig:f2fig4}}
\end{figure}

\par\end{center}

The transparent regime is less interesting in terms of an increase
of the signal since the geometry is now open. Yet whenever $Y$ is
larger than $1$, even if $X$ is in the transparent regime, one still
gets an increase of the signal at short distance (Fig. \ref{fig:f2fig7}).
But if both parameters are smaller than unity, the signal gets reduced
by several tens of percent when compared against the lateral setup
as can be seen on Fig. \ref{fig:f2fig9}-\ref{fig:f2fig10}. 

\begin{center}
\begin{figure}
\begin{centering}
\includegraphics[width=1\columnwidth]{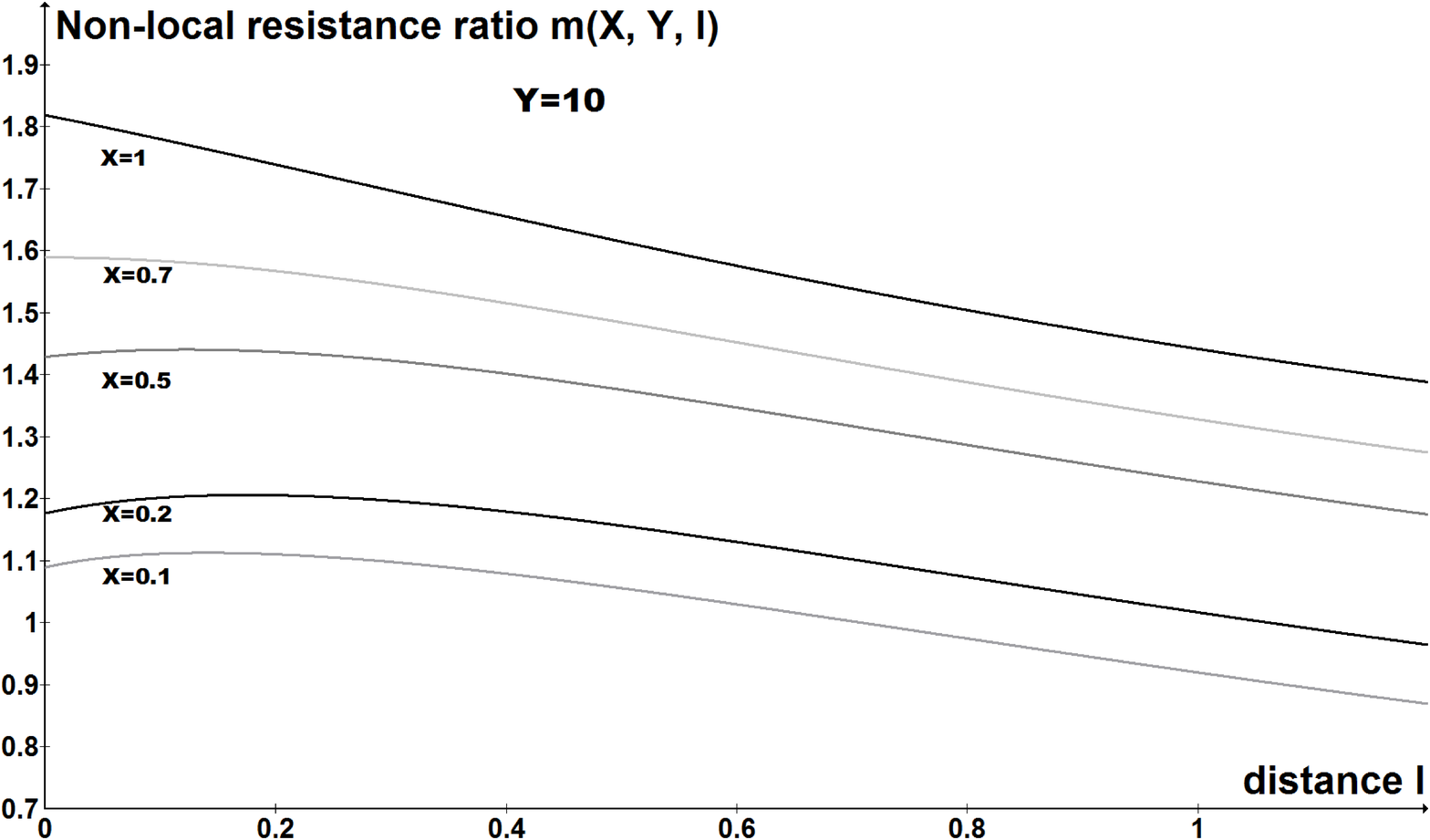}
\par\end{centering}

\caption{Ratio of non-local resistance variation in cross versus lateral setup
as a function of distance $l$ between injector $F1$ and detector
$F3$ for spin resistance mismatches $Y=10$ (at terminals $F2/F4$)
in the weak tunneling regime and $X=0.1-1$ (at terminals $F1/F3$)
in the transparent regime. \label{fig:f2fig7}}
\end{figure}

\par\end{center}

\begin{center}
\begin{figure}
\begin{centering}
\includegraphics[width=1\columnwidth]{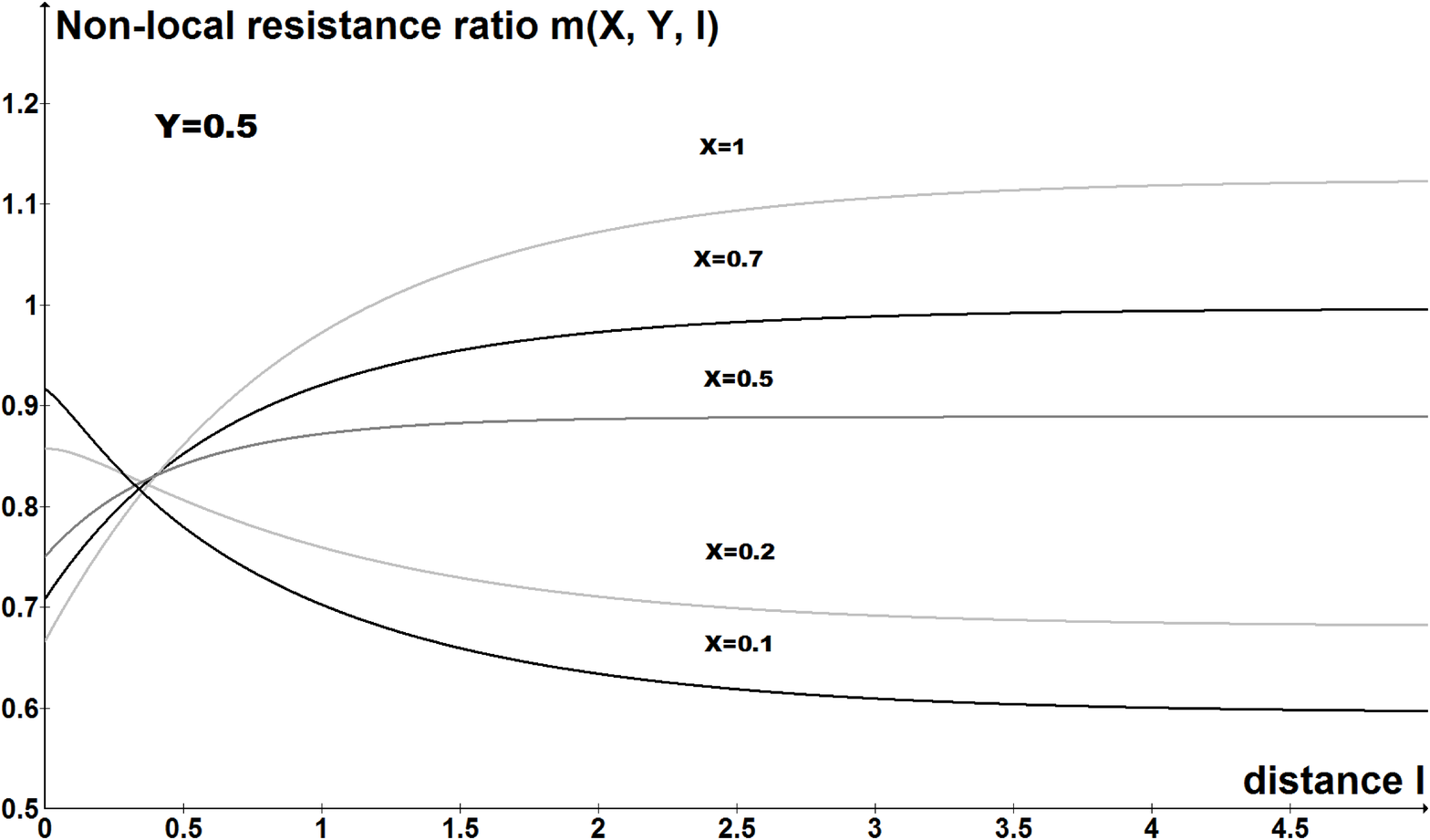}
\par\end{centering}

\caption{Ratio of non-local resistance variation in cross versus lateral setup
as a function of distance $l$ between injector $F1$ and detector
$F3$ for spin resistance mismatches $Y=0.5$ (at terminals $F2/F4$)
and $X=0.1-1$ (at terminals $F1/F3$). All electrodes are in the
transparent regime.\label{fig:f2fig9}}
\end{figure}

\par\end{center}

\begin{center}
\begin{figure}
\begin{centering}
\includegraphics[width=1\columnwidth]{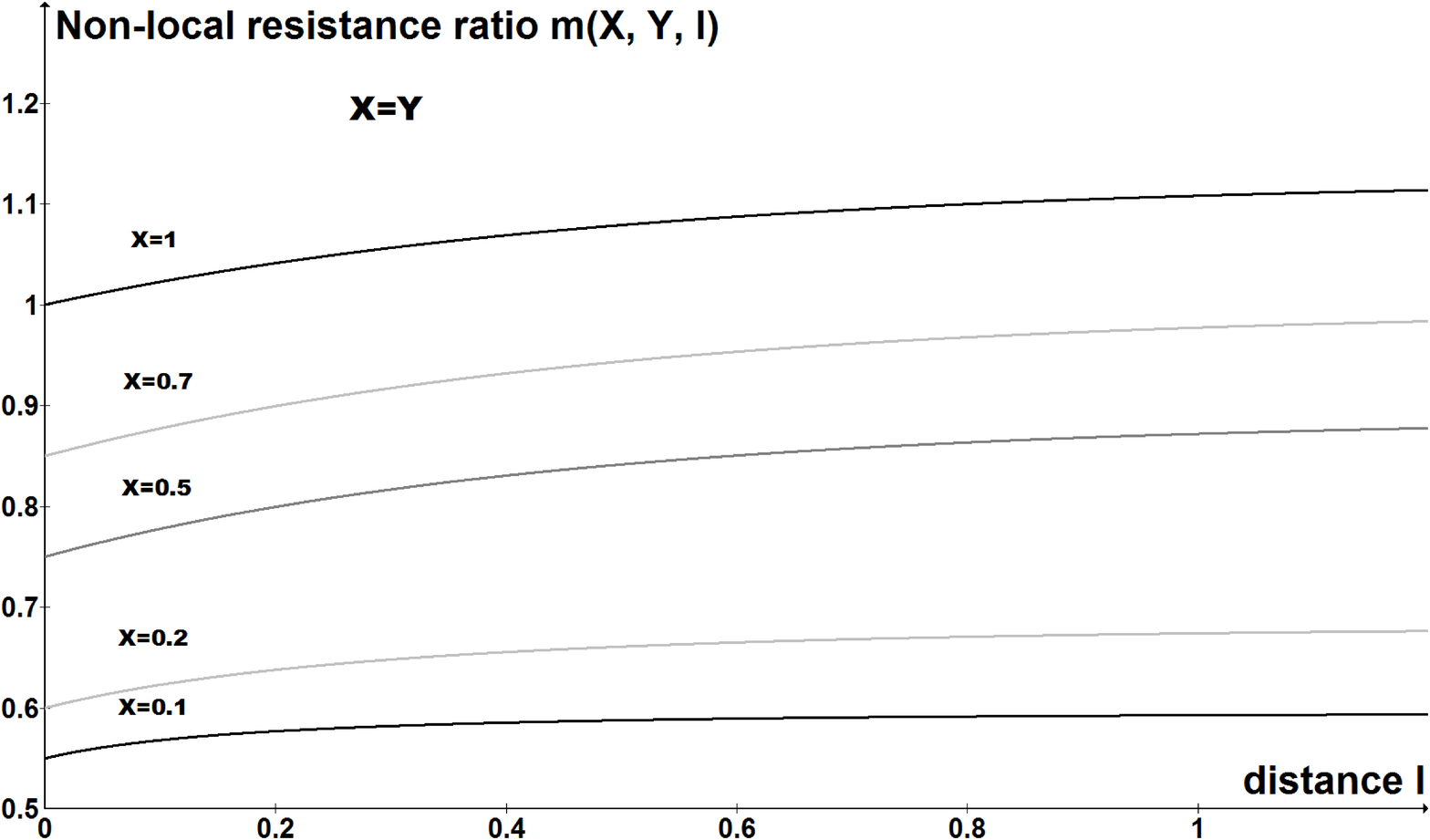}
\par\end{centering}

\caption{Ratio of non-local resistance variation in cross versus lateral setup
as a function of distance $l$ between injector $F1$ and detector
$F3$ for spin resistance mismatches $Y=0.1-1$ (at terminals $F2/F4$)
and $X=Y$ (at terminals $F1/F3$). All electrodes are in the transparent
regime.\label{fig:f2fig10}}
\end{figure}

\par\end{center}

To summarize this section devoted to the cross geometry with two ferromagnets
(one as charge source and spin injector, the other as spin accumulation
detector), we have confirmed the impact on spin confinement of tunnel
barriers even of moderate strength ($X=10$ might correspond to $R_{c}\sim10\;\Omega$
since $R_{N}$ is usually in the Ohm range). In the transparent limit,
the device under-performs when compared to the standard lateral spin
valve and should be avoided.

\subsection{Van der Pauw cross with three ferromagnets.\label{sub:3f}}

We now use two ferromagnets as spin injectors. When the (charge) source
and drain are in antiparallel orientation the signal is enhanced because
both ferromagnets acts as spin sources. But when they are parallel,
one is a spin source and the other a spin sink so that spin accumulation
is reduced. If the electrodes are identical ferromagnets, then the
signal will be doubled when compared with the case where there is
only one spin injector electrode. This is exactly what we found since
$R_{nl}\propto P_{eff,1}-P_{eff,2}=2P_{eff,1}$ if electrodes are
identical and antiparallel.

We assume that $F1-F2-F3$ are identical electrodes ($F1$ and $F2$
antiparallel while $F3$ can switch from one orientation to the other)
and $F4$ is a paramagnet. The spin resistance mismatches are set
as:
\begin{equation}
X_{1}=X_{2}=X_{3}=X;\; X_{4}=Y.
\end{equation}
We also assume $l_{1}=l_{2}=l_{3}=l_{4}$ and use the same definitions
as in previous section: 
\begin{eqnarray}
\widetilde{PR_{1}} & = & \widetilde{PR_{2}}=\widetilde{PR_{3}}=\widetilde{PR}\\
\widetilde{PR_{4}} & = & 0
\end{eqnarray}

Then:
\begin{equation}
R_{nl}=\frac{2\sigma_{1}\sigma_{3}\; R_{N}\;\left(\widetilde{PR}\right)^{2}}{\left[3\delta_{F}^{-}\delta_{F}^{+}+\frac{\delta_{N}^{-}}{\delta_{N}^{+}}\left(\delta_{F}^{+}\right)^{2}\right]}
\end{equation}
where $\sigma_{1}=\pm1$ and $\sigma_{3}=\pm$ refer to the majority
spin direction of ferromagnets $F1$ and $F3$ relative to an absolute
axis.

\begin{widetext}

Let us define
\begin{eqnarray}
\Delta_{3F}(X,\; Y,\; l) & = & \left[2\left(Y+1\right)\;\exp l_{0}+2\left(Y-1\right)\;\exp-l_{0}\right]^{-1}\nonumber \\
 & \times & \left\{ \left(Y+1\right)\;\exp l_{0}\;\right.\left[2\left(X+1\right)^{2}\;\exp2l_{0}\right.\left.-\left(X-1\right)^{2}\;\exp-2l_{0}+\left(X^{2}-1\right)\right]\nonumber \\
 & - & \left(Y-1\right)\;\exp-l_{0}\left[2\left(X-1\right)^{2}\;\exp-2l_{0}\right.\left.\left.-\left(X+1\right)^{2}\;\exp2l_{0}+\left(X^{2}-1\right)\right]\right\} 
\end{eqnarray}

\end{widetext}

Then: 
\begin{equation}
\delta R_{nl}=R_{P}-R_{AP}=\frac{4\; R_{N}\;\left(\widetilde{PR}\right)^{2}}{\Delta_{3F}}
\end{equation}
 where $P/AP$ refer to the direction of $F3$ relative to $F1$.
Therefore the non-local resistance roughly scales as:
\begin{equation}
\delta R_{nl}\sim\frac{C\; R_{N}\; X^{2}}{\Delta_{3F}(X,,\; Y,\; l)}
\end{equation}
(where $C$ is a constant; we have used upper and lower bounds on
$\left(\widetilde{PR}\right)^{2}$ as in section \ref{sub:Van-der-Pauw2f}
above).

The largest signal is found at small distance when:
\[
\delta R_{nl}\longrightarrow R_{N}\;\frac{XY}{3Y+X}.
\]
 We observe a characteristic quadratic dependence of the numerator
against a linear one in the denominator as a function of spin resistance
mismatch: this is what ensures that in the tunneling limit, $\delta R_{nl}$
can scale as $R_{c}$ (here as the smaller of $R_{N}X\gg R_{N}$ or
$R_{N}Y\gg R_{N}$. But as soon the geometry is open due to one or
both spin resistance mismatches in the transparent limit, $\delta R_{nl}$
scales as $R_{n}$ or below.

The ratio of the non-local resistance with regards to the standard
lateral setup is then:
\begin{equation}
m(X,\; Y,\; l)=\frac{\Delta_{0}}{\Delta_{3F}}
\end{equation}

By varying the spin resistance mismatch $Y$ at terminal $F4$ one
can interpolate between an open geometry ($Y\leq1$) and a closed
one ($Y\gg1$ while $X\gg1$).

Let us make some general comments comparing the three ferromagnets
cross setup with the two ferromagnets cross discussed in the previous
section \ref{subsec:Spin-confining-(closed)}.

When $X=Y$, one gets $\Delta_{2}=2\Delta_{3}$ so that 
\[
m_{3F}(X,\; Y=X,\; l)=2m_{2F}(X,\; Y=X,\; l).
\]
The non-local signal is exactly twice that found for the setup with
two ferromagnets. Indeed when all terminals have the same spin resistance
mismatch, the effective resistances $R_{eff}$ are identical in both
setups since the parameters are identical; but the total effective
polarization at injector $P_{eff,1}-P_{eff,2}=2P_{eff,1}$ is doubled
in the 3 ferromagnet setup ($P_{eff,2}=0$ in the two ferromagnet
setup). 

For $X>Y$ the effective resistance (see \ref{subsec:Spin-accumulation.})
for the 3 ferromagnet setup becomes larger than that of the 2 ferromagnet
setup (see \ref{subsec:Spin-confining-(closed)}; this is quite clear
since $R_{eff,i}(X_{i})$ is an increasing function of $X_{i}$ so
that 
\begin{eqnarray*}
R_{eff,3F} & = & \left[\frac{3}{R_{eff,1}(X)}+\frac{1}{R_{eff,1}(Y)}\right]^{-1}>\\
R_{eff,2F} & = & \left[\frac{2}{R_{eff,1}(X)}+\frac{2}{R_{eff,1}(Y)}\right].^{-1}
\end{eqnarray*}
 This means that the non-local signal will be more than doubled with
respect to the 2 ferromagnet setup studied in \ref{subsec:Spin-confining-(closed)}.
This effect does not require a totally closed geometry to occur.

For $X<Y$, the 3 ferromagnet geometry is by the same arguments less
spin confining since we have three terminals with a smaller spin resistance
against only two in the 2 ferromagnet setup. However the doubling
of effective polarizations remain which may mitigate the decrease
of effective spin resistance. But as a rule in order to achieve stronger
signals one should seek the condition $X\geq Y$.

\subsubsection{Open geometry ($Y=1$).\label{sub:Open-geometry-3f}}

We first consider the case when the paramagnetic terminal has no tunnel
barrier but is perfectly matched in terms of spin resistance to the
central cross. This is in the terminology of Ref. \cite{jaffres_spin_2010}
an open geometry. Nevertheless as in the 2 ferromagnet geometry studied
in \ref{sub:Van-der-Pauw2f}, we will find that spin accumulation
can still be enhanced even though not all the terminals are tunnel
barriers.

The spin resistance mismatch $Y$ is set to unity: $X_{4}=Y=1.$

\begin{widetext}

The non-local resistance ratio to the two ferromagnet cross (with
$Y=1)$ is then: 
\begin{eqnarray}
m(X,\; Y=1,\; l) & =2 & \left[\left(2X+1\right)^{2}\;\exp l-\exp-l\right]\\
\times & \left[2\left(X+1\right)^{2}\;\exp l\right. & \left.-\left(X-1\right)^{2}\;\exp-l+\left(X^{2}-1\right)\right]^{-1}
\end{eqnarray}

\end{widetext}

At large distance its limit is: 
\begin{equation}
m(X,\; Y=1,\; l)\longrightarrow\left(\frac{2X+1}{X+1}\right)^{2}.
\end{equation}
which is always larger than $1$ whether $X$ is in the spin confining
($X>1$) or the transparent regime ($X<1$). 

At small distance the ratio tends to:
\begin{equation}
m(X,\; Y=1,\; l)\longrightarrow4\frac{X+1}{X+3}
\end{equation}
 which is larger than $2$ and can be as large as $4$ in the tunneling
regime (provided $X>1$) and is comprised in the interval $\left[4/3;\;2\right]$
for $X\leq1$ (transparent regime).

So although the geometry is open, the use of tunneling junctions does
have some spin confining effect, which here is reinforced by the use
of two injector electrodes resulting in an enhancement of the non-local
resistance which can be up to $300\%$ increase in the tunneling regime.
For a distance $l=0.2$ and $X=2-10$, $m=2.45-3.30$ ($145-230\;\%$
increase). As can be seen in the next figure (Fig. \ref{fig:f3-fig0})
the enhancement can be seen at all distances so that this geometry
is already better than the lateral one in terms of spin confinement.

\begin{center}
\begin{figure}
\includegraphics[width=1\columnwidth]{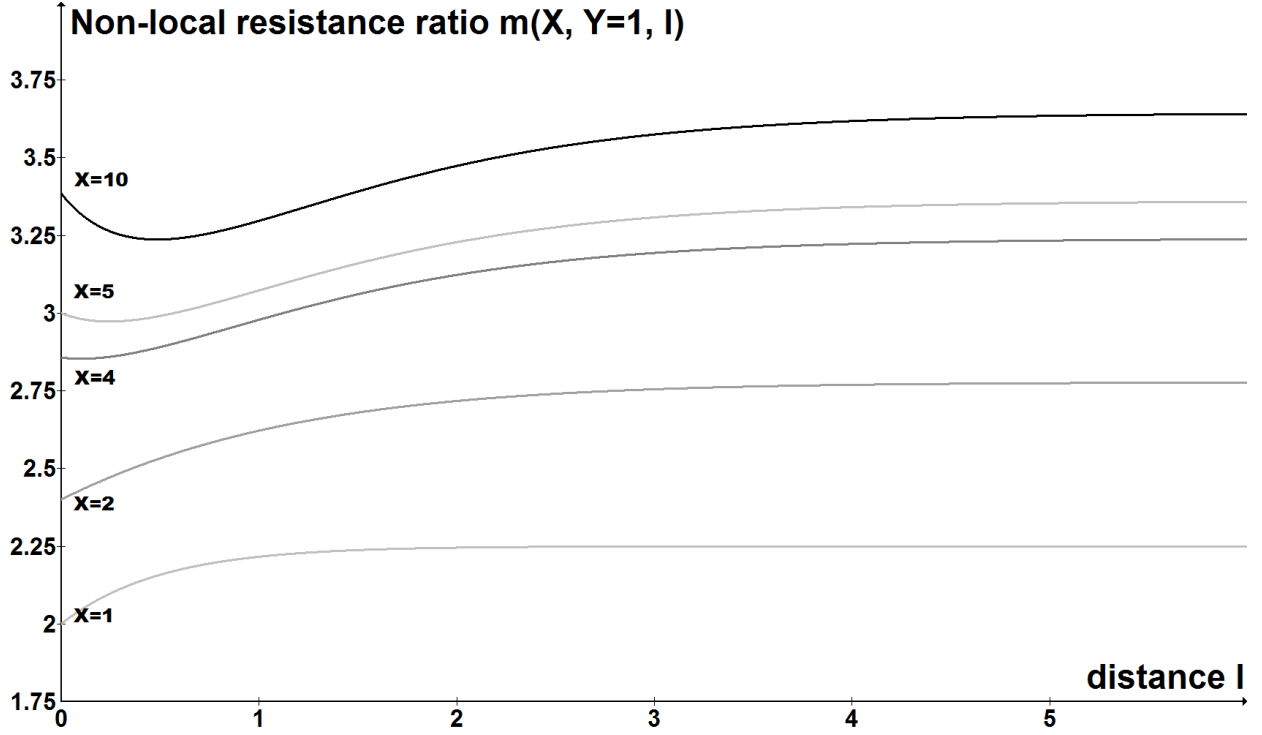}

\caption{Ratio of non-local resistance in cross geometry (with three ferromagnets)
versus lateral setup as a function of distance $l$ between injector
and detector. Spin resistance mismatch is $X=1-10$ at terminals $F1-F2-F3$
(weak tunneling regime). \label{fig:f3-fig0}}

\end{figure}

\par\end{center}

In the transparent regime ($X<1$), the doubling effect coming from
the two spin injectors compensate partly the spin leakage due to transparent
junctions so that there is always at least a $33\;\%$ increase at
short distance. The enhancement is still present at large distance
and can then reach up to $125\;\%$ as can be seen in Fig.\ref{fig:f3-fig0b}.

\begin{center}
\begin{figure}
\includegraphics[width=1\columnwidth]{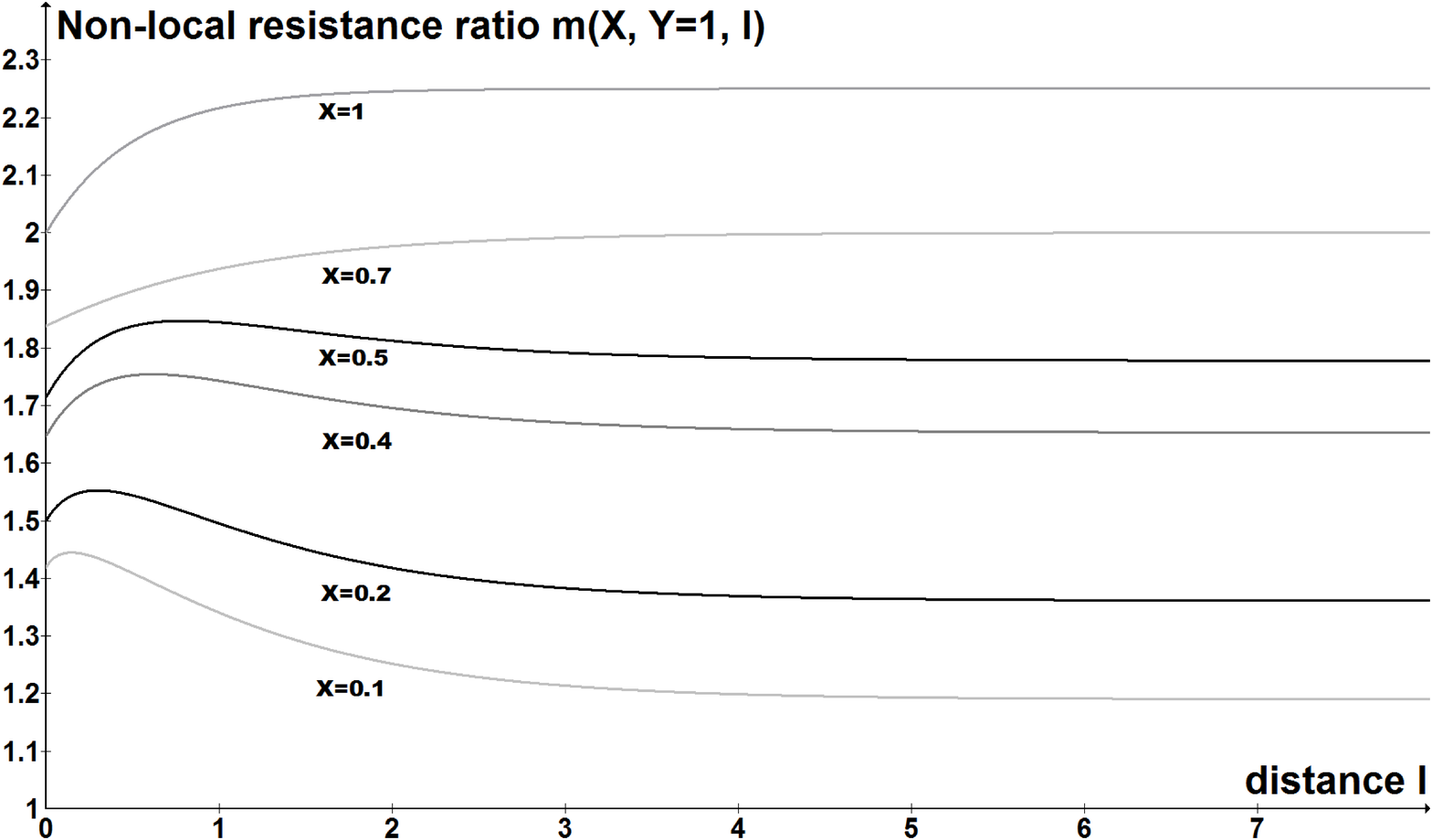}\caption{Ratio of non-local resistance in cross geometry (with three ferromagnets)
versus lateral setup as a function of distance $l$ between injector
and detector. Spin resistance mismatch is $X=0.1-1$ at terminals
$F1-F2-F3$ (transparent regime). There is still an enhancement of
the signal. \label{fig:f3-fig0b}}
\end{figure}

\par\end{center}

\subsubsection{General case.\label{sub:General-case.3f}}

We now conjugate the effects of spin confinement by tunnel barriers
and the doubling of injector terminals. At small distance one gets:
\begin{equation}
m(X,\; Y=1,\; l)\longrightarrow4Y\frac{X+1}{X+3Y}
\end{equation}
while at large distances: 
\begin{equation}
m(X,\; Y=1,\; l\longrightarrow\infty)\longrightarrow\left(\frac{2X+1}{X+1}\right)^{2}<4.
\end{equation}

Large values are achieved whenever both parameters $X$ and $Y$ are
large (both in the tunneling regime). 

Let us focus first on that tunneling regime.

There are then three interesting limits: 

(i) $X\gg Y\gg1$, which implies $m(X,\; Y,\; l=0)\sim4Y$; (note
also that the spin confinement is more pronounced in that limit than
in the 2 ferromagnet setup due to a larger effective spin resistance); 

(ii) $Y\gg X\gg1$ implying $m(X,\; Y,\; l=0)\sim\frac{4}{3}\left(X+1\right)$; 

(iii) finally for $X\sim Y\gg1$, $m(X,\; Y,\; l=0)\sim X+1$ (see
Fig. \ref{fig:f3-fig1}-\ref{fig:f3-fig3}). 

For a realistic value $X=Y\sim10$ (see section \ref{sub:Discussion.})
this yields an enhancement by a factor up to $11$ reaching therefore
an order of magnitude. At a finite distance the enhancement remains
considerable ranging from $m=3.3-6.6$ at $l=0.2$ for $X=10$ and
$Y=1-10$ (Fig. \ref{fig:f3-fig1}).

\begin{center}
\begin{figure}
\centering{}\includegraphics[width=1\columnwidth]{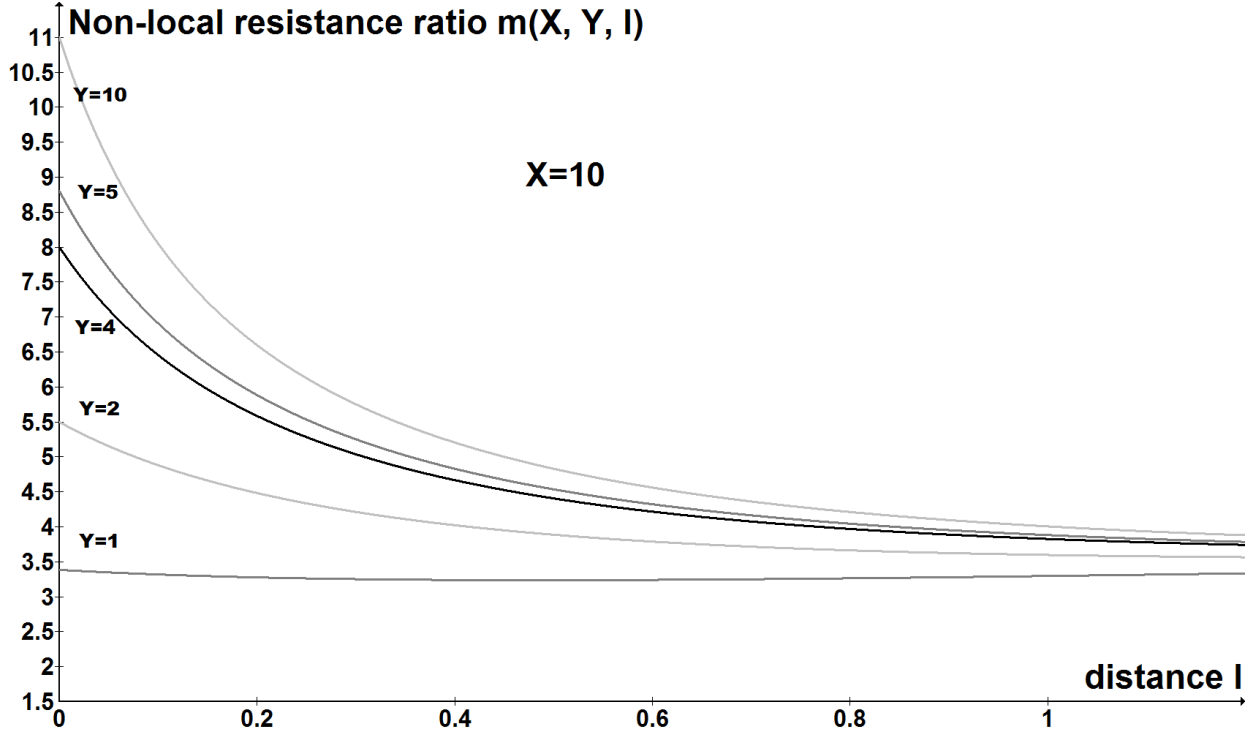}\caption{Ratio of non-local resistance in cross geometry (with three ferromagnets)
versus lateral setup as a function of distance $l$ between injector
and detector. Spin resistance mismatch is $X=10$ at terminals $F1-F2-F3$
and $Y=1-10$ at paramagnet $F4$. All electrodes are in the tunneling
regime.\label{fig:f3-fig1}}
\end{figure}

\par\end{center}

If one exchanges $X$ and $Y$ (with large $Y=10$ and $X=1-10$),
one gets weaker signals. This is normal, since when $Y$ is larger
than $X$, spin confinement is disfavored since there are then three
terminals with spin resistance mismatch smaller than the fourth (Fig.
\ref{fig:f3-fig2}). 

\begin{center}
\begin{figure}
\includegraphics[width=1\columnwidth]{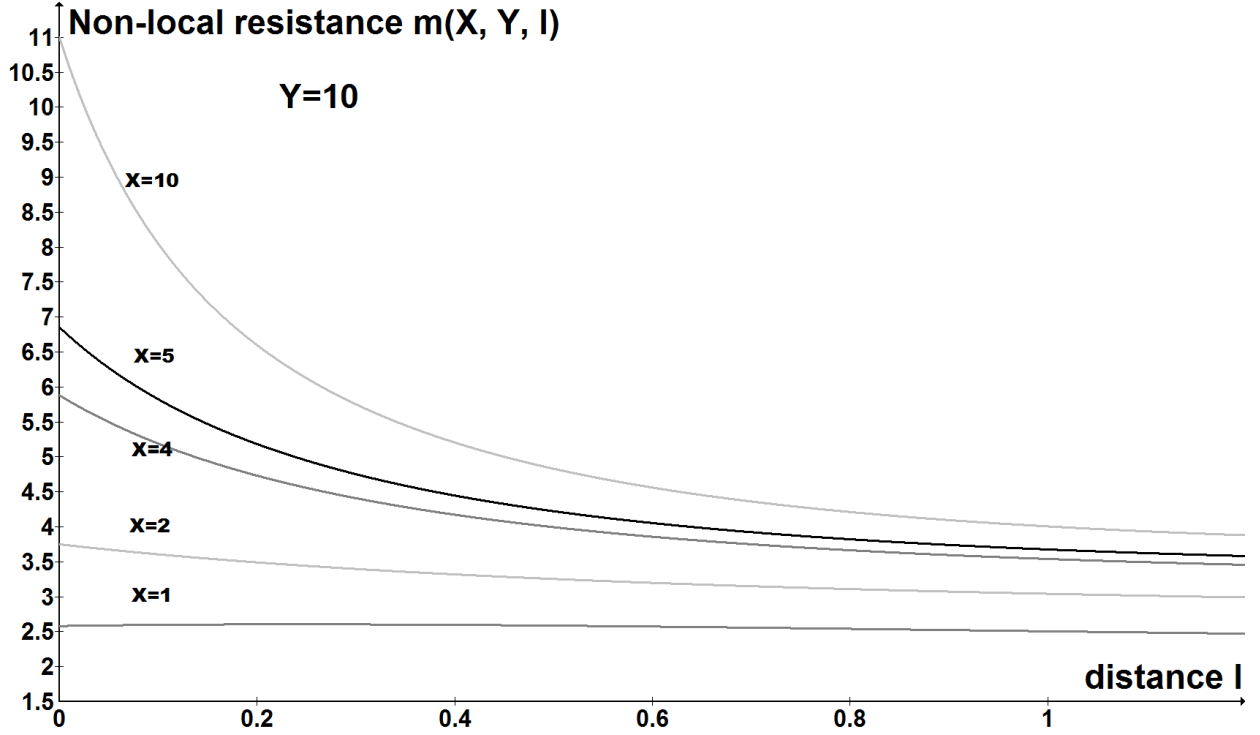}

\centering{}\caption{Ratio of non-local resistance in cross geometry (with three ferromagnets)
versus lateral setup as a function of distance $l$ between injector
and detector. Spin resistance mismatch is varied ($X=1-10$) at terminals
$F1-F2-F3$ and set at $Y=10$ at paramagnet $F4$. All electrodes
are in the tunneling regime.\label{fig:f3-fig2}.}
\end{figure}

\par\end{center}

When $X$ and $Y$ are about equal and large, $X\sim Y\gg1$ the relative
increase ($\sim X+1$) for a given mismatch $X$ is smaller than when
$Y\gg X\gg1$ ($m\sim\frac{4}{3}\left(X+1\right)$); however the latter
condition might be more inconvenient to realize for a modest gain
(for instance, if we aim at an order of magnitude increase, $X\sim10$
this would imply $Y\sim100$ (in the strong tunneling limit) yielding
$m\sim14$ at $l=0$; while already for $X=Y\sim10$, $m\sim11$ at
$l=0$ (see Fig. \ref{fig:f3-fig3}).

\begin{center}
\begin{figure}
\centering{}\includegraphics[width=1\columnwidth]{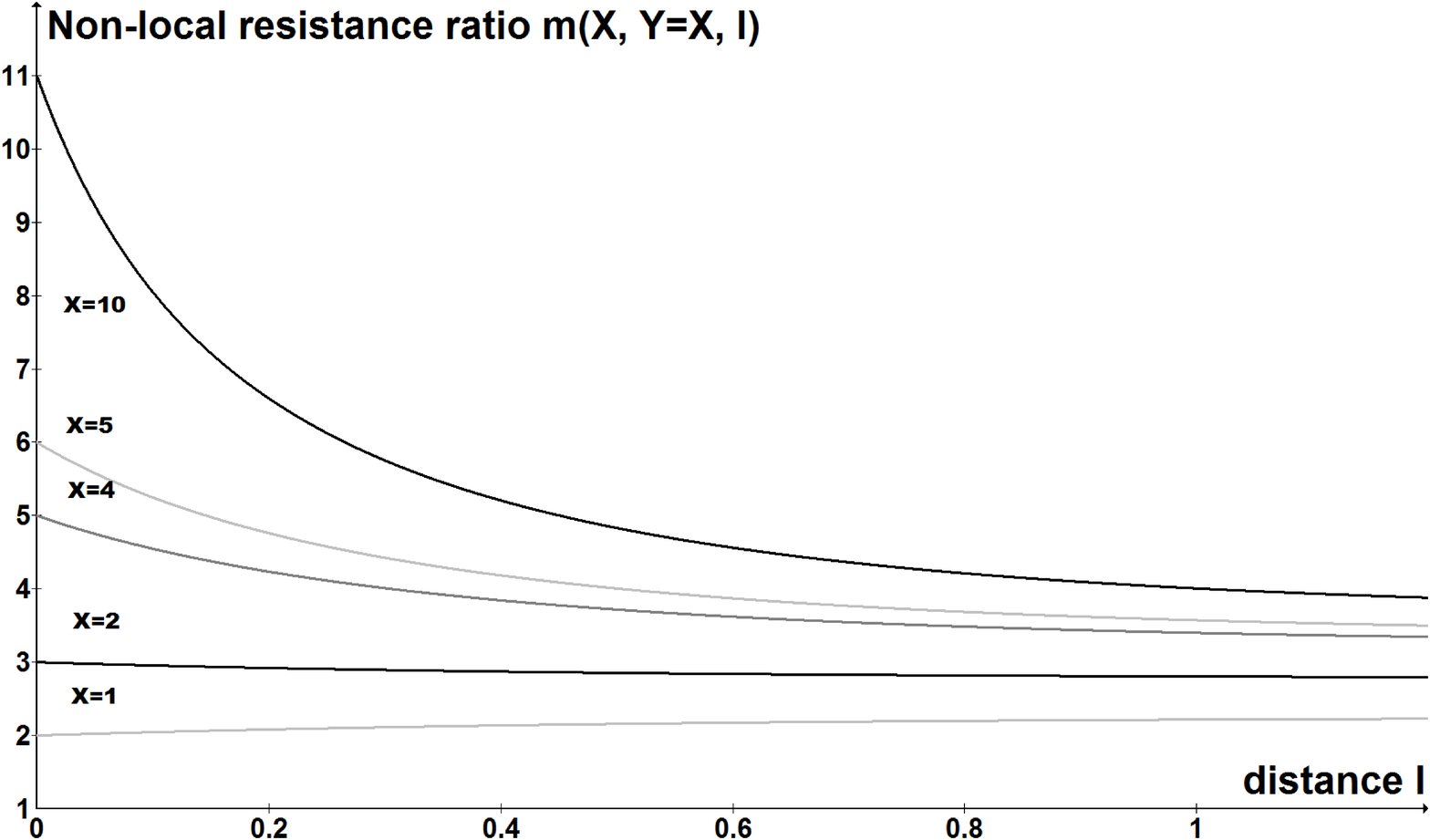}\caption{Ratio of non-local resistance in cross geometry (with three ferromagnets)
versus lateral setup as a function of distance $l$ between injector
and detector. Spin resistance mismatch is varied $X=1-10$ at terminals
$F1-F2-F3$ but $Y=X$ at paramagnet $F4$. All electrodes are in
the tunneling regime.\label{fig:f3-fig3}.}
\end{figure}

\par\end{center}

In practice the distance between injector and collector can not be
reduced arbitrarily. Nevertheless the ratio $m\left(X,\; Y,\; l\right)$
may remain large as can be seen in the next figure (Fig. \ref{fig:f3-fig4})
which shows the relative increase of signal at distances $l=0.1-0.3$.
For $X\sim Y=10$, $m\sim5.7-8.0$ for $l=0.1-0.3$ (which are quite
reachable at low temperature for $l=0.1$ or even room temperature
for $l=0.3$). This is a doubling of the signal when compared to the
values we found for $X\sim Y=10$ in the two ferromagnet closed geometry
(of section \ref{subsec:Spin-confining-(closed)}). (As previously
mentioned, this stems from the fact that the total effective polarization
is doubled due to the use of two spin injectors while the effective
resistances are identical.)

\begin{center}
\begin{figure}
\begin{centering}
\includegraphics[width=1\columnwidth]{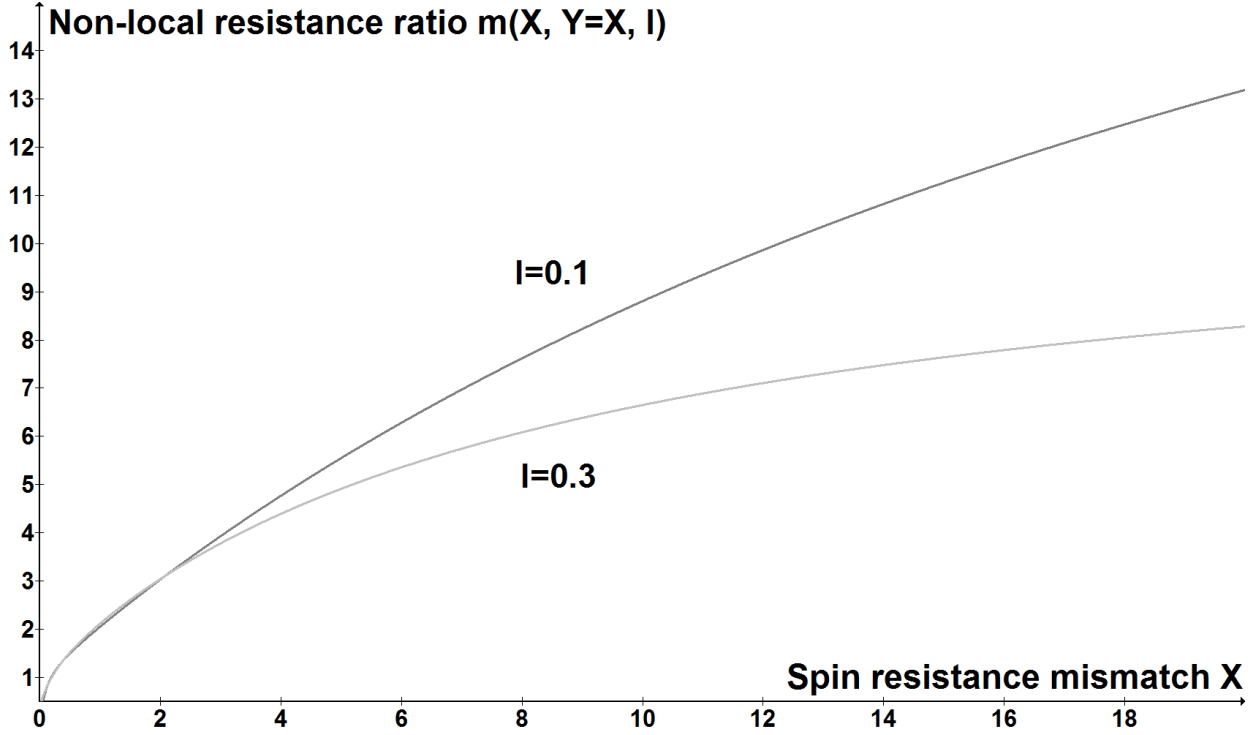}
\par\end{centering}

\caption{Ratio of non-local resistance in cross geometry (with three ferromagnets)
versus lateral setup at distances $l=0.1-0.3$ between injector and
detector. Spin resistance mismatches are set to identical values at
all terminals $X=Y$ \label{fig:f3-fig4}}

\end{figure}

\par\end{center}

Turning now to the transparent regime, we still observe an enhancement
of the signal (see Fig.\ref{fig:f3-fig6}): as before this is entirely
due to the doubling of spin injector terminals since there is spin
leakage at all terminals. For $X=Y\leq1$, the ratio $m$ varies between
$X+1$ and $2$ for an increase up to $125\;\%$ at large distance.

\begin{center}
\begin{figure}
\begin{centering}
\includegraphics[width=1\columnwidth]{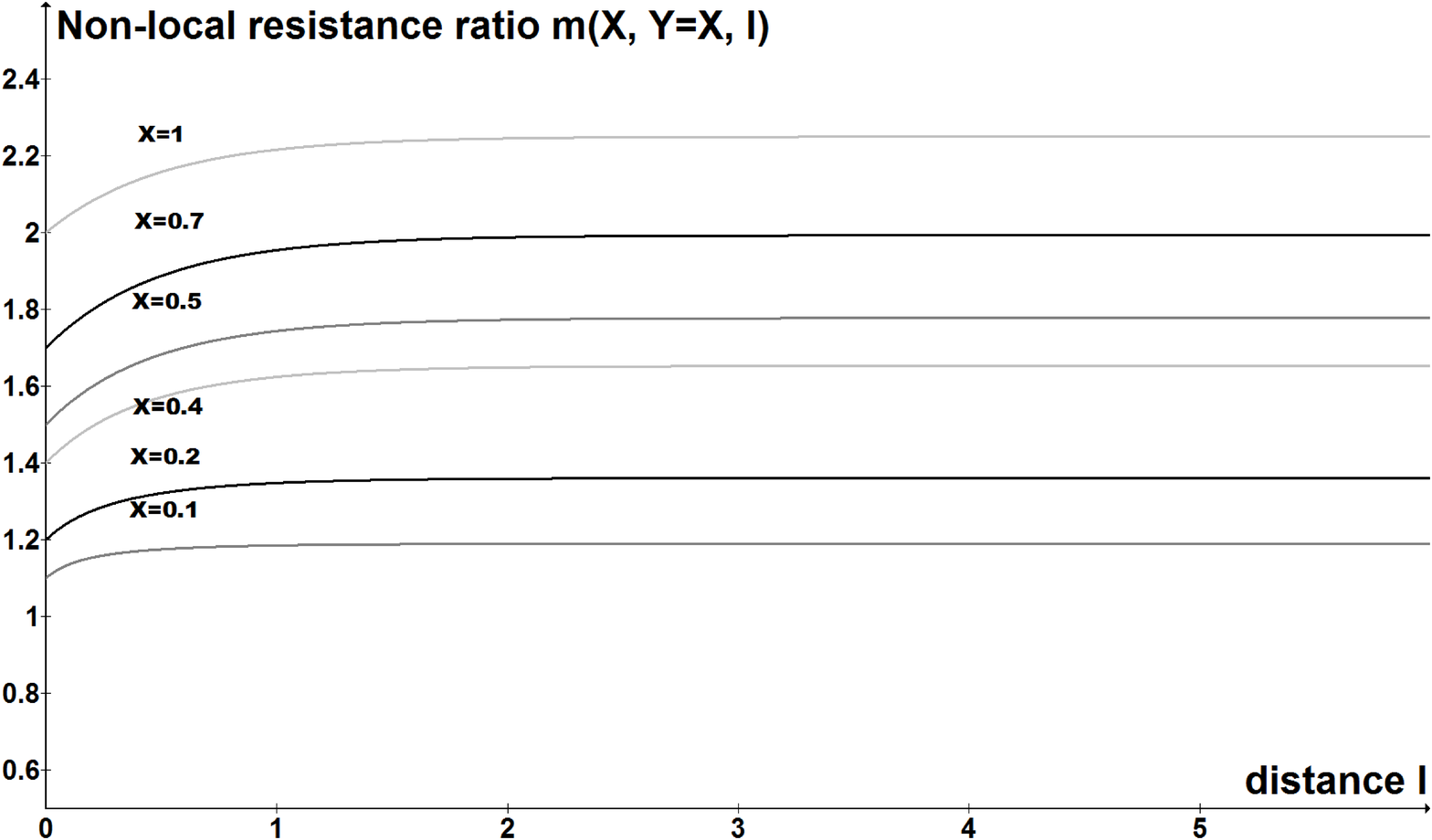}
\par\end{centering}

\caption{Ratio of non-local resistance in cross geometry (with three ferromagnets)
versus lateral setup as a function of distance $l$ between injector
and detector. Spin resistance mismatches are set to identical values
at all terminals $X=Y$ in the transparent regime ($X=0.1-1$).\label{fig:f3-fig6}}
\end{figure}

\par\end{center}

To summarize this section, we have seen the impact of doubling the
number of spin injectors conjugated to spin confinement resulting
from large spin resistance mismatches. Although this geometry with
three ferromagnets is quite interesting in terms of the large signals
which can be achieved, in practice it is better to consider a four
ferromagnet setup which will have all the qualities of the 3 ferromagnet
Van der Pauw cross with some additional properties: protection from
voltage offsets (which may be detrimental to SNR) and infinite $GMR$
ratio for the non-local resistance.

\section{Van der Pauw cross with Four ferromagnets.\label{sec:4f}}

We now consider now the main device of this paper, a four ferromagnet
Van der Pauw setup with the following symmetries: (i) identical injector
and collector electrodes ($F1$ and $F2$), (ii) identical detector
electrodes ($F3$ and $F4$). Besides an enhancement of the spin voltage
such a device is protected against voltage offsets and offers an infinite
non-local GMR ratio. Other general properties are described in section
\ref{sub:Properties}: notably such a setup will also display an ON-OFF
switch effect (for the basic 1-bit basic sensing function).

\subsection{Non-local resistance and infinite GMR.}

We consider a rhombus geometry for the cross: $OA=OB=l_{1}$ while
$OC=OD=l_{3}$. The resistance mismatch parameters for the device
are set as:

\begin{equation}
X_{1}=X_{2}=X;\; X_{3}=X_{4}=Y
\end{equation}
while: 
\begin{eqnarray}
\widetilde{PR_{1}} & = & \widetilde{PR_{2}}=\widetilde{PR},\\
\widetilde{PR_{3}} & = & \widetilde{PR_{4}}=\widetilde{PR}'.
\end{eqnarray}

The non-local resistance {[}Eq. (\ref{eq:non-local res}){]} becomes:
\begin{equation}
R_{nl}=R_{eff}\;\left(\sigma_{1}-\sigma_{2}\right)\:\left(\sigma_{3}-\sigma_{4}\right)\: P_{eff,1}\; P_{eff,3}
\end{equation}
where $\sigma_{i}=\pm1$ refer to the majority spin direction of terminal
$i$ relative to an absolute axis.

Eventually: 
\begin{equation}
R_{nl}=\frac{R_{n}\;\left(\sigma_{1}-\sigma_{2}\right)\:\left(\sigma_{3}-\sigma_{4}\right)\:\widetilde{PR}\;\widetilde{PR}'}{\left[\left(X+1\right)\left(Y+1\right)\exp l-\left(X-1\right)\left(Y-1\right)\exp-l\right]}
\end{equation}
where 
\begin{equation}
l=l_{1}+l_{3}.
\end{equation}
Upon magnetization switching of one detector electrode the variation
in non-local resistance is therefore:
\[
\delta R_{nl}=\frac{4R_{n}\;\widetilde{PR}\;\widetilde{PR}'}{\left[\left(X+1\right)\left(Y+1\right)\exp l-\left(X-1\right)\left(Y-1\right)\exp-l\right]}
\]

\subsection{Amplitude of non-local resistance. }

The signal is largest at small distance where the non-local resistance
obeys:
\begin{equation}
\inf\left(P_{c},\; P_{F}\right)\,\inf\left(P_{c}',\; P_{F}'\right)\,4R_{n}\;\frac{XY}{\left(X+Y\right)}\leq\delta R_{nl}(l\longrightarrow0)
\end{equation}
with a similar upper bound. Again one recovers the characteristic
quadratic dependence on spin resistance mismatches at the numerator
(linear at denominator), which imply that if both $X$ and $Y$ are
large, $\delta R_{nl}$ will scale with them, much beyond $R_{N}$.
If one of the spin mismatches (say $X$) is in the transparent regime,
it will set the scale, below or at most at $R_{N}$.

Let us compare in details with the standard lateral setup; we will
set $\widetilde{PR}=\widetilde{PR}'$ and $X=Y$ for simplicity. (When
$X\neq Y$ the same trends against the case $X=Y$ are seen as in
the three ferromagnet geometry.)

Then the non-local resistance ratio becomes: 
\[
m(X,\; l)=\frac{\delta R_{nl}}{\delta R_{nl,lateral}}=\frac{\left[\left(2X+1\right)^{2}\exp l-\exp-l\right]}{\left[\left(X+1\right)^{2}\exp l-\left(X-1\right)^{2}\exp-l\right]}.
\]
At small distance, one gets the enhancement:
\begin{equation}
m(X,\; l)\longrightarrow X+1.
\end{equation}
 This is the same enhancement as in the 3 ferromagnet setup when $X=Y$
(see \ref{sub:General-case.3f}).

We first consider the tunneling regime ($X>1$). We plot in the following
figures (Fig. (\ref{fig:f4-fig1}-\ref{fig:f4-fig2}) ) the length
and resistance mismatch dependence of the ratio $m(X,\; l)$. For
$X=1$, we observe that $m\approx2$ at all distances: in the case
of open cross geometry with two ferromagnets (see section \ref{subsec:Standard-(open)-geometry})
already one had $m\sim1$; the doubling here stems therefore from
the two spin injectors. For $X=10$, the enhancement remains strong
at increasing distance: for instance for $l=0.1-0.3$, $m\sim500-770\;\%$.

\begin{center}
\begin{figure}
\begin{centering}
\includegraphics[width=1\columnwidth]{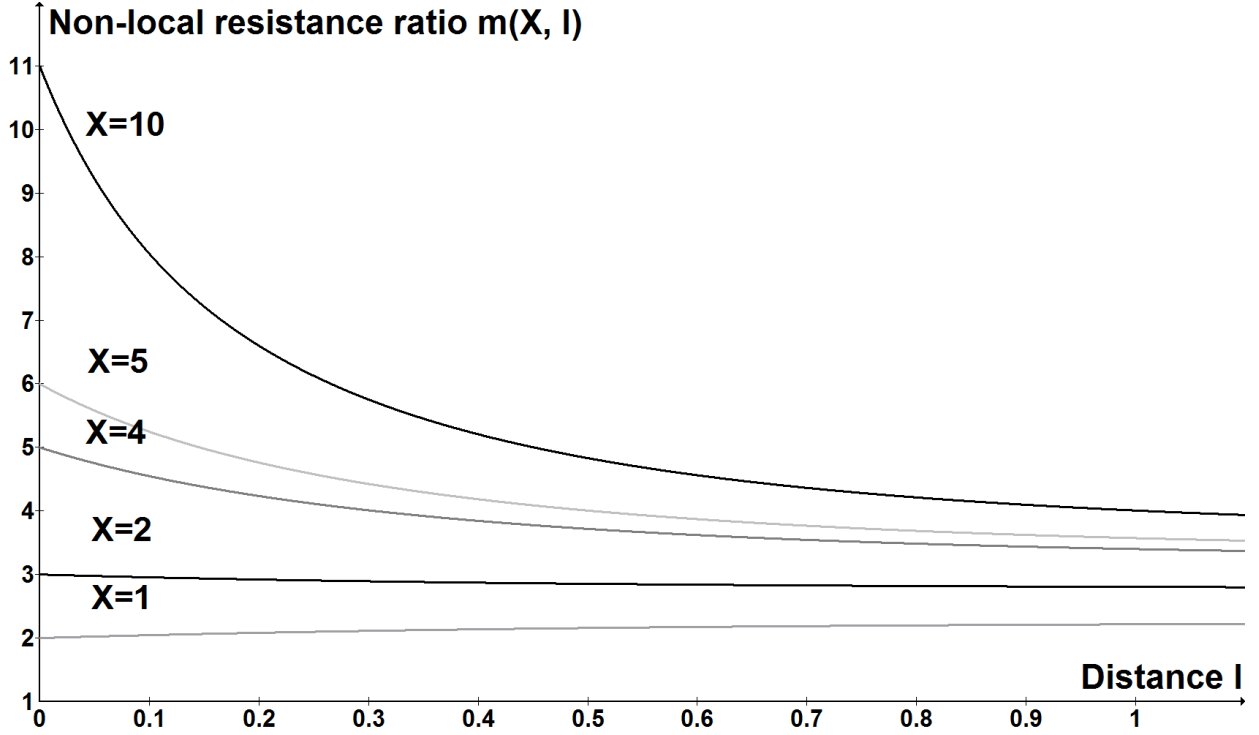}
\par\end{centering}

\caption{Ratio of non-local resistance in cross setup (with four ferromagnets)
versus standard lateral setup, as a function of distance $l$ between
injector and detector. Spin resistance mismatch $X$ is in the spin
confining regime ($X=1-10$). \label{fig:f4-fig1}}
\end{figure}

\par\end{center}

\begin{center}
\begin{figure}
\begin{centering}
\includegraphics[width=1\columnwidth]{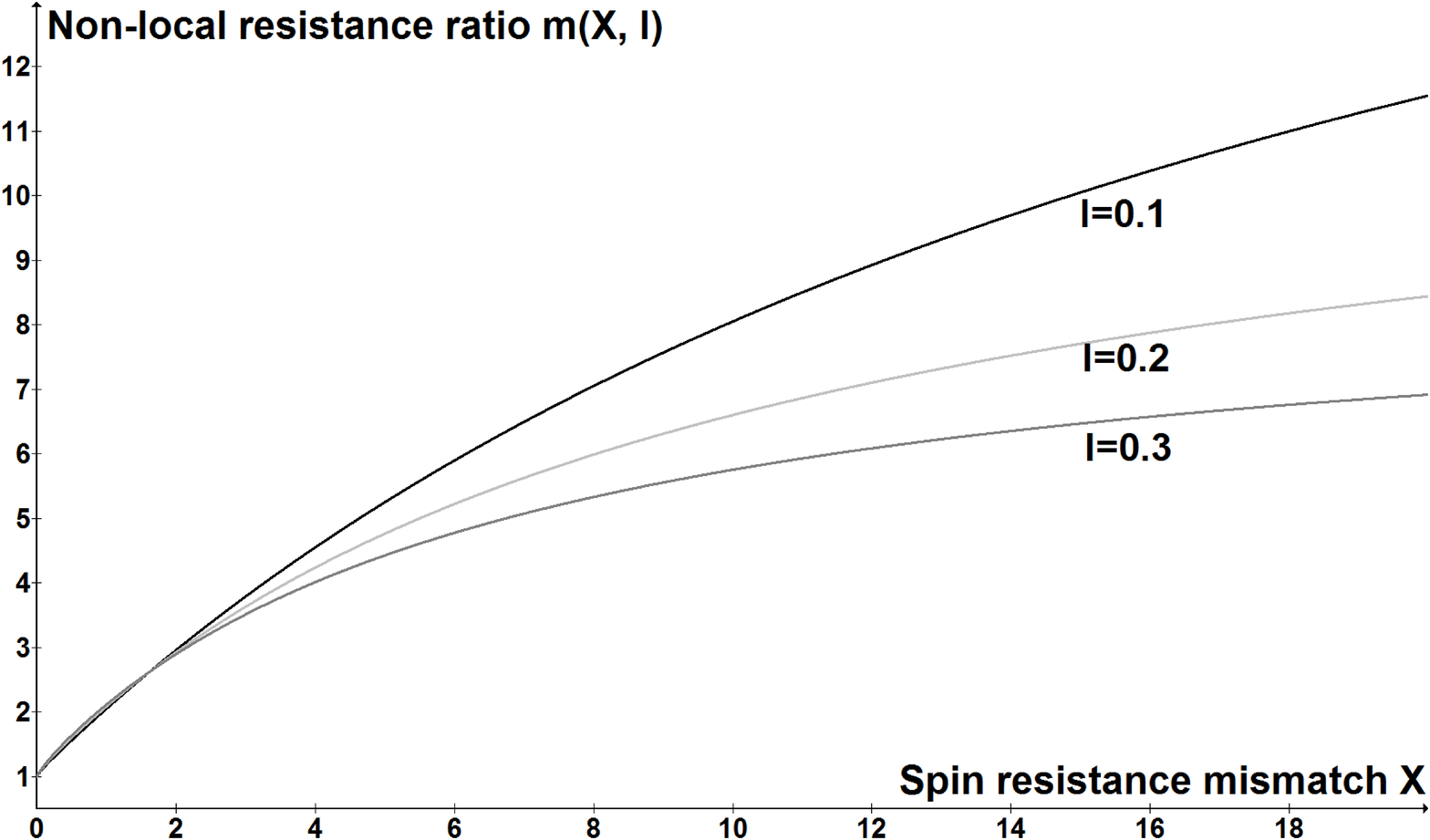}
\par\end{centering}

\caption{Ratio of non-local resistance in cross setup (with four ferromagnets)
versus standard lateral setup, at fixed distance $l=0.1-0.3$ between
injector and detector. Spin resistance mismatch $X\geq1$ is in the
spin confining regime.\label{fig:f4-fig2}}
\end{figure}

\par\end{center}

For completeness we now turn to the transparent regime ($X<1$) although
it is less interesting than the tunneling regime if one is to achieve
large signals. We observe the same trend in Fig. \ref{fig:f4-fig3}
as before with an increase in several tens of $\%$ and up to $100\;\%$
at large distance, which stems from the double spin injectors. At
shorter distance, the spin current leakage is more pronounced.

\begin{center}
\begin{figure}
\begin{centering}
\includegraphics[width=1\columnwidth]{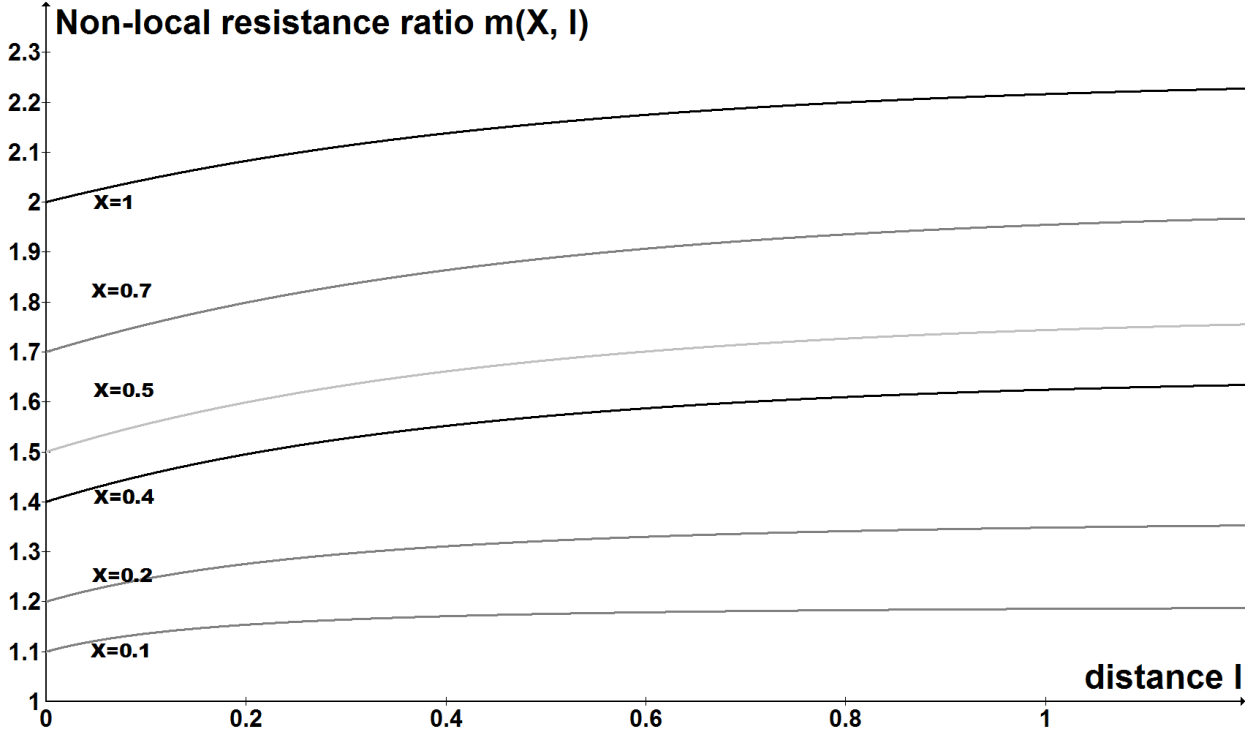}
\par\end{centering}

\caption{Ratio of non-local resistance in cross setup (with four ferromagnets)
versus standard lateral setup, as a function of distance $l$ between
injector and detector. Spin resistance mismatch $X$ is in the transparent
regime ($X=0.1-1$). All terminals are identical in this plot. \label{fig:f4-fig3}}
\end{figure}

\par\end{center}

\subsection{Discussion and comparison to experiments\label{sub:Discussion.}.}

Can we expect signals in the $mV$ range for non-local voltages? Already
the state-of-art has reached $\sim100-200\;\mu V$ in lateral spin
valves\cite{fukuma_giant_2011}; the peculiarity of these lateral
spin valves is that they use thin tunnel $MgO$ barriers ($1-6\; nm$)
between the ferromagnets (Permalloy $NiFe$) and the paramagnetic
channel which is silver $Ag$. Due to the thinness of $MgO$ the interface
resistances are much smaller than in usual tunnel junctions, with
an interface resistance times cross section product in the range $R_{I}A\sim0.2\;\Omega\mu m^{2}$.
While for as deposited samples the signals are already higher than
usual ($\sim10\;\mu V$), after annealing the signals can jump by
an order of magnitude; this has been accounted on oxygen vacancies
which increase after annealing.

Let us see how much we can get in similar conditions but with the
cross geometry with four ferromagnetic electrodes: we will borrow
the parameters found of Fukuma and coll.\cite{fukuma_giant_2011}
for lateral spin valves based on $Py-MgO-Ag$ junctions; the size
of the junctions is $A=0.022\;\mu m^{2}$; the separation between
detector and injector is $L=300\; nm$ . 

At $10\; K$, for a $1\; mA$ current, the largest spin voltage is
measured at $112\;\mu V$ with the following parameters: $P_{c}=0.44$
for the interface conductance polarization (noted $P_{I}$ by Fukuma
and coll.\cite{fukuma_giant_2011} ), $P_{F}=0.35$ for $Py$ electrodes,
$l_{F}=5\; nm$, $l_{N}=1100\; nm$ for the spin relaxation lengths
of $Py$ and silver $Ag$, $R_{N}=\rho_{N}^{*}l_{N}/A=0.89\;\Omega$
for $Ag$ spin resistance, $R_{F}=\rho_{F}^{*}l_{F}/A=0.09\;\Omega$
($R_{NiFe}=0.08\;\Omega$ in Fukuma and coll.\cite{fukuma_giant_2011}
is related to our spin resistance $R_{F}$ by $R_{NiFe}=R_{F}\left(1-P_{F}^{2}\right)$
), $R_{c}=12.13\;\Omega$ (our interface spin resistance $R_{c}$
is related to $R_{I}$ of Fukuma and coll. by $R_{c}=R_{I}/(1-P_{I}^{2})$
with $R_{I}A=0.2152\;\Omega\mu m^{2}$ and $A=0.022\;\mu m^{2}$).

Then the spin impedance mismatch is as large as: 
\[
X=\frac{R_{F}+R_{c}}{R_{N}}=13.73
\]
while: 
\[
l=L/l_{N}=0.27.
\]

For these values and using the measured $\Delta V_{nl}=110\;\mu V$,
$\Delta R_{nl}=110\; m\Omega$, $I=1\; mA$ this implies:
\[
m(X,\; l)=660\;\%
\]
which would imply a spin voltage variation as large as:
\[
\Delta V\sim725\;\mu V
\]
or: 
\[
\Delta R_{nl}\sim725\; m\Omega
\]
 (since $I=1\; mA$); if we use the symmetric four-terminal cross
device, the baseline signal vanishes (for parallel orientation of
detector electrodes), so that 
\[
R\; A=\Delta R\; A=16\; m\Omega\mu m^{2}
\]
which compared with CPP GMR spin valves is large in terms of $\Delta R\; A$
product (a few $m\Omega\mu m^{2}$) but small in terms of customary
$R\; A$ values ($10-100\; m\Omega\mu m^{2}$)\cite{nagasaka_cpp-gmr_2009}.

Still at $10\; K$, the highest voltage reported is $200\;\mu V$
for a larger current $I=3.5\; mA$ so that the non-local resistance
is smaller $\Delta R_{nl}=63\; m\Omega$; using our setup we predict
one would have achieved signals in the range: 
\begin{eqnarray*}
\Delta V_{nl} & \sim & 1.320\; mV\\
\Delta R_{nl} & \sim & 415\; m\Omega\\
\Delta R\; A & \sim & 9.1\; m\Omega\mu m^{2}\\
j & \sim & 1.6\;10^{7}A/cm^{2}.
\end{eqnarray*}
The current density is therefore reasonable although the $mV$ mark
for the spin voltage is reached! The non-local resistance variation
is however smaller (Fukuma and coll.\cite{fukuma_giant_2011} attributes
the decrease to an enhanced spin relaxation in $Ag$ due to phonons). 

If we put into perspective these predictions, we may remember that
the first measurements of non-local resistance in bulk metal wires\cite{johnson_interfacial_1985}
by Johnson and Silsbee yielded $\sim n\Omega$, were thereafter improved
in thin films\cite{johnson_spin_1993} to about $10\;\mu\Omega$;
Fukuma and coll.\cite{fukuma_giant_2011} have pushed to $\sim100\; m\Omega$.
The largest non-local resistance variation observed is however already
in the Ohm range in lateral tunnel junctions\cite{valenzuela_spin-polarized_2004,valenzuela_spin_2005},
with one major catch: currents must remain small ($\sim\mu A$) due
to spin depolarization at larger currents, which results in a spin
voltage still in the customary $\mu V$ range. In contrast, in metallic
lateral spin valves larger currents can be achieved\cite{fukuma_giant_2011,fukuma_enhanced_2010,ji_enhanced_2006,ji_non-local_2007,ji_spin_2004,kimura_enhancement_2006,kimura_large_2007,van_staa_spin_2008}. 

Turning now to room temperature, there is few change in spin impedance
mismatch $X$ 
\[
X=13.44
\]
but due to a much smaller $l_{N}\sim300\; nm$ 
\[
l=1.
\]
One still finds a sizable enhancement:
\[
m(X,\; l)=415\;\%
\]
 which would lead to the following numbers (using original experimental
values of Fukuma and coll.\cite{fukuma_giant_2011} $\Delta V_{nl}=51\;\mu V$,
$\Delta R_{nl}=51\; m\Omega$, $I=1\; mA$): 
\begin{eqnarray*}
\Delta V_{nl} & \sim & 210\;\mu V\\
\Delta R_{nl} & \sim & 210\; m\Omega\\
\Delta R\; A & \sim & 4.5\; m\Omega\mu m^{2}\\
j & \sim & 5\;10^{6}A/cm^{2}.
\end{eqnarray*}

There is still room for improvement at room temperature before reaching
the $mV$ mark. One obvious way would be to decrease the distance
between injector and detector; for instance suppose $L=100\; nm$
which is quite reasonable in terms of lithography, then $l=0.33$;
taking the previous room-temperature value for $X$ ($X=13.44$) would
yield:

\[
m(X,\; l)=605\;\%
\]
so that 
\begin{eqnarray*}
\Delta V_{nl} & \sim & 310\;\mu V\\
\Delta R_{nl} & \sim & 310\; m\Omega\\
\Delta R\; A & \sim & 6.8\; m\Omega\mu m^{2}\\
j & \sim & 5\;10^{6}A/cm^{2}.
\end{eqnarray*}
If we lower to a still achievable $50\; nm$ for which $l=0.17$ then:
\[
m(X,\; l)=795\;\%
\]
and: 
\begin{eqnarray*}
\Delta V_{nl} & \sim & 405\;\mu V\\
\Delta R_{nl} & \sim & 405\; m\Omega\\
\Delta R\; A & \sim & 8.9\; m\Omega\mu m^{2}\\
j & \sim & 5\;10^{6}A/cm^{2}.
\end{eqnarray*}
(at such length $L=50\; nm$ $m=1150\%$ at $10K$ so that $\Delta V_{nl}\sim1.250\; mV$
again in the $mV$ mark at $I=1\; mA$).

Are there any other ways to further improve the signal at room-temperature?
We will show elsewhere\cite{kv} that with an all-lateral structure
even stronger enhancements are found due to better spin confinement.
In the latter example one would get as much as $m\sim1070\;\%$ at
room temperature (or $m\sim1320\;\%$ at $10\; K$).

\textbf{Application to sensor and read-heads technology?}

Reaching the $1\; Tbit/inch^{2}$ mark up to $10\; Tbit/inch^{2}$
areal density requires new recording methods to push beyond the super-paramagnetic
limit; bit patterning and thermally assisted magnetic recording are
currently investigated\cite{wood_2009}. But such large areal densities
are also very challenging for current sensor technologies. The transition
from AMR, CIP GMR, TMR and now CPP GMR has accompanied dramatic and
steady increase of the areal density of hard-drives up to about $0.5-0.7\; Tbit/inch^{2}$
(as of late 2011). Further increase to reach the $1\; Tbit/inch^{2}$
has motivated the latest transition from TMR read-heads to CPP metallic
spin valves because TMR is limited to $R\; A\sim1\;\Omega\mu m^{2}$
which will not be small enough to accommodate the high data rates
required for the desired areal densities \cite{nagasaka_cpp-gmr_2009}.
In contrast CPP GMR read-heads can have $R\; A<0.1\;\Omega\mu m^{2}$
which is the minimum required for an areal density $1\; Tbit/inch^{2}$.
An additional requirement for CPP spin valves is to have larger $\Delta R\; A>5\; m\Omega\mu m^{2}$
instead of the usual$<1\; m\Omega\mu m^{2}$ which are too small to
ensure good $SNR$ (signal to noise ratio). The source of noises are
usually Johnson thermal noise in metallic spin valves (both CIP and
CPP) or shot noise in TMR valves. But at the shrinking sizes relevant
for larger areal densities additional relevant sources of noise appear
to be STT (spin transfer torque) noise, mag-noise (noise due to thermal
fluctuations in the free layer) and amplifier noise\cite{katine_device_2008,nagasaka_cpp-gmr_2009,takahashi_spin_2003}.

Non-local devices have been barred for practical use due to small
signals and have mostly been interesting for fundamental and pedagogical
use. As we have shown this may prove less of a concern by using geometries
with better spin confinement and with multiple injectors. Such non-local
devices used as sensors would hold several advantages over their standard
counterparts while retaining the small $RA$ of CPP metallic spin
valves: 

- smaller thickness of the sensing detector (two or three layers since
the reference layer is located elsewhere and can have a different
coercive field by playing on its geometry); 

- diminished sensitivity to STT noise since the largest momentum transfer
is between source and drain electrodes; 

- no self-field since no current flows through the detectors (Oersted
fields are at injectors). 

- In the case of a symmetric cross geometry, the possibility to achieve
$100\;\%$ or maximum MR ratio (using the pessimistic ratio) so that
$R=\Delta R$ is also advantageous regarding Johnson and shot noise
providing a $10-20\; dB$ boost in SNR (as explained above in section
\ref{sub:Properties}). 

In addition the maximum MR ratio property allows to satisfy both requirements
sought in CPP spin valves: small enough $R\; A<0.1\;\Omega\mu m^{2}$
for high data rates, but large enough $\Delta R\; A>5\; m\Omega\mu m^{2}$
for good SNR. (The latter figure would need to be adjusted to take
into account the peculiarities of the non-local device but we use
it as a reasonable rule-of-thumb). 

The main constraints for a large SNR will remain mag-noise, amplifier
noise as well as the ability to scale down the non-local setup to
par with the $20-30\; nm$ track width relevant\cite{takagishi_magnetoresistance_2010,katine_device_2008}
for areal densities above $1\; Tbit/inch^{2}$. Although we expect
STT noise to be less of a nuisance, angular momentum transfer still
occurs from injectors to detectors and a quantification of critical
currents adapted to the geometry would be interesting. The angular
response is thus an important aspect to explore in view of a potential
use as a sensor and is discussed elsewhere\cite{kv}.

Although the geometry studied in this paper might seem quite peculiar,
there is actually a lot of flexibility in terms of design, provided
the following recommendations are enforced: 

(i) confine spin with large interface resistances (when compared with
the paramagnet spin resistance) wherever spin current can leak; 

(ii) in the smallest volume for the paramagnet connecting the ferromagnetic
terminals (when compared to $l_{N}^{3}$); 

(iii) to ensure infinite MR (optimistic) ratio, use two identical
ferromagnetic electrodes as detectors within a symmetric arrangement
with respect to injectors (the geometry of the ferromagnetic electrodes
need not be completely identical: they just need to be identical to
within a few spin diffusion lengths $l_{F}$ of the contact to the
paramagnet; this is an important practical remark since this allows
to have different coercive fields for the various electrodes while
keeping the properties stemming from symmetry); 

(iv) for further gain in the signal, use two antiparallel ferromagnetic
electrodes. 

With these prescriptions non-local devices will perhaps be able to
reach the realm of applications.

\section{Conclusion and prospects.\label{sec:Conclusion-and-prospects.}}

We have presented a variant of non-local spin valves based on a Van
der Pauw cross setup: its basic characteristic is its reliance on
four collinear ferromagnetic terminals. By playing on the numerous
magnetization configurations of the four ferromagnets and on their
symmetries several functionalities have been described: (i) an improved
non-local spin valve (when used to read 1 bit) with an infinite GMR
for the non-local resistance; (ii) ON-OFF switch effect where magnetization
at the injectors can control the read-off at detectors; (iii) 3-bit
storage or sensing and offset-free 2-bit reading; (iv) direct Spin
Hall measurement; (v) use as programmable magneto-logic gates. 

In addition to these functionalities the amplitude of the non-local
resistance has the potential to be much increased in such a setup
due to the conjunction of (i) spin confinement by tunnel barriers;
(ii) the use of two electrodes as spin injectors instead of only one.
We have also studied the separate or combined impact of both features
by considering setups with two or three ferromagnets: the use of two
injectors basically doubles the signal while the use of tunnel barriers
hinder spin leaking from the Van der Pauw cross, thus increasing spin
accumulation within it. The latter effect is already observed even
when not all terminals are connected by tunnel barriers to the Van
der Pauw cross. An important parameter is the spin resistance mismatch
at each ferromagnetic-paramagnetic interface (at the four terminals),
which is the ratio between the spin resistances of the ferromagnet
and the paramagnet. The larger this parameter, the better the spin
confinement in the Van der Pauw cross.

Additional functionalities will be discussed elsewhere: non-collinear
properties which are essential for sensor applications; as well as
caloritronic ones. A related non-local device in the lateral geometry
for which signals are much stronger will be also investigated\cite{kv}.

\appendix

\section{Computing the Spin Voltage.\label{sec:Computing-the-Spin}}

We derive in this Appendix the general expression of the spin voltage
by relying on the one-dimensional drift-diffusion equations\cite{johnson_thermodynamic_1987,valet_theory_1993,rashba_diffusion_2002}.

\subsection{Basic equations.}

\subsubsection{Spin accumulation vectors.}

We orient spin currents away from the origin $O$ of the cross.

In each electrode ($F1-F4$) the spin accumulation and spin currents
are then:

$F1$:
\begin{eqnarray}
\Delta\mu_{F1}(z) & = & \Delta\mu_{F1}(A)\:\exp-\frac{z-L_{1}}{l_{F1}}\\
I_{s,F1}(z) & = & -P_{F1}I_{c}-\frac{\Delta\mu_{F1}(A)}{R_{F1}}\:\exp-\frac{z-L_{1}}{l_{F1}}
\end{eqnarray}
so that: 
\begin{equation}
I_{s,F1}(A)=-P_{F1}I_{c}-\frac{\Delta\mu_{F1}(A)}{R_{F1}}.\label{eq:k1a}
\end{equation}

$F2$: without loss of generality we assume $F2$ antiparallel to
$F1$ which allows to have symmetric equations (for the parallel case
one should take opposite polarizations) 
\begin{equation}
I_{s,F2}(B)=-P_{F2}I_{c}-\frac{\Delta\mu_{F2}(B)}{R_{F2}}
\end{equation}
where $I_{s,F2}(B)$ is counted positive when flowing in the direction
$z'=-z$; 

$F3$: 
\begin{equation}
I_{s,F3}(C)=-\frac{\Delta\mu_{F3}(C)}{R_{F3}}
\end{equation}
where $I_{s,F3}(B)$ is oriented in the direction $x$.

$F4$:
\begin{equation}
I_{s,F4}(D)=-\frac{\Delta\mu_{F4}(D)}{R_{D4}}
\end{equation}

where $I_{s,N4}(D)$ is oriented in the direction $x'=-x$.

In the arms ($I-IV$) we define spin accumulation vectors which are
related to the spin accumulation and spin currents by:

I: 

\begin{eqnarray}
\Delta\mu_{I}(z) & = & \mathbf{K_{I}}\:.\:\left(\exp\frac{z}{l_{N}};\:\exp-\frac{z}{l_{N}}\right),\\
I_{s,I}(z) & = & \frac{\mathbf{K_{I}}}{R_{N}}\:.\:\left(\exp\frac{z}{l_{N}};\:-\exp-\frac{z}{l_{N}}\right)
\end{eqnarray}
where $\mathbf{K_{I}}$ is a constant two-component vector. 

II: following our convention of orienting away from the origin 
\begin{eqnarray}
\Delta\mu_{II}(z') & = & \mathbf{K_{II}}\:.\:\left(\exp\frac{z'}{l_{N}};\:\exp-\frac{z'}{l_{N}}\right),\\
I_{s,II}(z') & = & \frac{\mathbf{K_{II}}}{R_{N}}\:.\:\left(\exp\frac{z'}{l_{N}};\:-\exp-\frac{z'}{l_{N}}\right)
\end{eqnarray}

III: 
\begin{eqnarray}
\Delta\mu_{III}(x) & = & \mathbf{K_{III}}\:.\:\left(\exp\frac{x}{l_{N}};\:\exp-\frac{x}{l_{N}}\right),\\
I_{s,III}(x) & = & \frac{\mathbf{K_{III}}}{R_{N}}\:.\:\left(\exp\frac{x}{l_{N}};\:-\exp-\frac{x}{l_{N}}\right)
\end{eqnarray}

IV: 
\begin{eqnarray}
\Delta\mu_{IV}(x') & = & \mathbf{K_{IV}}\:.\:\left(\exp\frac{x'}{l_{N}};\:\exp-\frac{x'}{l_{N}}\right),\\
I_{s,IV}(x') & = & \frac{\mathbf{K_{IV}}}{R_{N}}\:.\:\left(\exp\frac{x'}{l_{N}};\:-\exp-\frac{x'}{l_{N}}\right)
\end{eqnarray}

\subsubsection{Conditions at ferromagnetic - paramagnetic interfaces and at cross
center.}

\textbf{Interfaces.} At interfaces we will neglect all spin flip but
assume interface resistance is spin dependent. This implies that spin
current at points $A-D$ is continuous: 
\begin{equation}
I_{s,F1}(A)=I_{s,I}(A)\label{eq:k1b}
\end{equation}
and so forth at points $B-D$.

At points $A-D$ the interface resistance induces a discontinuity
in spin accumulations:
\begin{equation}
\Delta\mu_{F1}(A)-\Delta\mu_{I}(A)=R_{c1}\: I_{s}(A)+R_{c1}\: P_{c1}\: I_{c}\label{eq:k1c}
\end{equation}
\begin{equation}
\Delta\mu_{F2}(B)-\Delta\mu_{II}(B)=R_{c2}\: I_{s}(B)+R_{c2}\: P_{c2}\: I_{c}
\end{equation}
 
\begin{eqnarray}
\Delta\mu_{F3}(C)-\Delta\mu_{III}(C) & = & R_{c3}\: I_{s}(C)\\
\Delta\mu_{F4}(D)-\Delta\mu_{IV}(D) & = & R_{c4}\: I_{s}(D)
\end{eqnarray}

\textbf{Cross center.} At $O$, the spin accumulations are continuous:

\begin{equation}
\Delta\mu_{I}(O)=\Delta\mu_{II}(B)=\Delta\mu_{III}(C)=\Delta\mu_{IV}(O)
\end{equation}
and spin current continuity implies: 
\begin{equation}
0=I_{s,I}(O)+I_{s,II}(O)+I_{s,III}(O)+I_{s,IV}(O)\label{eq:cont3}
\end{equation}
 so that for $\alpha=I-IV$: 
\begin{eqnarray}
\mathbf{K_{\alpha}}\:.\:\left(1;\:1\right) & \equiv\Delta\mu(O) & ;\label{eq:cont1}\\
\sum_{\alpha}\mathbf{K_{\alpha}}\:.\:\left(1;\:-1\right) & = & 0.\label{eq:cont2}
\end{eqnarray}

\subsection{Solution of the equations.}

\subsubsection{Spin accumulation vectors $\mathbf{K_{\alpha}}$}

We first solve for all four spin accumulation vectors in the arms
($I-IV$).

From the equations (\ref{eq:k1a}, \ref{eq:k1b}, \ref{eq:k1c}) at
interface $F1$ - $I$ one gets: 
\begin{equation}
\Delta\mu_{I}(A)+\left(R_{c1}+R_{F1}\right)\: I_{s}(A)=-\left(P_{F1}R_{F1}+P_{c1}R_{c1}\right)\: I_{c}
\end{equation}

After substitution in terms of spin accumulation vector $\mathbf{K_{I}}$:
\begin{eqnarray}
\mathbf{K_{I}}\:.\: & \left(\left(1+X_{1}\right)\;\exp l_{1};\:\left(1-X_{1}\right)\;\exp-l_{1}\right)\nonumber \\
= & -\left(P_{F1}R_{F1}+P_{c1}R_{c1}\right)\: I_{c}
\end{eqnarray}
where: 
\begin{equation}
X_{1}=\frac{R_{F1}+R_{c1}}{R_{N}}
\end{equation}
 and 
\begin{equation}
l_{1}=\frac{L_{1}}{l_{N}}.
\end{equation}
$\mathbf{K_{II}}$ obeys a similar equation by changing the index
$1$ with index $2$ since we have assumed that the magnetization
of $F2$ is antiparallel to that of $F1$.

\begin{eqnarray}
\mathbf{K_{II}}\:.\: & \left(\left(1+X_{2}\right)\;\exp l_{2};\:\left(1-X_{2}\right)\;\exp-l_{2}\right)\nonumber \\
= & -\left(P_{F2}R_{F2}+P_{c2}R_{c2}\right)\: I_{c}
\end{eqnarray}

while the equations for arms $III-IV$ obtain by cancelling the charge
current:

\begin{eqnarray}
\mathbf{K_{III}}\:.\:\left(\left(1+X_{3}\right)\;\exp l_{3};\:\left(1-X_{3}\right)\;\exp-l_{3}\right) & = & 0,\\
\mathbf{K_{IV}}\:.\:\left(\left(1+X_{4}\right)\;\exp l_{4};\:\left(1-X_{4}\right)\;\exp-l_{4}\right) & = & 0.
\end{eqnarray}
with $X_{2/3/4}$ and $l_{2/3/4}$ defined as $X_{1}$ and $l_{1}$. 

Using the continuity of spin accumulation and spin current at $O$
yields a second equation for each spin accumulation vector:

\begin{eqnarray}
\mathbf{K_{\alpha}}\:.\:\left(1;\:1\right) & =\Delta\mu(O) & ,\;\;\left(\alpha=\mathrm{I,..,IV}\right)
\end{eqnarray}
Solving for the $\mathbf{K_{\alpha}}$ vectors in terms of $\Delta\mu(O)$
leads to the following expressions:
\begin{eqnarray}
\mathbf{K_{I}} & = & \frac{\Delta\mu(O)}{2\delta_{1}^{+}}\;\left(\begin{array}{c}
\left(X_{1}-1\right)\;\exp-l_{1}\\
\left(X_{1}+1\right)\;\exp l_{1}
\end{array}\right)\nonumber \\
 & + & \frac{\left(P_{F1}R_{F1}+P_{c1}R_{c1}\right)}{2\delta_{1}^{+}}\; I_{c}\;\left(\begin{array}{c}
-1\\
1
\end{array}\right)\\
\mathbf{K_{II}} & = & 1\longleftrightarrow2
\end{eqnarray}
while: 
\begin{eqnarray}
\mathbf{K_{III}} & = & \frac{\Delta\mu(O)}{2\delta_{3}^{+}}\;\left(\begin{array}{c}
\left(X_{3}-1\right)\;\exp-l_{3}\\
\left(X_{3}+1\right)\;\exp l_{3}
\end{array}\right),\label{eq:k3}\\
\mathbf{K_{IV}} & = & 3\longleftrightarrow4,\label{eq:k4}
\end{eqnarray}
where:
\begin{equation}
\delta_{i}^{\pm}=\frac{\left(X_{i}+1\right)}{2}\;\exp l_{i}\pm\frac{\left(X_{i}-1\right)}{2}\;\exp-l_{i}.
\end{equation}

\subsubsection{Spin currents at cross center\label{sub:Spin-currents}.}

The relation between spin currents at $O$ on each arm and the spin
accumulation $\Delta\mu(O)$ is derived using:

\begin{equation}
I_{s,I}(O)=\frac{1}{R_{N}}\;\mathbf{K_{I}}\:.\:\left(1;\:-1\right)
\end{equation}
so that:
\begin{equation}
I_{s,I}(O)=-\frac{\Delta\mu(O)}{R_{eff,1}}-P_{eff,1}\; I_{c}
\end{equation}
with a similar relation for arm $II$ where:

\begin{equation}
R_{eff,i}=R_{N}\;\frac{\delta_{i}^{+}}{\delta_{i}^{-}}
\end{equation}

\begin{equation}
P_{eff,i}=\frac{\widetilde{PR_{i}}}{\delta_{i}^{+}}
\end{equation}
and:

\begin{eqnarray}
\widetilde{PR_{i}} & = & \left[P_{Fi}R_{Fi}+P_{ci}R_{ci}\right]/R_{N}
\end{eqnarray}

For arm $III$, one has simply: 
\begin{equation}
I_{s,III}(O)=-\frac{\Delta\mu(O)}{R_{eff,3}}
\end{equation}
and a similar relation for arm $IV$.

\subsubsection{Spin accumulation at cross center\label{sub:Spin-accumulation-appendix}.}

Plugging the expressions of the spin currents into the spin current
continuity equation at $O$ {[}Eq. (\ref{eq:cont3}){]} allows determination
of $\Delta\mu(O)$:
\begin{equation}
\Delta\mu(O)=-\left(P_{eff,1}+P_{eff,2}\right)\;\frac{1}{\sum_{i=1-4}\; R_{eff,i}^{-1}}\; I_{c}.
\end{equation}

Observe that the expression of $\Delta\mu(O)$ does not depend on
the polarizations of the ferromagnetic electrodes used as detectors
($F_{3}$ and $F_{4}$): it is fully determined by the injectors only
($F1$ and $F2$). The previous expression has been derived by assuming
$F1$ and $F2$ antiparallel, which reflects into the polarizations
signs. In the more general case of arbitrary collinear orientation,
one has: 
\begin{eqnarray}
\Delta\mu(O) & = & -\left(P_{eff,1}-P_{eff,2}\right)\; R_{eff}\; I_{c}
\end{eqnarray}
where positive polarizations are defined relative to an absolute axis
(negative polarizations in the opposite direction). The largest spin
accumulation in magnitude is achieved for antiparallel injector and
collector electrodes.

This can be plugged in the expressions for the spin current:
\begin{equation}
I_{s,I}(O)=\left[\frac{\left(P_{eff,1}-P_{eff,2}\right)\; R_{eff}}{R_{eff,1}}-P_{eff,1}\right]\; I_{c}
\end{equation}
with a similar expression for arm $II$ while for arm $III$ (or $IV$
with appropriate substitutions):
\begin{equation}
I_{s,III}(O)=\left[\frac{\left(P_{eff,1}-P_{eff,2}\right)\; R_{eff}}{R_{eff,3}}\right]\; I_{c}.
\end{equation}

\subsubsection{Spin voltage\label{sub:Spin-voltage-appendix}.}

The voltage is measured as:
\begin{equation}
V_{nl}=-\left[\mu_{F3}(+\infty)-\mu_{F4}(+\infty)\right]/e.
\end{equation}

This voltage depends on the orientations of the four ferromagnetic
terminals $F1-F4$ so that in the most general case it can assume
sixteen distinct values ($2^{4}$); but since the setup is invariant
under a joint flipping of all the magnetizations this number is reduced
to eight. 

We will need the chemical potential variation at interfaces:

\begin{eqnarray*}
\mu_{F3}(C)-\mu_{III}(C) & = & -R_{c3}\: P_{c3}\: I_{s}(C);\\
\mu_{IV}(D)-\mu_{F4}(D) & = & R_{c4}\: P_{c4}\: I_{s}(D).
\end{eqnarray*}

Leaving out the factor $e$: 
\begin{eqnarray*}
-V_{nl} & = & \left[\mu_{F3}(+\infty)-\mu_{F3}(C)\right]+\left[\mu_{F3}(C)-\mu_{III}(C)\right]\\
 & + & \left[\mu_{III}(C)-\mu_{III}(O)\right]+\left[\mu_{IV}(O)-\mu_{IV}(D)\right]\\
 & + & \left[\mu_{IV}(D)-\mu_{F4}(D)\right]+\left[\mu_{F4}(D)-\mu_{F4}(+\infty)\right]\\
 & = & P_{F3}\:\Delta\mu_{F3}(C)-R_{c3}\: P_{c3}\: I_{s}(C)\\
 & - & P_{F4}\:\Delta\mu_{F4}(D)+R_{c4}\: P_{c4}\: I_{s}(D)
\end{eqnarray*}
so that:
\begin{eqnarray*}
-V_{nl} & = & \left[\frac{P_{F3}R_{F3}+P_{c3}R_{c3}}{R_{F3}}\:\Delta\mu_{F3}(C)\right]\\
 & + & \left[\frac{P_{F4}R_{F4}+P_{c4}R_{c4}}{R_{F4}}\:\Delta\mu_{F4}(D)\right]\\
 & = & \left[\frac{P_{F3}R_{F3}+P_{c3}R_{c3}}{R_{F3}+R_{c3}}\:\Delta\mu_{III}(C)\right]\\
 & - & \left[\frac{P_{F4}R_{F4}+P_{c4}R_{c4}}{R_{F4}+R_{c4}}\:\Delta\mu_{IV}(D)\right]
\end{eqnarray*}

The spin accumulations $\Delta\mu_{III}(C)$ and $\Delta\mu_{IV}(D)$
are derived easily from the spin accumulation vectors $\mathbf{K_{III}}$
and $\mathbf{K_{IV}}$ {[}Eq. (\ref{eq:k3}-\ref{eq:k4}){]}: 
\begin{eqnarray}
\Delta\mu_{III}(C) & = & \mathbf{K_{III}}\:.\:\left(\exp l_{3};\:\exp-l_{3}\right)\nonumber \\
 & = & \frac{\Delta\mu(O)}{\delta_{3}^{+}}\; X_{3}\label{eq:mu_c}
\end{eqnarray}
and:
\[
\Delta\mu_{IV}(D)=3\longleftrightarrow4
\]
One thus gets a voltage drop:
\begin{eqnarray}
V_{nl} & = & -\Delta\mu(O)\;\left(\frac{\widetilde{PR_{3}}}{\delta_{3}^{+}}-\frac{\widetilde{PR_{4}}}{\delta_{4}^{+}}\right)\\
 & = & -\Delta\mu(O)\;\left(P_{eff,3}-P_{eff,4}\right)
\end{eqnarray}

\subsubsection{Spin currents and spin accumulations at detector terminals\label{sub:spin_currents-detectors}.}

The spin current at point $C$ is given by:
\begin{equation}
I_{s}(C)=\frac{1}{R_{N}}\;\mathbf{K_{III}}\:.\:\left(\exp l_{3};\:-\exp-l_{3}\right)
\end{equation}
so that:
\begin{equation}
I_{s}(C)=-\frac{\Delta\mu(O)}{R_{N}\;\delta_{3}^{+}}=\frac{\left(P_{eff,1}-P_{eff,2}\right)\; R_{eff}\; I_{c}}{R_{N}\;\delta_{3}^{+}}.
\end{equation}
Let us study $\left|I_{s}(C)\right|$as a function of $X_{3}$ while
keeping all other parameters fixed. 
\begin{eqnarray*}
\left|I_{s}(C)\right| & \propto & \frac{1}{a\delta_{3}^{+}+\delta_{3}^{-}}\\
 & \propto & \frac{1}{2\left\{ X_{3}\,\left[a\,\cosh l_{3}+\sinh l_{3}\right]+\left[a\,\sinh l_{3}+\cosh l_{3}\right]\right\} }
\end{eqnarray*}
where: 
\[
a=\sum_{i\neq3}\; R_{eff,i}^{-1}>0.
\]
The spin current is therefore (in magnitude) a decreasing function
of $X_{3}$: in order to have large spin currents it is therefore
better to have transparent junctions rather than large tunnel barriers.

The spin accumulation at $C$ behaves in an opposite manner:
\begin{eqnarray*}
\left|\Delta\mu_{III}(C)\right| & = & \frac{\left|\Delta\mu(O)\right|}{\delta_{3}^{+}}\; X_{3}\\
 & \propto & \frac{X_{3}}{a\delta_{3}^{+}+\delta_{3}^{-}}\\
 & \propto & \frac{X_{3}}{X_{3}\,\left[a\,\cosh l_{3}+\sinh l_{3}\right]+\left[a\,\sinh l_{3}+\cosh l_{3}\right]}
\end{eqnarray*}
which is an increasing function of $X_{3}$. A large accumulation
at the interface with detector $F3$ is favored by large spin impedance
mismatch $X_{3}$.

The spin currents and accumulations at source and drain can be studied
in the same manner with identical conclusions: large $X$ leads to
larger spin accumulations but smaller spin currents; with an opposite
result for small $X$.

\subsection{Local resistance and non-local charge current\label{sub:Local-resistance-appendix}.}

The resistance can also be measured locally along the charge current
path from source to drain; if additionally we close the circuit between
the two detector electrodes a charge current is generated.

\subsubsection{Local resistance at injectors\label{sub:rlocal}.}

The voltage probes at source and drain are positioned at $z_{1}$
and $z_{2}$ which we will assume to be very large ($\gg l_{Fi}$
the spin relaxation lengths in the ferromagnets).

\begin{equation}
V_{local}=-\left[\mu_{F1}(z_{1})-\mu_{F2}(z_{2})\right].
\end{equation}

(we have left out the factor $e$).

We will need the dependence of the chemical potentials as well as
the interface discontinuities:

\begin{eqnarray}
\mu_{F1}(z) & =- & \rho_{F1}^{*}\,\left(1-P_{F1}^{2}\right)\: I_{c}\: z+C_{1}-P_{F1}\Delta\mu_{F1}(z)\\
\mu_{F2}(z') & = & \rho_{F2}^{*}\,\left(1-P_{F2}^{2}\right)\: I_{c}\: z'+C_{2}-P_{F2}\Delta\mu_{F2}(z')
\end{eqnarray}

where $C_{1-2}$ are constants and:

\begin{eqnarray}
\mu_{F1}(A)-\mu_{I}(A) & = & -R_{c1}\left(I_{c}+P_{c1}I_{s}(A)\right)\\
\mu_{F2}(B)-\mu_{II}(B) & = & R_{c2}\left(I_{c}-P_{c2}I_{s}(B)\right)
\end{eqnarray}
Since: 
\begin{eqnarray}
-V_{local} & = & \left[\mu_{F1}(z_{1})-\mu_{F1}(A)\right]+\left[\mu_{F1}(A)-\mu_{I}(A)\right]\nonumber \\
 & + & \left[\mu_{I}(A)-\mu_{II}(B)\right]+\left[\mu_{II}(B)-\mu_{F2}(B)\right]\nonumber \\
 & + & \left[\mu_{F2}(B)-\mu_{F2}(z_{2})\right]
\end{eqnarray}
eventually:
\begin{eqnarray}
-V_{local} & = & \left[\frac{P_{F1}R_{F1}+P_{c1}R_{c1}}{R_{F1}+R_{c1}}\:\Delta\mu_{I}(A)\right]\nonumber \\
 & - & \left[\frac{P_{F2}R_{F2}+P_{c2}R_{c2}}{R_{F2}+R_{c2}}\:\Delta\mu_{II}(B)\right]\nonumber \\
 & - & R_{0}\; I_{c}
\end{eqnarray}
where $R_{0}\left(\left\{ P_{1-2}\right\} \right)$ is an even function
of polarizations $\left\{ P_{1-2}\right\} =P_{Fi},\: P_{ci},\: i=1-2$:
\begin{eqnarray*}
R_{0}\left\{ P_{1-2}\right\}  & = & \sum_{i=1-2}\rho_{Fi}^{*}\,\left(1-P_{Fi}^{2}\right)\: z_{i}+\rho_{N}^{*}\: l_{i}\\
+ & R_{ci} & +\frac{\left[P_{Fi}^{2}R_{ci}R_{Fi}-P_{ci}^{2}R_{ci}^{2}-2P_{Fi}P_{ci}R_{Fi}R_{ci}\right]}{R_{Fi}+R_{ci}}.
\end{eqnarray*}
(It can be checked that $R_{0}\geq0$ as it should be.)

Since: 
\begin{equation}
\Delta\mu_{I}(A)=\mathbf{K_{I}}\:.\:\left(\exp l_{1};\:\exp-l_{1}\right)=\frac{\Delta\mu(O)}{\delta_{1}^{+}}\; X_{1}
\end{equation}
(with a similar expression for $\Delta\mu_{II}(B)$)
\begin{eqnarray}
-V_{local} & = & \Delta\mu(O)\:\left[P_{eff,1}-P_{eff,2}\right]-R_{0}I_{c}\nonumber \\
 & = & -R_{eff}\;\left[P_{eff,1}-P_{eff,2}\right]^{2}\; I_{c}-R_{0}I_{c}
\end{eqnarray}
The resistance variation when one flips terminal $F2$ is therefore:
\begin{eqnarray}
\Delta R_{local}= & R_{AP}-R_{P} & =\frac{V_{AP}-V_{P}}{I_{c}}\nonumber \\
= & 4\: P_{eff,1} & \: P_{eff,2}\: R_{eff}
\end{eqnarray}

\subsubsection{Local resistance at detectors and non-local charge current\label{sub:nonlocalcurrent}.}

The previous expressions will help us to derive the expression of
the non-local charge current flowing through the detectors $F3$ and
$F4$ when they are in closed circuit instead of being in open circuit.

Suppose one drives a current $I_{c}'$ through terminals $F3$ and
$F4$ but not through $F1$ and $F2$ exchanging the roles of injectors
and detectors ($I_{c}=0$); 
\begin{equation}
V_{34}=R_{eff}\:\left[P_{eff,3}-P_{eff,4}\right]^{2}\; I_{c}'+R_{0}\left\{ P_{3-4}\right\} \; I_{c}'
\end{equation}
We can define the local conductance at detectors: 
\begin{equation}
I_{c}'=g_{34}V_{34}
\end{equation}
 where 
\begin{equation}
g_{34}=\left[R_{eff}\:\left[P_{eff,3}-P_{eff,4}\right]^{2}+R_{0}\left\{ P_{3-4}\right\} \right]^{-1}.
\end{equation}
We now switch on the charge current $I_{c}$ at source and drain $F1$
and $F2$; when $I_{c}'=0$, 
\begin{eqnarray}
V_{34} & =V_{nl} & =R_{nl}\; I_{c}\nonumber \\
=R_{eff} & \;\left(P_{eff,1}-P_{eff,2}\right)\; & \left(P_{eff,3}-P_{eff,4}\right)\; I_{c}.
\end{eqnarray}
 When both $I_{c}$ and $I_{c}'$ are switched on, by superposition
one gets the general expression of the current flowing through detector
electrodes:
\begin{equation}
I_{c}'=g_{34}\left(V_{34}-R_{nl}\; I_{c}\right).
\end{equation}
 If we short $F3$ and $F4$, $V_{34}=0$ so that the non-local charge
current is:
\begin{equation}
I_{nl}=-g_{34}\; R_{nl}\; I_{c}
\end{equation}
or:
\begin{equation}
I_{nl}=-\frac{R_{eff}\;\left(P_{eff,1}-P_{eff,2}\right)\;\left(P_{eff,3}-P_{eff,4}\right)}{R_{eff}\:\left[P_{eff,3}-P_{eff,4}\right]^{2}+R_{0}\left\{ P_{3-4}\right\} }\; I_{c}
\end{equation}

\section{Revisiting the bipolar spin switch\label{sec:Revisiting-the-bipolar}.}

The original theory by Johnson of the bipolar spin switch was refined
by Hershfield and Zhao\cite{hershfield_charge_1997} who included
spin relaxation in the ferromagnets, and also by Fert and Lee\cite{fert_theory_1996}
who studied the (adverse) impact of interface spin flips. If we discard
the effect of spin flips, the non-local resistance variation was found
by Hershfield and Zhao to be:
\begin{equation}
\Delta R_{nl}=\frac{4R_{n}\;\widetilde{PR}\;\widetilde{PR}'}{\left[\left(X+1\right)\left(Y+1\right)\exp l-\left(X-1\right)\left(Y-1\right)\exp-l\right]}\label{eq:bipolar}
\end{equation}
where we have renamed parameters according to the definitions used
in this paper: $X$ and $\widetilde{PR}=\left(P_{c}R_{c}+P_{F}R_{F}\right)/R_{N}$
are defined at injector while $Y$ and $\widetilde{PR}'$ pertain
to the detector. Fert and Lee have the same expression in the case
$\widetilde{PR}=\widetilde{PR}'$and $X=Y$ (identical ferromagnets
at injector and detector). 

What is noteworthy is that this expression is actually identical to
that we found for the four ferromagnetic terminal cross geometry.
In the limit of large $X$ and $Y$ , $\Delta R_{nl}$ might therefore
be quite large as has been discussed repeatedly in this paper due
to spin confinement.

\begin{figure}
\begin{centering}
\includegraphics[width=1\columnwidth]{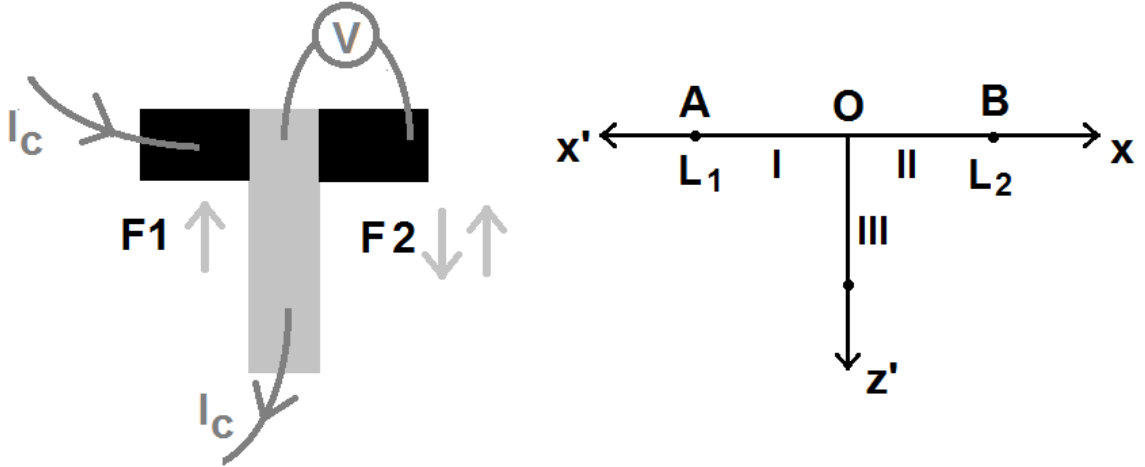}
\par\end{centering}

\caption{Left: geometry used by Hershfield-Zhao and Fert-Lee for the bipolar
spin switch calculations. Right: geometry in our simplified one-dimensional
treatment which includes spin leaking to the bottom.\label{fig:bipolar}}

\end{figure}

But a look at the geometry may cast doubt on this result (see Fig.
\ref{fig:bipolar} ): there should be some spin leakage in the setup
even if both injector and detector are in tunnel contacts with the
central paramagnet, because spin is leaving along with charge on the
bottom paramagnetic arm used as (charge) current drain. Hershfield
and Zhao argued that spin leakage could be neglected in the bottom
arm since in Johnson's original experiment the \emph{voltage} pads
are much larger than the spin relaxation length (the pads area was
$0.01\; mm^{2}\gg l_{N}^{2}$) but it is difficult to see how this
can be relevant for the bottom arm which is the \emph{current drain}.
The argument however applies to the spin leaking through the voltage
probes.

An expression which takes into account spin flow in the bottom arm
is actually easy to derive in line with the calculations done with
the cross geometry.

Referring to Fig. \ref{fig:bipolar} we consider a three arm geometry
(arms $I-III$) with two ferromagnetic terminals $F1$ and $F2$ with
current flowing from $F1$ to $III$ through $I$. We allow a finite
size for arms $I$ and $II$ ($L_{1}=l_{1}\; l_{N}$ and $L_{2}=l_{2}\; l_{N}$)
for a more general expression. We find:

\begin{equation}
\Delta R_{nl}=2R_{eff}\; P_{eff,1}\; P_{eff,2}
\end{equation}
where:
\begin{equation}
R_{eff}^{-1}=\left[\frac{1}{R_{N}'}+\frac{1}{R_{eff,1}}+\frac{1}{R_{eff,2}}\right]
\end{equation}
 The effective polarizations $P_{eff,i}$ for terminals $F1/F2$ and
spin resistances $R_{eff,i}$ for arms $I$ and $II$ are defined
as in the bulk of the paper (with $X_{1}=X$ and $X_{2}=Y$):
\begin{equation}
R_{eff}^{-1}=R_{N}^{-1}\;\left[\frac{R_{N}}{R_{N}'}+\frac{\delta_{1}^{-}}{\delta_{1}^{+}}+\frac{\delta_{2}^{-}}{\delta_{2}^{+}}\right]
\end{equation}

We have allowed for different cross sections $A_{N}$ and $A_{N}'$
along $x'Ox$ and $Oz'$ so that the spin resistance along $x'Ox$
is $R_{N}=\rho_{N}^{*}l_{N}/A_{N}$ but along $Oz'$ is 
\[
R_{N}'=\rho_{N}^{*}l_{N}/A_{N}'.
\]
The total spin resistance is easy to interpret: it corresponds to
parallel addition of the spin resistance of the three arms with spin
resistance $R_{N}'$ for arm $III$ and spin resistances $R_{eff,1}$
and $R_{eff,2}$ for arms $I$ and $II$ respectively.

Suppose we neglect arm $III$ spin resistance so that: 
\begin{equation}
R_{eff}^{-1}=\frac{1}{R_{eff,1}}+\frac{1}{R_{eff,2}}.
\end{equation}
One then recovers Fert-Lee and Hershfield-Zhao expression {[}Eq. (\ref{eq:bipolar}){]}
when $l_{1}=l_{2}=l/2$, which therefore does imply a neglect of spin
leakage to the bottom terminal.

When is it allowed to do so? The condition is: 
\begin{equation}
R_{N}'\gg(R_{eff,1},\; R_{eff,2})
\end{equation}
The effective resistance varies between $R_{N}$ and $R_{F}+R_{c}$;
the condition then reduces to:
\begin{equation}
R_{N}'\gg\left(R_{F1}+R_{c1},\; R_{F2}+R_{c2},\; R_{N}\right).\label{eq:condition1}
\end{equation}
which is favored by small cross-section $A_{N}'\ll A_{N}$ . Note
that the condition $R_{N}\ll R_{N}'$ is however not enough to recover
Eq. (\ref{eq:bipolar}).

\begin{widetext}

So more generally when the condition in Eq. (\ref{eq:condition1})
is not met, one gets (when $l_{1}=l_{2}=l/2$):
\begin{eqnarray}
\Delta R_{nl} & = & 8R_{n}\;\widetilde{PR}\;\widetilde{PR}'\nonumber \\
 & \times & \left\{ 2\left(X+1\right)\left(Y+1\right)\exp l-2\left(X-1\right)\left(Y-1\right)\exp-l\right.\nonumber \\
+ & \frac{R_{N}}{R_{N}'} & \left.\left[\left(X+1\right)\left(Y+1\right)\exp l+\left(X-1\right)\left(Y-1\right)\exp-l+2\left(XY-1\right)\right]\right\} ^{-1}\label{eq:bipolar2}
\end{eqnarray}

\end{widetext}which is significantly different from Eq. (\ref{eq:bipolar})
. Let's define the spin mismatch $Z=R_{N}'/R_{N}$. At small distance
where the signal will be largest: 
\[
\Delta R_{nl}\sim\frac{XY}{X^{-1}+Y^{-1}+Z^{-1}}.
\]

The scale will be set by the smaller of the spin impedance mismatches.
If anyone of them is in the transparent regime, one will recover the
scaling characteristic of open systems: $\Delta R_{nl}\leq R_{N}$
as it should be. 

We have neglected in the above spin leakage in the voltage probes
but it is easy to generalize the present considerations to include
it, should it become experimentally relevant.

\end{document}